\documentclass[prd, aps, amsfonts, amssymb, amsmath, showpacs, nofootinbib, superscriptaddress, preprintnumbers, onecolumn, notitlepage ]{revtex4-1}
\usepackage{graphicx, multirow,soul}
\usepackage{hyperref}
\usepackage[all]{hypcap}
\usepackage{url}
\usepackage{minibox}
\usepackage{color}
\usepackage{subfigure}
\usepackage{slashed}
\usepackage{float}
\usepackage{amssymb}
\usepackage{tikz}
\usepackage{comment}
\usepackage{todonotes}
\usepackage{multirow}
\usepackage{orcidlink}
\usetikzlibrary{arrows.meta}
\usepackage{tikz-feynman}
\graphicspath{{./figs/}}

\newcommand{\order}[1]{\mathcal{O}\left(\alpha_s^{#1}\right)}
\def\nn{\nonumber} 

\def\ito{\leftarrow} 

\definecolor{darkblue}{rgb}{0.0,0,0.5}

\begin{document}

\preprint{FERMILAB-PUB-23-322-T, MSUHEP-23-017}

\title{Improving ResBos for the precision needs of the LHC}

\author{Joshua Isaacson\,\orcidlink{0000-0001-6164-1707}}\email{isaacson@fnal.gov}
\affiliation{Theoretical Physics Department, Fermi National Accelerator Laboratory, P.O. Box 500, Batavia, IL 60510, USA.}
\author{Yao Fu}
\affiliation{Department of Modern Physics, University of Science and Technology of China, Jinzhai Road 96, Hefei, Anhui, 230026, China}
\author{C.-P. Yuan}
\affiliation{Department of Physics and Astronomy, Michigan State University, East Lansing, MI 48823, USA.}

\begin{abstract}
The resummation calculation (ResBos) is a widely used tool for the simulation of single vector boson production at colliders. In this work, we develop a significant improvement over the ResBos code by increasing the accuracy from NNLL+NLO to N${}^3$LL+NNLO and release the ResBos v2.0 code. Furthermore, we propose a new non-perturbative function that includes information about the rapidity of the system (IFY). The IFY functional form was fitted to data from fixed target experiments, the Tevatron, and the LHC. We find that the non-perturbative function has mild rapidity dependence based on the results of the fit. Finally, we investigate the effects that this increased precision has on the measurement of the $W$ boson by CDF and impacts on future LHC measurements.
\end{abstract}

\maketitle

\section{Introduction} \label{sec:intro}

At the collider experiments, the $W$ and $Z$ boson channels are considered standard candle processes.
Specifically, the lepton pair production cross sections are among the most precise observables
measurable at the Large Hadron Collider (LHC)~\cite{ATLAS:2016fij,CMS:2019raw}.
Additionally, they appear as backgrounds to
all beyond the Standard Model (BSM) searches. Therefore, it is of vital importance that the theoretical calculations
for these processes are calculated to the highest precision possible. In the case of the
total inclusive rate, these processes have been calculated to N${}^3$LO accuracy in
Quantum Chromodynamics (QCD)~\cite{Duhr:2020seh, Duhr:2020sdp, Duhr:2021vwj},
NLO accuracy in the Electroweak theory 
(EW)~\cite{CarloniCalame:2007cd,Dittmaier:2009cr,Barze:2013fru},
and the first order for the mixed QCD-EW corrections~\cite{Dittmaier:2015rxo,Bonciani:2019nuy,
Dittmaier:2020vra,Bonciani:2021zzf,Armadillo:2022bgm,Buccioni:2022kgy}.

In the case of differential distributions, the calculations are starting to be completed
at N${}^3$LO for the rapidity and transverse mass distributions in Ref.~\cite{Chen:2022cgv} and
the transverse momentum of the vector boson and the lepton in Ref.~\cite{Chen:2022cgv}.
Furthermore, there was an investigation of the effects of higher order corrections to
triple-differential Drell-Yan observables in Ref.~\cite{Gehrmann-DeRidder:2023urf}.
Care has to be taken
when calculating observables sensitive to low transverse momentum $W$ and $Z$ bosons.
In this region, each order in the fixed-order calculation diverges in the limit that the
transverse momentum goes to zero. However, it was first demonstrated by Collins, Soper, and
Sterman~\cite{Collins:1984kg} that the form of these terms forms a series that can be 
formally summed. This approach is known as transverse momentum resummation. The current
state-of-the-art resummation calculation is at N${}^4$LL${}_p$+N${}^3$LO~\cite{Neumann:2022lft},
with most other codes at N${}^3$LL'+N${}^3$LO 
accuracy~\cite{Chen:2022cgv,Bizon:2018foh,Bizon:2019zgf,Re:2021con}.

In this work, we present the next version of the ResBos 
program~\cite{Balazs:1997xd,Landry:2002ix}.
The ResBos version 2 (henceforth ResBos2) code was developed to address
some major concerns from the original
ResBos code~\cite{Isaacson:2017hgb}.
One such concern was the precision of the code. Previously, the accuracy was
only at NNLL+NLO with an approximate correction to NNLL+NNLO neglecting the corrections to
the angular functions. The ResBos2 code is at N${}^3$LL+NNLO accuracy, including the correct
angular functions. These concerns were some of the major theory criticisms levied against
the recent CDF $W$ mass measurement. The CDF experiment measured the $W$ mass as
80,433 $\pm$ 9 MeV~\cite{CDF:2022hxs}, which is the most precise direct measurement.
This result disagrees with the Standard Model (SM) electroweak global fit result of
80,359.1 $\pm$ 5.2 MeV~\cite{deBlas:2021wap}.
The impact of these updates are evaluated in Ref.~\cite{Isaacson:2022rts}.

The rest of this paper is organized as follows. Firstly, we review the two different
$b$-space resummation formalisms implemented in the ResBos2 code in Sec.~\ref{sec:resummation}.
Section~\ref{sec:improvements} discuss the improvements implemented into the ResBos2 code
over the previous version of the code. We investigate the rapidity dependence of the
non-perturbative Sudakov factor in Sec.~\ref{sec:non-pert}. The comparison LHC data are given 
for the $Z$ boson in Sec.~\ref{sec:z_obs} and results for the $W$ boson mass at the Tevatron
and the LHC in Sec.~\ref{sec:w_mass}. Finally, conclusions and a future outlook are given in 
Sec.~\ref{sec:conclusions}

% Discussion on Resummation and its importance
\section{Resummation Formalisms} \label{sec:resummation}

When performing fixed-order calculations, the calculation is organized by the power of $\alpha_s$. Fixed-order calculations make sense when each term in the series is smaller than the previous term. However, there are certain phase space points that result in each subsequent term being larger than the previous one, causing the breakdown of the fixed-order calculation. To resolve this, resummation is introduced. Resummation reorganizes that calculation by noticing that there are certain terms that appear at every order in $\alpha_s$~\cite{Collins:1984kg}. These terms that appear in a specific form at each order are logarithms of two scales, e.g. $\log\left(\frac{Q^2}{p_T^2}\right)$. The number of logarithmic terms included in the calculation is denoted by leading log for having only the leading term, and adding next-to for each additional log term included. The organization into different orders of precision are summarized in Tab.~\ref{tab:ResumOrders}.
\begin{table}[!b]
    \begin{center}
    \begin{tabular}{| c | c | c c | c |}
    \hline
    & & \multicolumn{2}{c |}{Anomalous Dimension} & \\
    Order & Boundary Condition ($C$) & $\gamma_i$ (non-cusp, $B$) & $\Gamma_{cusp}, \beta$ ($A$) & Fixed Order Matching ($Y$) \\
    \hline
    LL & 1 & - & 1-loop & - \\
    NLL & 1 & 1-loop & 2-loop & - \\
    NLL' (+ NLO) & $\alpha_s$ & 1-loop & 2-loop & $\alpha_s$ \\
    NNLL (+ NLO) & $\alpha_s$ & 2-loop & 3-loop & $\alpha_s$ \\
    NNLL' (+ NNLO) & $\alpha_s^2$ & 2-loop & 3-loop & $\alpha_s^2$ \\
    N${}^3$LL (+ NNLO) & $\alpha_s^2$ & 3-loop & 4-loop & $\alpha_s^2$ \\
    N${}^3$LL' (+ N${}^3$LO) & $\alpha_s^3$ & 3-loop & 4-loop & $\alpha_s^3$ \\
    N${}^4$LL (+ N${}^3$LO) & $\alpha_s^3$ & 4-loop & 5-loop & $\alpha_s^3$ \\
    \hline
    \end{tabular}
    \caption{The different components needed for different orders of resummation.}
    \label{tab:ResumOrders}
    \end{center}
\end{table}

The dynamics of multiple soft-gluon radiation in scattering processes is treated through the use of the resummation formalism ~\cite{Sterman:1986aj,Catani:1989ne,Catani:1990rp,Catani:1992ua,Berger:1993yp}. There are many applications of resummation at modern colliders. In this work, the focus will be on the treatment of transverse momentum resummation. The formalism was originally shown to be possible for all the large logarithms (leading and subleading) to all orders by Collins, Soper, and Sterman~\cite{Collins:1984kg}. The formalism developed in their work will be referred to as the CSS Formalism. A more recent formalism was developed by Catani, de Florian, and Grazzini, which is known as the CFG Formalism~\cite{Catani:2013tia}. The details of the two formalisms are explained in Sec.~\ref{subsubsec:CSS} and Sec.~\ref{subsubsec:CFG} respectively. The differences between the two formalisms are highlighted in Sec.~\ref{subsubsec:CSSvCFG}. The remainder of this section will focus on the general outline of $p_T$ resummation.

\begin{figure}
    \begin{center}
        \begin{tikzpicture}
            \begin{feynman}
                \vertex at (0, 0) (f);
                \vertex at (2, 0) (c);
                \vertex at (4, 0) (d);
                \vertex at (0, 2) (a);
                \vertex at (0, -2) (b);
                \vertex at (-2, -2) (i2);
                \vertex at (-2, 2) (i1);
            
    \diagram*{
        (c) -- [photon] (d),
        (i1) -- [fermion] (a) [blob, label=\(C\)] -- [fermion] (c) [blob, label=\(H\)] -- [fermion] (b) [blob, label=\(C\)] -- [fermion] (i2) ,
        (i1) -- [draw=none] (f) [blob, label=\(S\)],
    }; 
    \node[circle, fill=black, label={$C$}, minimum size=0.75cm] at (a) {};
    \node[circle, fill=black, label={$C$}, minimum size=0.75cm] at (b) {};
    \node[circle, fill=black, label={$H$}, minimum size=0.75cm] at (c) {};
    \node[circle, fill=black, label={$S$}, minimum size=0.75cm] at (f) {};
    \end{feynman}
\end{tikzpicture}
    \end{center}
    \caption{A diagrammatic representation of the factorized cross-section for Drell-Yan, broken into a soft, collinear, and hard factor. The soft factor is labeled by the $S$, the collinear factors are labeled by the $C$'s, and the hard factor is labeled by the $H$}
    \label{fig: FactorizationDY}
\end{figure}
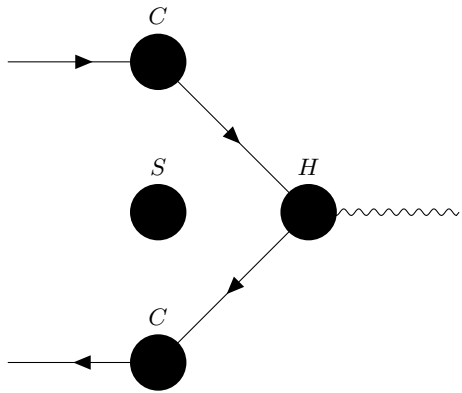
Firstly, resummation is a means to relate the different scales of a multi-scale process to a single scale, which also removes the large logarithms that result from the large difference between the scales. Therefore, the first step is to factorize the cross-section calculation into the different scale regions that are involved in the calculation. The regions that are important to this work are known as the hard factor, the soft factor, and the collinear or jet factors. A diagrammatic representation of each piece for the Drell-Yan process can be seen in Fig.~\ref{fig: FactorizationDY}, and can be expressed as:

\begin{equation}
    \frac{d\sigma}{dQ^2dydp_T^2}\propto H(\mu,\mu_R)S(\mu,\mu_{\rm Res})C_1(\mu,\mu_F)C_2(\mu,\mu_F)J(\mu,R\mu')\,,
\end{equation}
where $H$ is the hard factor, $S$ is the soft factor, $C_1$ and $C_2$ are the collinear factors for each incoming hadron, and $J$ is the jet factor. In this work, we are inclusive on the number of jets so the factor $J(\mu,R\mu') = 1$. However, it plays an important role for example in Higgs+jet resummation~\cite{Sun:2016kkh}. Additionally, the various scales are given as the hard scale $\mu$, the renormalization scale $\mu_R$, the resummation scale $\mu_{\rm Res}$, the factorization scale $\mu_F$, and the jet radius ($R$) scaled by the jet scale $\mu'$.

The remainder of the section is to discuss the calculation of the soft factor, and derive the well known Sudakov factor. Starting from the fixed-order calculation up to the $n^{\text{th}}$ order in $\alpha_s$, the result can be split into a singular piece, and a regular piece. The singular piece are terms that are proportional to $\frac{1}{p_T^2}\log^m\left(\frac{Q^2}{p_T^2}\right)$ ($m=0,1,...,2n-1$) and $\delta(p_T)$, and the regular terms are less singular than those previously mentioned. This calculation breaks down when $\alpha_s^n \frac{1}{p_T^2}\log^m\left(\frac{Q^2}{p_T^2}\right)$ becomes large. 

To resolve this issue, the logarithms need to be summed to all orders to obtain a finite result in the limit $p_T\rightarrow 0$, and remove all large logarithms from the final result. In order to perform the resummation correctly, the cross-section needs to be Fourier transformed into impact parameter ($b$) space. In the impact parameter space, the total transverse momentum is explicitly conserved~\cite{Parisi:1979se}. After the Fourier transform, the cross section can be expressed as:
\begin{equation}
    \frac{d\sigma}{dQ^2dp_T^2dy}=\frac{1}{\left(2\pi\right)^2}\int d^2b e^{i\vec{q}_T\cdot \vec{b}} \tilde{W}\left(b,Q,x_1,x_2\right) + Y(p_T,Q,x_1,x_2),
\end{equation}
where $\tilde{W}$ contains the resummation of the singular pieces of the cross section, and $Y$ contains the regular pieces of the cross section defined by taking the fixed-order calculation and subtracting the corresponding asymptotic piece. The asymptotic piece contains the terms that are at least as singular as $\frac{1}{p_T^2}$ in the fixed-order calculation in the limit $p_T\rightarrow 0$. 

By studying the form of the singular piece, the $x_1$ and $x_2$ dependence in $\tilde{W}$ can be factorized into:
\begin{equation}
    \tilde{W}\left(b,Q,x_1,x_2\right)=\sum_j C_j\left(b,Q,x_1\right)C_j\left(b,Q,x_2\right)\tilde{W}\left(b,Q\right),
\end{equation}
where $C_j$ is a convolution of the PDFs with a collinear Wilson coefficient, with the convolution defined as:
\begin{equation}
    C_j=\sum_a \int_x^1 \frac{dz}{z}C_{ja}\left(\frac{x}{z},b,\mu,Q\right)f_a\left(z,\mu\right),
\end{equation}
where $C_{ja}$ is the Wilson coefficient, $f_a$ is the PDF, the sum $a$ runs over all incoming partons, and $j$ represents the parton that enters into the hard cross section calculation. These functions are the collinear factors as previously mentioned. The remaining term contains the hard factor, and the soft factors. 

$\tilde{W}$ is determined by solving the evolution equation~\cite{Davies:1984hs}:
\begin{equation}
    \label{eq: WEvolve}
    \frac{\partial}{\partial\log Q^2}\tilde{W}\left(Q,b\right)=\left[K\left(b\mu,g_s\left(\mu\right)\right)+G\left(Q/\mu,g_s\left(\mu\right)\right)\right]\tilde{W}\left(Q,b\right),
\end{equation}
where $K\left(b\mu,g_s\left(\mu\right)\right)$ and $G\left(Q/\mu,g_s\left(\mu\right)\right)$ satisfy the renormalization group equations (RGEs),
\begin{align}
    \frac{d}{d\log\mu}K\left(b\mu,g_s\left(\mu\right)\right) &= -\gamma_K\left(g_s\left(\mu\right)\right), \\
    \frac{d}{d\log\mu}G\left(b/\mu,g_s\left(\mu\right)\right) &= \gamma_K\left(g_s\left(\mu\right)\right), 
\end{align}
where $\gamma_K$ is the anomalous dimension, calculated from the singular terms of the cross section~\cite{KORCHEMSKY1987342,Bern:2005iz,Henn:2019swt,vonManteuffel:2020vjv}. 
Through the RGE equations, $K\left(b\mu,g_s\left(\mu\right)\right)$ and $G\left(b/\mu,g_s\left(\mu\right)\right)$ can be evolved independently to scales of order $1/b$ and Q, respectively, removing all large logarithms from the calculation. After solving these equations, the $A$ and $B$ functions can be defined such that Eq.~\eqref{eq: WEvolve} can be rewritten as:
\begin{align}
    &\frac{\partial}{\partial\log Q^2}\tilde{W}\left(Q,b\right)=\\&-\left(\int_{C_1^2/b^2}^{C_2^2Q^2} \frac{d\mu^2}{\mu^2}\left(A\left(g_s\left(\mu\right),C_1\right)\log\frac{C_2^2Q^2}{\mu^2}+B\left(g_s\left(\mu\right), C_1, C_2\right)\right)\right) \tilde{W}\left(Q,b\right), \nonumber
\end{align}
where $C_1$ and $C_2$ are arbitrary constants of integration arising from solving the RGEs. It is possible to calculate the values for the $A$ and $B$ functions order by order in perturbation theory.

Finally, to obtain a result that can be used to make predictions of the cross section, the evolution equation of $\tilde{W}$ needs to be solved. The solution can be written as
\begin{equation}
    \tilde{W}\left(Q,b\right) = e^{-\mathcal{S}\left(Q,b\right)}\tilde{W}\left(\frac{C_1}{C_2b},b\right),
\end{equation}
where $\mathcal{S}$ is known as the Sudakov factor, and is given by
\begin{equation}
    \mathcal{S}\left(Q,b\right) = \int_{C_1^2/b^2}^{C_2^2Q^2} \frac{d\mu^2}{\mu^2}\left(A\left(g_s\left(\mu\right),C_1\right)\log\frac{C_2^2Q^2}{\mu^2}+B\left(g_s\left(\mu\right), C_1, C_2\right)\right).
\end{equation}
Putting all of the results above together, the resummed cross section can by written as:
\begin{align}
    \frac{d\sigma}{dQ^2dp_T^2dy} &= \frac{H}{\left(2\pi\right)^2}\int d^2b e^{i\vec{q}_T\cdot \vec{b}}e^{-\mathcal{S}\left(Q,b\right)} \\ \times &\sum_jC_j\left(\frac{C_1}{C_2b},Q,x_1\right)C_j\left(\frac{C_1}{C_2b},Q,x_2\right) + Y(p_T,Q,x_1,x_2). \nonumber
\end{align}

This is the general form for transverse momentum resummation. However, this form is not the final form used in calculations, due the fact that when the impact parameter becomes large, the scale of resummation goes below $\Lambda_{QCD}$. Therefore the calculation becomes non-perturbative. To prevent using a scale below $\Lambda_{QCD}$, the $b^{*}$ prescription is introduced, where:
\begin{equation}
    b^*=\frac{b}{\sqrt{1+\frac{b^2}{b_{max}^2}}},
\end{equation}
where $b_{max}$ is chosen such that $1/b_{max}$ is of order $\Lambda_{QCD}$. The lower bound of the Sudakov integral is then modified from $C_1^2/b^2$ to $C_1^2/b_*^2$. This functional form prevents $b^*$ from ever being large than $b_{max}$, preventing scales below $\Lambda_{QCD}$. However, this causes the prediction to be inaccurate at low $p_T$, since a piece is removed by the $b^*$ prescription. To resolve this, a non-perturbative function needs to be introduced.

There are many different proposals for the form of the non-perturbative function~\cite{Landry:2002ix,Konychev:2005iy,Qiu:2000hf,Korchemsky:1994is}. In this section, the general concepts of the non-perturbative function will be covered. The method of obtaining this function is through fits to data. It is believed that the non-perturbative function should be universal, and only depend on the color structure of the initial states. This then gives the final form of the resummation formalism in a scheme independent way as:
\begin{equation}
    \frac{d\sigma}{dQ^2dp_T^2dy}=\sum_{i,j}\frac{H}{(2\pi)^2}\int d^2b e^{i\vec{p_T}\cdot\vec{b}}e^{-S_{\text{pert}}}e^{-S_{NP}}C\otimes f_i C\otimes f_j,
\end{equation}
where $S_{\text{pert}}$ is the Sudakov factor, while $S_{NP}$ is the non-perturbative Sudakov factor. Finally, up to this point the integration coefficients ($C_1,C_2$, and $C_3$) were left to be arbitrary. The canonical choice for these scales are given by $C_1=b_0$, $C_2=1$, and $C_3=b_0$, where $b_0=2e^{-\gamma_E}$. In Appendix~\ref{app:ScaleVariation}, the relationship between the canonical scale choice and any arbitrary choice is calculated up through $\alpha_s^3$. The theory uncertainty due to the missing higher order corrections can be estimated by modifying the values of $C_1, C_2$, and $C_3$.

\subsection{Collins-Soper-Sterman Formalism}
\label{subsubsec:CSS}

So far, the resummation formalism has been developed in a resummation scheme independent way. Here, the Collins-Soper-Sterman Formalism is introduced~\cite{Collins:1984kg}. In this formalism, the hard matrix element, $H$, is taken to be 1, with no corrections as a function of $\alpha_s$, and the $B$ and $C$ coefficients become process dependent. The $A$, $B$, and $C$ coefficients can be expanded as a series in $\alpha_s$ as:
\begin{align}
    A=\sum_{n=1}^\infty \left(\frac{\alpha_s}{\pi}\right)^n A^{(n)}, \\
    B=\sum_{n=1}^\infty \left(\frac{\alpha_s}{\pi}\right)^n B^{(n)}, \\
    C_{ij}=\delta_{ij}+\sum_{n=1}^\infty \left(\frac{\alpha_s}{\pi}\right)^n C_{ij}^{(n)}.
\end{align} 
For Drell-Yan, the coefficients for $A$ up to $\alpha_s^3$, $B$ up to $\alpha_s^2$ and $C$ up to $\alpha_s$ are given with the canonical scale choice as~\cite{KODAIRA198266,CATANI1988335,Zhu:2012ts,deFlorian:2000pr,deFlorian:2001zd,Becher:2010tm}:   
\begin{align}
    A^{(1)} &= C_F, \\
    A^{(2)} &= \frac{1}{2}C_F\left(\left(\frac{67}{18}-\frac{\pi^2}{6}\right)C_A-\frac{5}{9}N_f\right), \\
    A^{(3)} &= C_F\left(\frac{C_F N_f}{2}\left(\zeta_3-\frac{55}{48}\right)-\frac{N_f^2}{108}\right.\nonumber\\ 
    &\left.+C_A^2\left(\frac{11\zeta_3}{24}+\frac{11\pi^4}{720}-\frac{67\pi^2}{216}+\frac{245}{96}\right) \right.\nonumber \\
    &\left.+C_A N_f\left(\frac{-7\zeta_3}{12}+\frac{5\pi^2}{108}-\frac{209}{432}\right)\right), \\
    B^{(1)} &= -\frac{3}{2}C_F, \\
    B^{(2)} &= C_F^2\left(\frac{\pi^2}{4}-\frac{3}{16}-3\zeta_3\right)+C_F C_A\left(\frac{11}{36}\pi^2-\frac{193}{48}+\frac{3}{2}\zeta_3\right)\nonumber \\
    &+C_F N_f\left(\frac{17}{24}-\frac{\pi^2}{18}\right), \\
    C_{qq}^{(1)}(z) &= \frac{1}{2}C_F(1-z) + \delta(1-z)\frac{1}{4}C_F\left(\pi^2-8\right),\\
    C_{qg}^{(1)}(z) &= \frac{1}{2}z(1-z), \\
    C_{q\bar{q}}^{(1)}(z) &= C_{qq'}^{(1)}(z) = C_{q\bar{q}'}^{(1)}(z) = 0,
\end{align}
where $C_F=4/3$, $C_A=3$, and $N_f$ is the number of active quarks.
The results for $B^{(3)}$ can be found in Ref.~\cite{Li:2016ctv}, and for $C^{(2)}$ can be found in Ref.~\cite{Catani:2012qa}.

\subsection{Catani-deFlorian-Grazzini Formalism}
\label{subsubsec:CFG}

Catani, deFlorian, and Grazzini realized that the behavior of soft gluons is independent of the hard process, and developed a resummation formalism in which the hard factor which is process dependent can be pulled out of the Fourier transform~\cite{Catani:2000vq}. This then leads to the calculation in impact parameter space only depending on the initial state partons, and not the hard factor. Like in CSS, the $A$, $B$, and $C$ functions can be expanded as a series in $\alpha_s$. However, in addition to these three, the hard factor $H$ is not fixed to one, but can also be expanded as a series in $\alpha_s$. In the CFG formalism, the $A$, $B$, and $C$ coefficients are given by:
\begin{align}
    A^{(1)} &= C_F, \\
    A^{(2)} &= \frac{1}{2}C_F\left(\left(\frac{67}{18}-\frac{\pi^2}{6}\right)C_A-\frac{5}{9}N_f\right), \\
    A^{(3)} &= C_F\left(\frac{C_F N_f}{2}\left(\zeta_3-\frac{55}{48}\right)-\frac{N_f^2}{108}\right.\nonumber\\ 
    &\left.+C_A^2\left(\frac{11\zeta_3}{24}+\frac{11\pi^4}{720}-\frac{67\pi^2}{216}+\frac{245}{96}\right) \right.\nonumber \\
    &\left.+C_A N_f\left(\frac{-7\zeta_3}{12}+\frac{5\pi^2}{108}-\frac{209}{432}\right)\right), \\
    B^{(1)} &= -\frac{3}{2}C_F, \\
    B^{(2)} &= \left(\left(-3+24\zeta_2-48\zeta_3\right)C_F^2+\left(\frac{-17}{3}-\frac{88}{3}\zeta_2+24\zeta_3\right)C_F C_A \right. \nonumber \\
    &\left.+\left(\frac{2}{3}+\frac{16}{3}\zeta_2\right)C_F N_f\right)/16+C_F\beta_0\zeta_2, \\
    C_{qq}^{(1)}(z) &= \frac{1}{2}C_F(1-z), \\
    C_{gq}^{(1)}(z) &= \frac{1}{2}C_F z, \\
    C_{qg}^{(1)}(z) &= \frac{1}{2}z(1-z), \\
    C_{q\bar{q}}^{(1)}(z) &= C_{qq'}^{(1)}(z) = C_{q\bar{q}'}^{(1)}(z) = 0.
\end{align}

The relationship for obtaining the $A$, $B$, and $C$ coefficients in the CFG formalism from the coefficients in the CSS formalism can be found in Sec.~\ref{subsubsec:CSSvCFG}. The hard factor is process dependent, and for Drell-Yan are given as:
\begin{align}
    H^{DY(1)}&=C_F\left(\frac{\pi^{2}}{2}-4\right), \\
    H^{DY(2)}&=C_F C_A\left(\frac{59\zeta_3}{18}-\frac{1535}{192}+\frac{215\pi^2}{216}-\frac{\pi^4}{240}\right)\nonumber \\
    &+\frac{1}{4}C_F^2\left(-15\zeta_3+\frac{511}{16}-\frac{67\pi^2}{12}+\frac{17\pi^4}{45}\right) \nonumber \\
    &+\frac{1}{864}C_F N_f\left(192\zeta_3+1143-152\pi^2\right),
\end{align}
up to $\mathcal{O}\left(\alpha_s^2\right)$ ~\cite{Catani:2013tia}.

\subsection{Comparison of CSS to CFG}
\label{subsubsec:CSSvCFG}

The conversion between the CSS and CFG Formalisms can be given using the all orders relations~\cite{Catani:2000vq}:
\begin{align}
    C^F_{ab}(z)&=\left[H_a^F\right]^\frac{1}{2}C_{ab}(z), \\
    B^F_{c}&=B_c-\beta \frac{d\ln H_c^F}{d\ln\alpha_s},
\end{align}
where the $F$ superscript is used to indicate which pieces are process dependent, and $\beta$ is the function that describes the running of $\alpha_s$. This can be expanded order by order to give a conversion between explicit CSS and CFG coefficients. It is important to note that the $A$ coefficients are always universal, and $B^{(1)}$ is also universal (only depends on the color structure of the initial state). The conversions up to N$^3$LL resummation are listed below:
\begin{align}
    \label{eq: CSStoCFG}
    C^{(1)F}_{ab}(z)&=C_{ab}^{(1)}(z)+\delta_{ab}\delta(1-z)\frac{1}{2}H_{a}^{(1)F}, \\
    C^{(2)F}_{ab}(z)&=C_{ab}^{(2)}(z)+\frac{1}{2}H_a^{(1)F}C_{ab}^{(1)}(z)\nonumber\\
    &+\delta_{ab}\delta(1-z)\frac{1}{2}\left(H_a^{(2)F}-\frac{1}{4}\left(H_a^{(1)F}\right)^2\right), \\
    B^{(2)F}_c&=B^{(2)}_c+\beta_0H_c^{(1)F}, \\
    B^{(3)F}_c&=B^{(3)}_c+\beta_1H_c^{(1)F}+2\beta_0\left(H_a^{(2)F}-\frac{1}{2}\left(H_a^{(1)F}\right)^2\right),
\end{align}
with $\beta_0=\frac{11C_A-2N_f}{12}$ and $\beta_1=\frac{17C_A^2-5C_AN_f-3C_FN_f}{24}$.

\section{ResBos2 Improvements}\label{sec:improvements}

In this section, the different physics improvements implemented into ResBos2 are
discussed below. For technical code improvements see App.~\ref{app:computational}. These
improvements include the changes to N${}^3$LL+NNLO accuracy and the inclusion of the 
NNLO accurate angular distributions. Both of these issues were a major concern in the usage of the
ResBos prediction for the CDF $W$ mass measurement (see Sec.~\ref{sec:w_mass} and Ref.~\cite{Isaacson:2022rts} for more details).

\subsection{Resummation at \texorpdfstring{N${}^3$LL}{N3LL} Matched to NNLO}

In order to match the resummed calculation to the fixed order calculation, the asymptotic expansion needs to be calculated
to the same order as the perturbative calculation. Additionally, a matching procedure needs to be defined to manipulate the transition from
the region described by the resummed calculation (small transverse momentum) to the region described by the fixed order calculation
(large transverse momentum). 

The asymptotic expansion can be calculated using two different methods. Firstly, since it should reproduce the singular structure
of the perturbative calculation, one can take the transverse momentum to zero limit of the fixed order calculation and keep terms
that are more divergent than $1/p_T$. The other approach is to take the resummed calculation and expand it to a fixed order in the
strong coupling constant. The two calculations should be identical and is a good validation of the calculations. In this work,
we analytically expand the resummed calculation to $\mathcal{O}\left(\alpha_s^3\right)$ to prepare for matching to an N${}^3$LO
prediction. Furthermore, we numerically validate that the asymptotic expansion and the perturbative calculation agree in the limit
of small transverse momentum. The expansion of the $A$, $B$, $C$, and $H$ coefficients to $\order{n}$ can be explicitly found up to 
$\order{3}$ in Section~\ref{subsubsec:CSS}, Section~\ref{subsubsec:CFG}, and Appendix~\ref{app:B2C3} for the CSS and CFG formalisms. The
expansion of both the CSS and CFG formalism result in the same singular and asymptotic piece, so it is sufficient to only consider the
CSS formalism. The lepton variables and angle between $\vec{b}$ and $\vec{q}_T$ are integrated out to simplify the discussion, but do
not modify the results. After these simplifications, the resummation formalism becomes: 
\begin{align}
\lim_{p_T\rightarrow 0} \frac{d\sigma}{dQ^2dydp_T^2} &\propto \frac{1}{2\pi p_T^2}\int_0^\infty d\eta \eta J_0\left(\eta\right) e^{-S\left(\eta/p_T,Q\right)} \nonumber \\
&\times C\otimes f_{j}\left(x_1,p_T^2/\eta^2\right)C\otimes f_{\bar{k}}\left(x_2,p_T^2/\eta^2\right)+ (j\leftrightarrow \bar{k}),
\end{align}
where terms that are not of importance in the derivation have been dropped, and terms that are less singular than $\frac{1}{p_T^2}$
or $\delta(p_T)$ have also been dropped.  The asymptotic piece is obtained by integrating over $\eta=bp_T$. Additional details can be found in App.~\ref{app:bessel_int}.

Calculating the results to $\order{}$ is fairly straightforward. However, the results at higher orders quickly become untractable.
Therefore, it is useful to introduce the following definition,
\begin{equation}
\frac{d\sigma}{dQ^2dydp_T^2}=\frac{\sigma_0}{S}\frac{1}{2\pi p_T^2}\sum_{i,j}\sum_{n=1}^{\infty}\sum_{m=0}^{2n-1}\left(\frac{\alpha_s(\mu^2)}{\pi}\right)^n {}_{n}C^{(i,j)}_{m}\ln^m\left(\frac{Q^2}{p_T^2}\right),
\end{equation}
which becomes very useful for organization beyond $\order{}$. The definition above differs from that found in 
Ref.~\cite{Arnold1991381} by expanding in factors of $\frac{\alpha_s}{\pi}$ instead of $\frac{\alpha_s}{2\pi}$, and the overall factor
for the $\frac{1}{p_T^2}$ term is $\frac{1}{2\pi}$ instead of $\frac{1}{\pi}$. Using these definitions the coefficients up to
$\order{2}$ are given as:
\begin{align*}
{}_{1}C^{(i,j)}_{1} &= 2A^{(1)}f_if_j,  \\
{}_{1}C^{(i,j)}_{0} &= 2B^{(1)}f_if_j+\left[f_j\left(P_{i\leftarrow b} \otimes f_b\right)f_i\left(P_{j\leftarrow a} \otimes f_a\right)\right], \\
{}_{2}C^{(i,j)}_{3} & = -2\left(A^{(1)}\right)^2f_if_j,\\
{}_{2}C^{(i,j)}_{2} & = \left(-6A^{(1)}B^{(1)}+2A^{(1)}\beta_0\right)f_if_j-3A^{(1)}\left(f_j\left(P_{i\leftarrow b} \otimes f_b\right)f_i\left(P_{j\leftarrow a} \otimes f_a\right)\right),\\
{}_{2}C^{(i,j)}_{1} & = \left(A^{(1)}\beta_0\ln\frac{\mu_R^2}{Q^2}+2A^{(2)}-2\left(B^{(1)}\right)^2+B^{(1)}\beta_0\right)f_if_j+4A^{(1)}\left(C^{(1)}\otimes f_i\right) f_j \\
&-2A^{(1)}\left(P^{(1)}\otimes f_i\right) f_j \ln\frac{\mu_F^2}{Q^2}-4B^{(1)}\left(P^{(1)}\otimes f_i\right) f_j \\
&+\beta_0\left(P^{(1)}\otimes f_i\right) f_j-\left(P^{(1)}\otimes P^{(1)}\otimes f_i\right) f_j - \left(P^{(1)}\otimes f_i\right)\left( P^{(1)}\otimes f_j\right) + i\leftrightarrow j,\\
{}_{2}C^{(i,j)}_{0} & = \left(4\left(A^{(1)}\right)^2\zeta(3)+B^{(1)}\beta_0\ln\frac{\mu_R^2}{Q^2}+2B^{(2)}\right)f_if_j \\
&+B^{(1)}\left(4\left(C^{(1)}\otimes f_i\right) f_j-2\left(P^{(1)}\otimes f_i\right)f_j\ln\frac{\mu_F^2}{Q^2}\right) \\
&-\beta_0\left(4\left(C^{(1)}\otimes f_i\right) f_j-2\left(P^{(1)}\otimes f_i\right)f_j\ln\frac{\mu_R^2}{Q^2}\right)\\
&+2\left(C^{(1)}\otimes P^{(1)}\otimes f_i\right) f_j+2\left(C^{(1)}\otimes f_i\right) \left(P^{(1})\otimes f_j\right)-\left(P^{(1)}\otimes P^{(1)}\otimes f_i\right) f_j\ln\frac{\mu_F^2}{Q^2} \\
&+\left(P^{(1)}\otimes f_i\right) \left(P^{(1)}\otimes f_j\right)\ln\frac{\mu_F^2}{Q^2}+\left(P^{(2)}\otimes f_i\right) f_j + i \leftrightarrow j,
\end{align*}
and the results for $\order{3}$ are given in App.~\ref{app:asym3}.

\subsection{NNLO Angular Distributions}

The decay of the $Z$ boson into a pair of leptons can be described by a set of angular
functions with an associated coefficient~\cite{MIRKES19923,Mirkes:1994eb,Mirkes:1994dp,Mirkes:1994nr}. These are given as
\begin{align*}
    \frac{d\sigma}{dp_TdydQ^2d\cos\theta d\phi} &= \mathcal{L}_0\left(1+\cos^2\theta\right) + A_0\left(1-3\cos^2\theta\right) \\
    & + A_1 \sin 2\theta \cos\phi + A_2\sin^2\theta \cos 2\phi \\
    & + A_3 \sin\theta \cos\phi + A_4 \cos \theta \\
    & + A_5 \sin^2\theta\sin 2\phi + A_6 \sin 2\theta \sin\phi \\
    & + A_7 \sin\theta\sin\phi\,,
\end{align*}
where $\mathcal{L}_0$ is an overall normalization factor, $A_i$ are the different
angular coefficients, and $\theta, \phi$ are the polar and azimuthal angles defined
in the Collins-Soper frame~\cite{PhysRevD.16.2219}, respectively.

At leading order, only $\mathcal{L}_0$ and $A_4$ are non-zero. At next-to-leading order, all the terms are
non-zero with the exception of $A_5$, $A_6$, and $A_7$, which are non-zero at next-to-next-to-leading order.
Furthermore, at next-to-leading order the values for $A_0$ and $A_2$ are equal and is known as the
Lam-Tung relation~\cite{Lam:1978pu}. This relation breaks down at NNLO and the first non-zero measurement of 
$A_0-A_2$ was done by the ATLAS experiment~\cite{ATLAS:2016rnf}. In this work, we will compare the ResBos2 predictions
to those from ATLAS, with the exception of $A_5$, $A_6$, and $A_7$, since these results are first non-zero at NNLO and are approximately zero.

\begin{figure}
    \centering
    \includegraphics[width=0.3\textwidth]{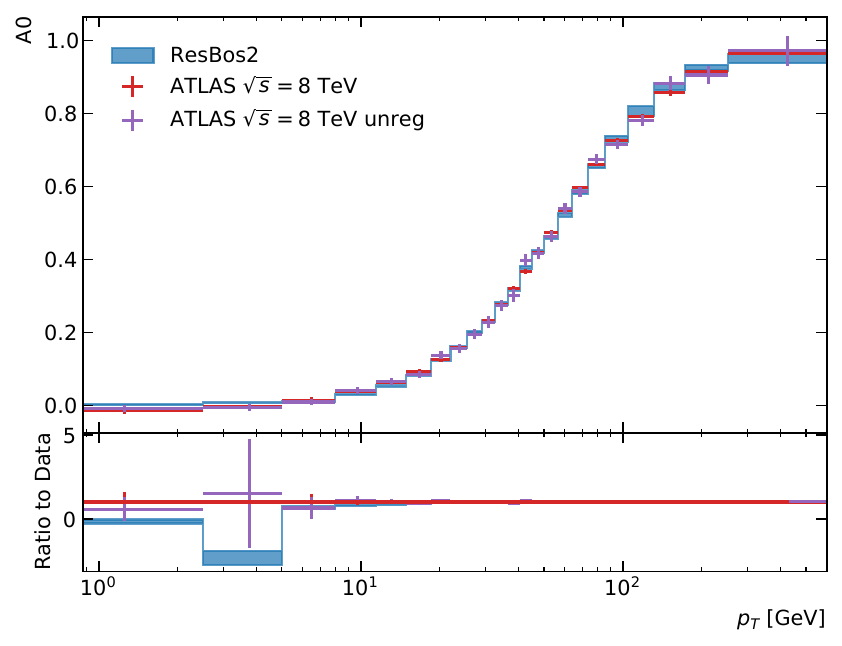}\hfill
    \includegraphics[width=0.3\textwidth]{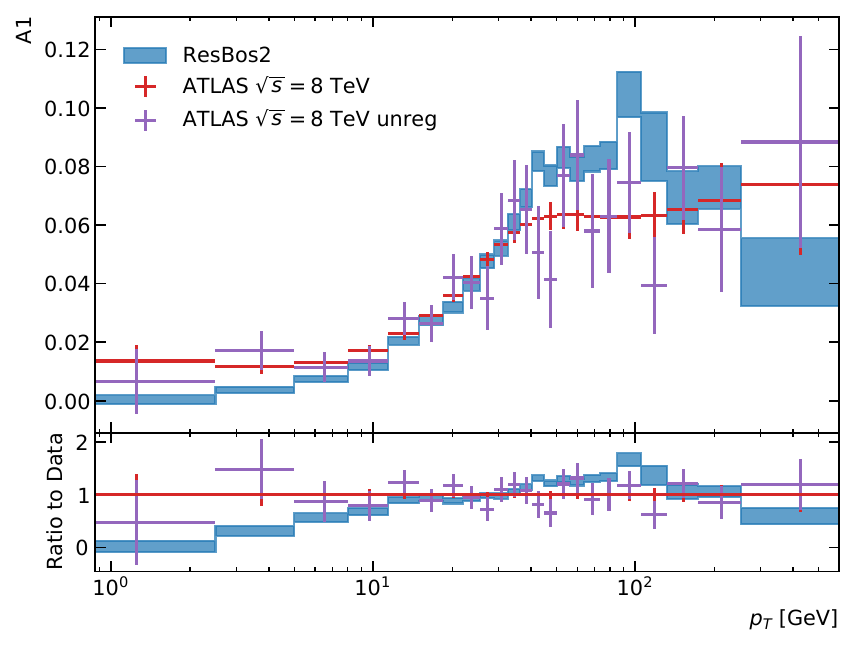}\hfill
    \includegraphics[width=0.3\textwidth]{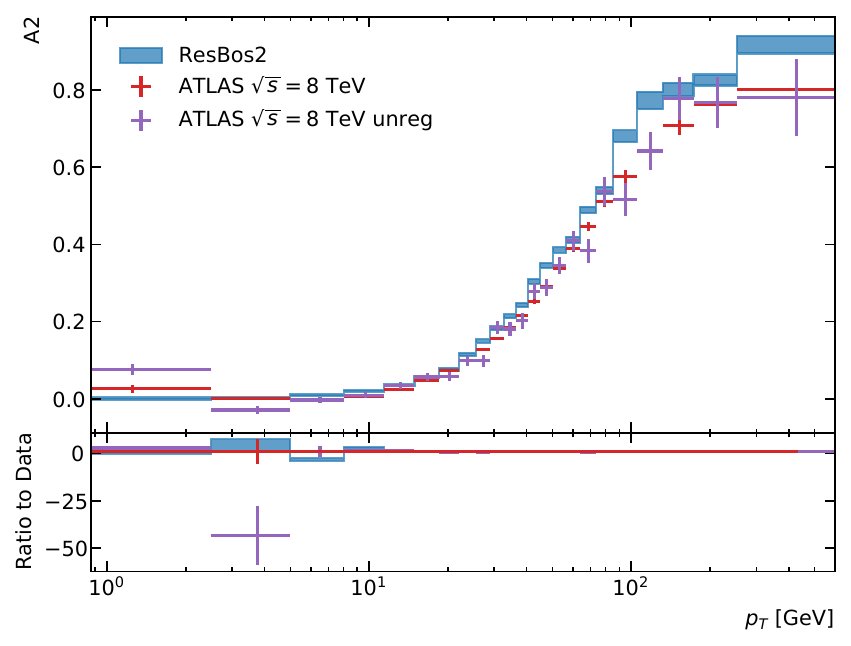}\\
    \includegraphics[width=0.3\textwidth]{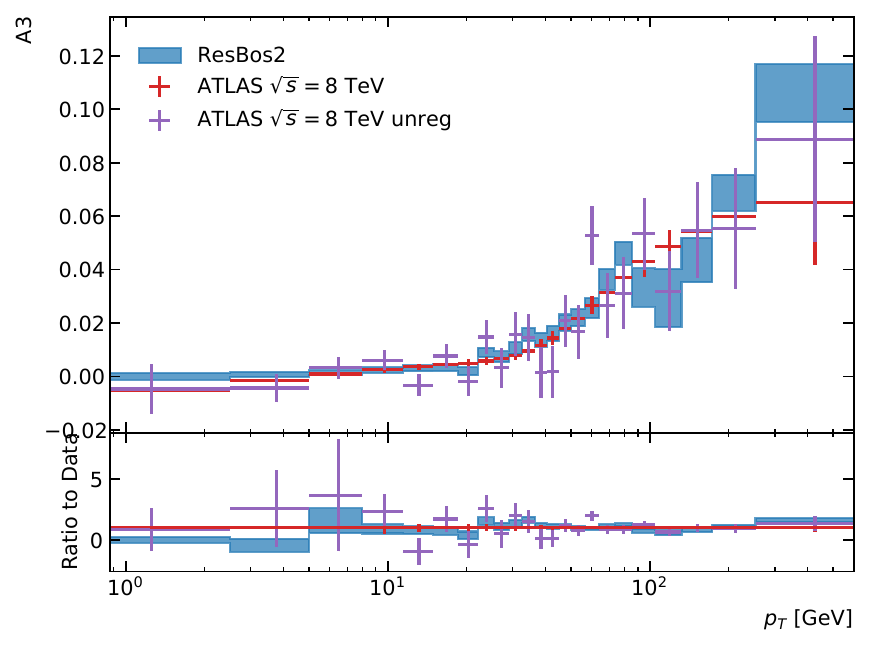}\hfill
    \includegraphics[width=0.3\textwidth]{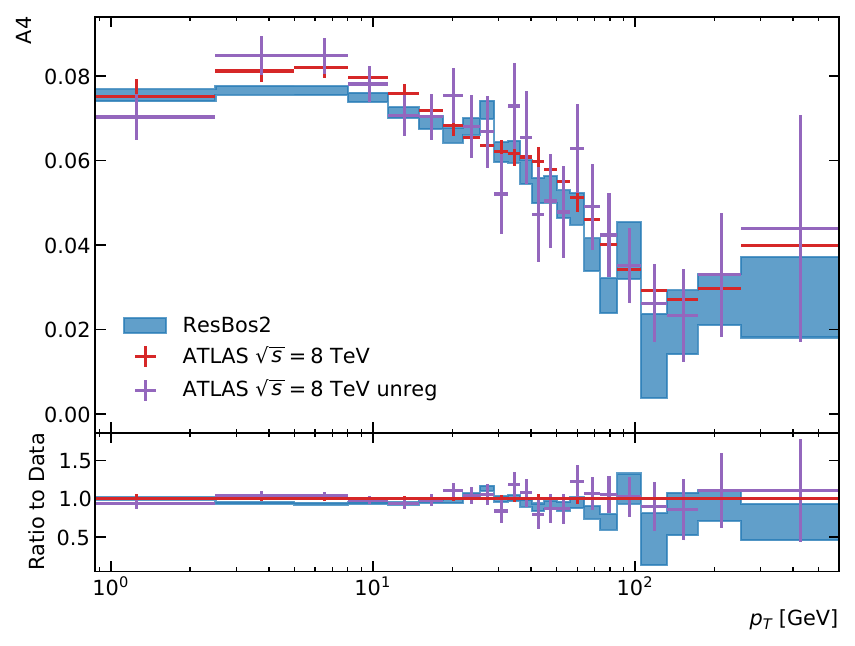}\hfill
    \includegraphics[width=0.3\textwidth]{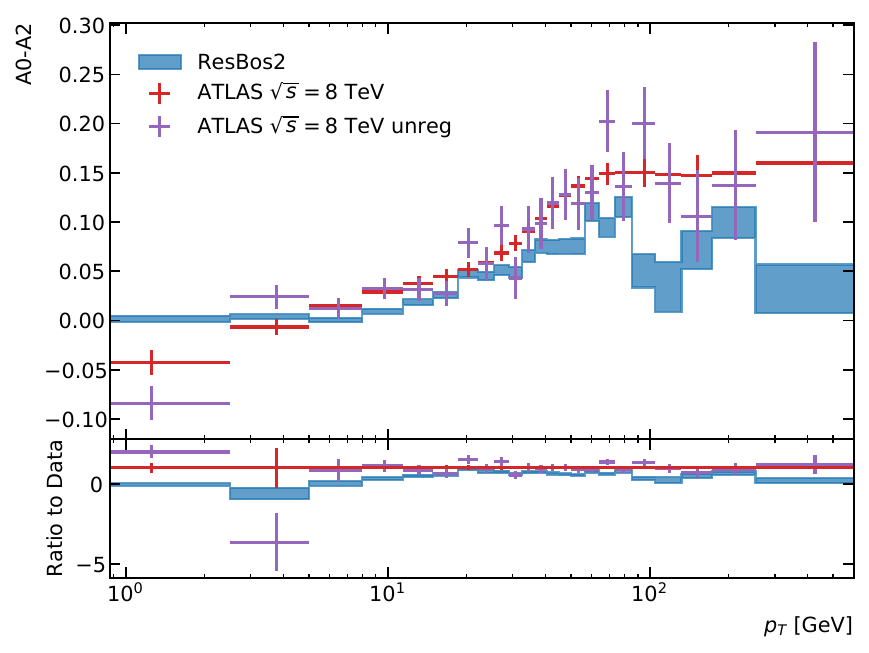}\\
    \caption{Comparison between the ResBos2 calculation assuming that the angular coefficients follow the same resummation as the overall rate and the ATLAS 8 TeV angular coefficients data from Ref.~\cite{ATLAS:2016rnf}. The red curve is the regularised data. The purple curve is the unregularised data.}
    \label{fig:ATLAS_Angular_new}
\end{figure}

In order to make these predictions in the ResBos code, we are required to match to a 
fixed order calculation at NNLO. This is achieved through scaling the ResBos2 NLO calculation by a set
of k-factors obtained in MCFM~\cite{Campbell:2015qma,Boughezal:2016wmq,Campbell:2019dru} for each coefficient independently. These k-factors are differential in
the invariant mass, transverse momentum, and rapidity of the lepton pair. The fixed order calculation
is then matched to the resummation calculation within ResBos2 and the results are compared to the ATLAS
data~\cite{ATLAS:2016rnf}. The results are shown in Fig.~\ref{fig:ATLAS_Angular_new}. Here we can see good agreement
between the ResBos2 prediction and the experimental data. Results are shown for both the regularized and unregularized results, which are described in detail in Ref.~\cite{ATLAS:2016rnf}.

 \begin{figure}
     \centering
     \includegraphics[width=0.3\textwidth]{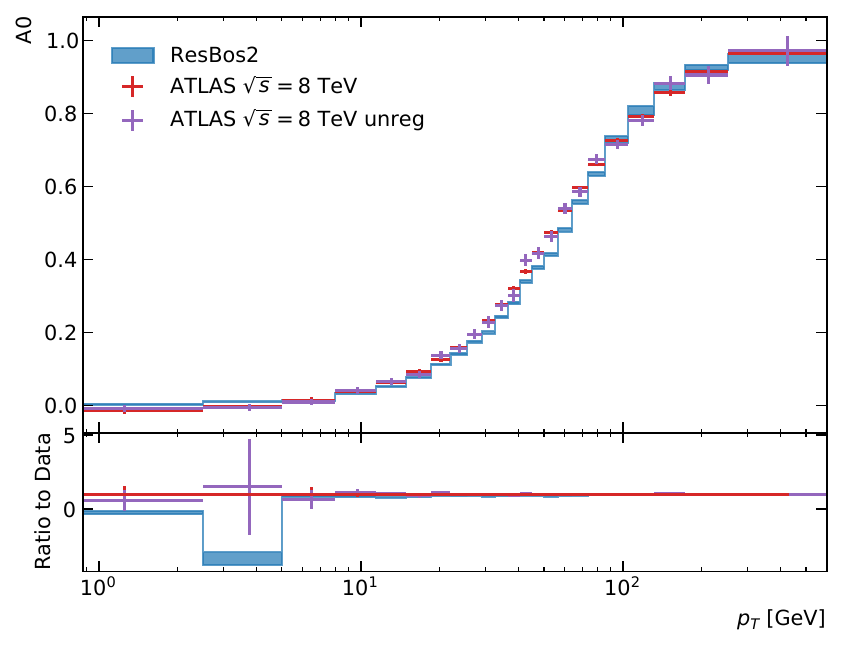}\hfill
     \includegraphics[width=0.3\textwidth]{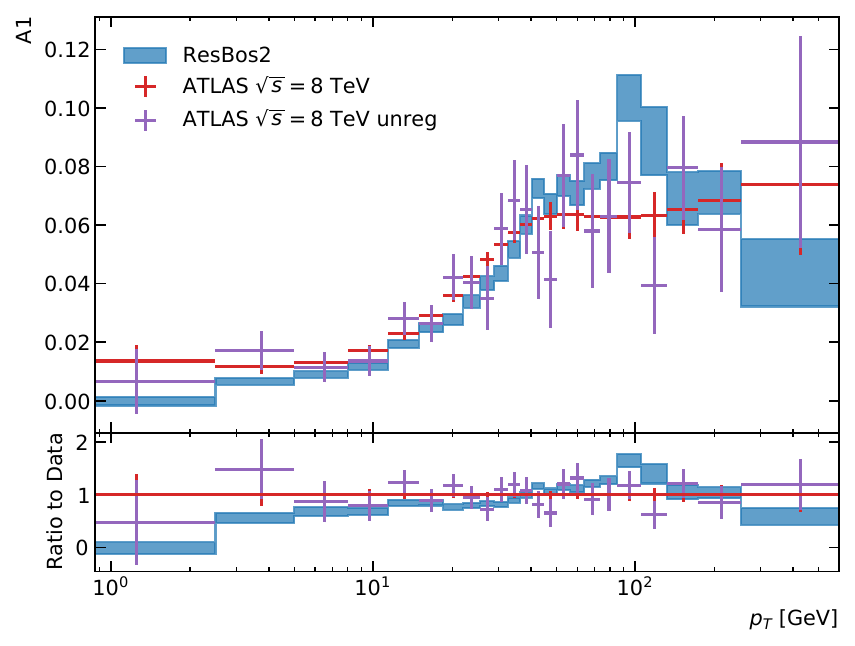}\hfill
     \includegraphics[width=0.3\textwidth]{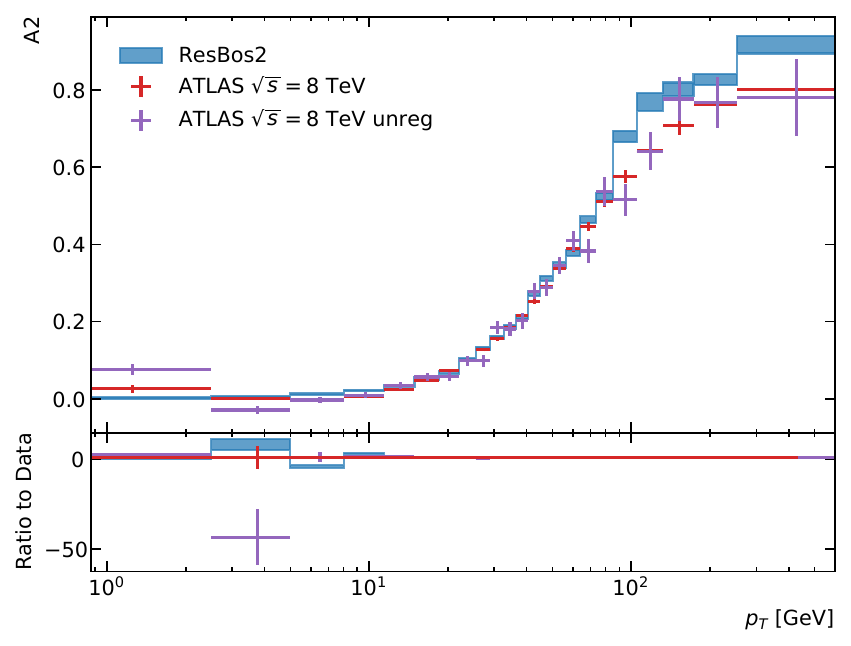}\\
     \includegraphics[width=0.3\textwidth]{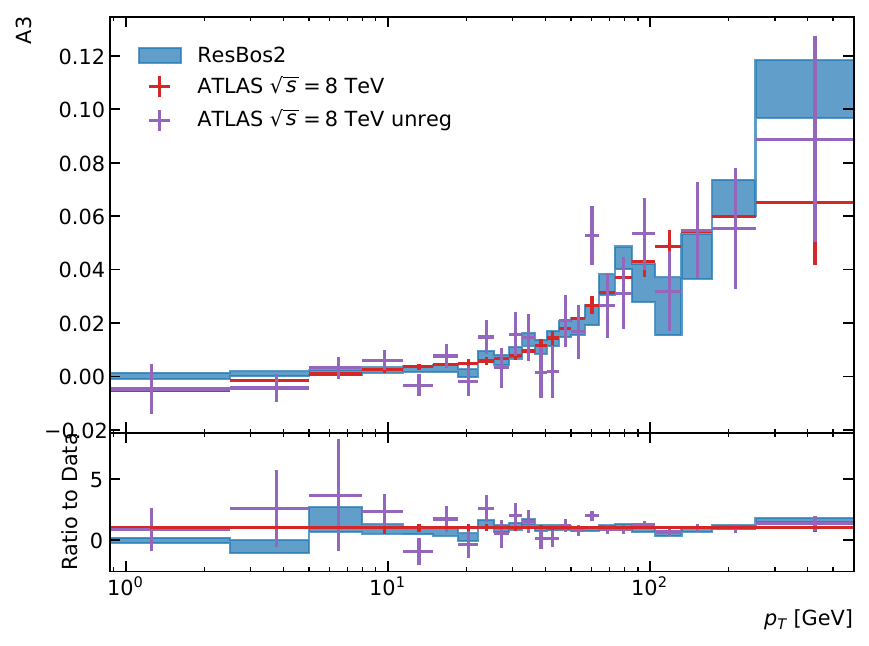}\hfill
     \includegraphics[width=0.3\textwidth]{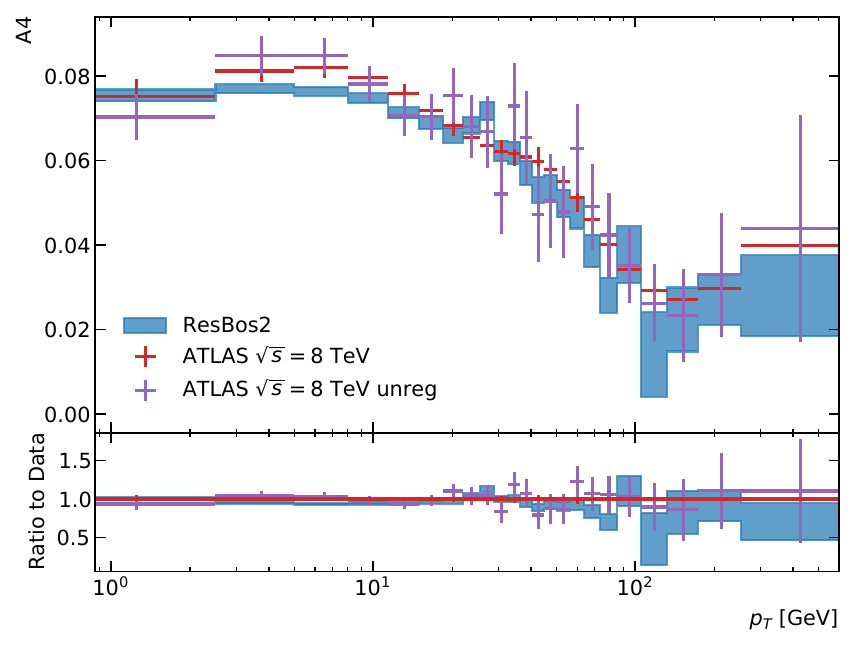}\hfill
     \includegraphics[width=0.3\textwidth]{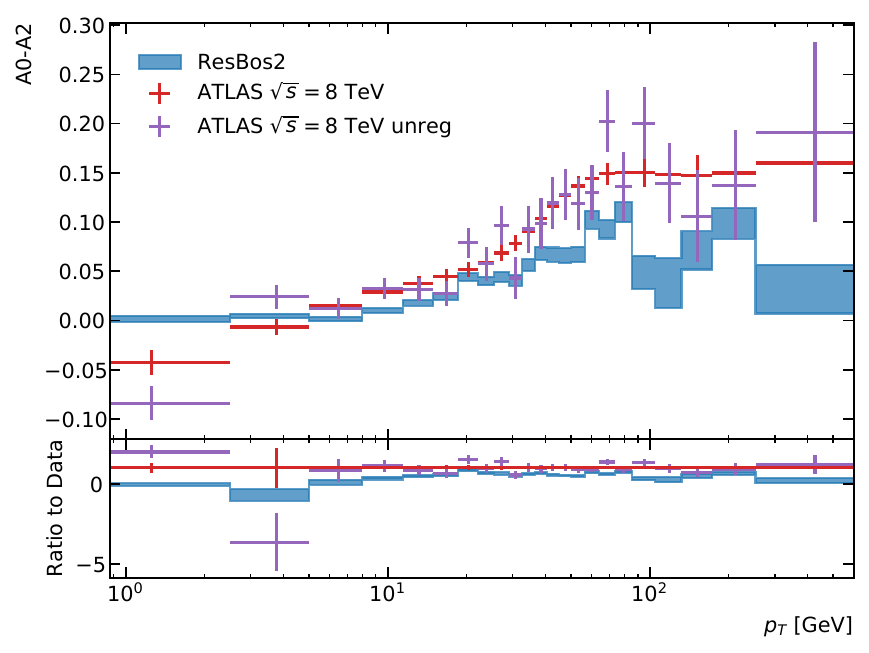}\\
     \caption{Comparison between the ResBos2 calculation assuming the angular coefficients are resummed separately from the overall rate. and the ATLAS 8 TeV angular coefficients data from Ref.~\cite{ATLAS:2016rnf}. The red curve is the regularised data. The purple curve is the unregularised data.}
     \label{fig:ATLAS_Angular_old}
 \end{figure}

An open question in the calculation of the angular coefficients is in the handling of the resummation effects
on all of the coefficients with the exception of $\mathcal{L}_0$ and $A_4$. Here we propose two schemes to
estimate the possible size of the effects of resummation on the coefficients. The first scheme is to assume 
that the coefficients are modified in the same manner as $\mathcal{L}_0$ and $A_4$ due to resummation. The
second scheme is to assume that the coefficients are not modified at all by resummation. These two schemes
should encompass the true effects of resummation. The effects of choosing these two schemes can be seen
in Fig.~\ref{fig:ATLAS_Angular_new} and Fig.~\ref{fig:ATLAS_Angular_old}, respectively.
We leave further investigation into these effects to a future work.

\section{New Non-Perturbative Fit}\label{sec:non-pert}

In this work, we propose a new non-perturbative fit containing information about the rapidity($y$) of the system given by:
\begin{equation}
    \label{eq:IFY}
    S_{IFY} = g_1 + (g_2 + g_3 b^2) \log\left(\frac{Q}{M_Z}\right)+g_4 \log\left(\frac{1960}{\sqrt{s}}\right) + g_5\left(\tanh\left(g_6 y_{\rm Max}\right)+\tanh\left(g_6 (|y| - y_{\rm Max})\right)\right),
\end{equation}
where $g_1$ through $g_6$ are parameters to be fit, $y_{\rm Max}$ is fixed to be 5 in this study, and $\sqrt{s}$ is the center of mass energy in GeV. 
This form is chosen such that at the Tevatron, the dominate contribution comes
to the non-perturbative function comes from $g_1$. Furthermore, the term proportional to $g_5$ is chosen
such that for $y=0$ the contribution from this term vanishes. Hence, to fit the experimental data set in which the $y$ dependence has been integrated out, we set $y$ to be zero in the above equation.

For the fit, we include data from CDF~\cite{Affolder:1999jh,Aaltonen:2012fi},
D0~\cite{Abbott:1999wk,Abazov:2007ac}, E288~\cite{Ito:1980ev},
E605~\cite{Moreno:1990sf}, R209~\cite{Antreasyan:1981uv}, and a rapidity separated measurement from 
ATLAS~\cite{ATLAS:2015iiu} and CMS~\cite{CMS:2019raw}, and high rapidity measurement from LHCb~\cite{LHCb:2021huf}
We leave the other LHC measurements discussed in Sec.~\ref{sec:z_obs} as validation of the non-perturbative fit.
Additional details on the kinematics of each experiment are given in Tab.~\ref{tab:nonpert_datasets}.

\begin{table}[ht]
    \centering
    \begin{tabular}{|c|c|c|c|}
    \hline
    Experiment & $\sqrt{s}$ & Cuts & $N_{pts}$ \\
    \hline
    CDF Run 1~\cite{Affolder:1999jh} & 1800 GeV & $66$ GeV $ < M_{\ell\ell} < 116$ GeV & 32 \\
    CDF Run 2~\cite{Aaltonen:2012fi} & 1960 GeV & $66$ GeV $ < M_{\ell\ell} < 116$ GeV & 41 \\
    D0 Run 1~\cite{Abbott:1999wk} & 1800 GeV & $66$ GeV $ < M_{\ell\ell} < 116$ GeV & 15 \\
    D0 Run 2~\cite{Abazov:2007ac} & 1960 GeV & $66$ GeV $ < M_{\ell\ell} < 116$ GeV & 8 \\
    E288 200~\cite{Ito:1980ev} & 19.4 GeV & 4 GeV $ < M_{\ell\ell} < $ 8 GeV, y = 0.4 & 28 \\
    E288 300~\cite{Ito:1980ev} & 23.8 GeV & 4 GeV $ < M_{\ell\ell} < $ 8 GeV, 11 GeV $ < M_{\ell\ell} < $ 12 GeV, y = 0.21 & 35 \\
    E288 400~\cite{Ito:1980ev} & 27.4 GeV & 5 GeV $ < M_{\ell\ell} < $ 8 GeV, 11 GeV $ < M_{\ell\ell} < $ 14 GeV, y = 0.03 & 42 \\
    E605~\cite{Moreno:1990sf} & 38.8 GeV & 7 GeV $ < M_{\ell\ell} < $ 9 GeV, 10.5 GeV $ < M_{\ell\ell} < $ 14 GeV, $E_z$ = 0.1 GeV  & 35 \\
    R209~\cite{Antreasyan:1981uv} & 62 GeV & 5 GeV $ < M_{\ell\ell} < $ 11 GeV, 0.1 $< x < 0.8$ & 10 \\
    ATLAS~\cite{ATLAS:2015iiu} & 8000 GeV & $66$ GeV $ < M_{\ell\ell} < 116$ GeV, $p_{T_\ell} > 20$ GeV, $|\eta_\ell|<2.4$ & 48 \\
    CMS~\cite{CMS:2019raw} & 13000 GeV & $76.1876$ GeV $ < M_{\ell\ell} < 106.1876$ GeV, $p_{T_\ell} > 25$ GeV, $|\eta_\ell| < 2.4$ & 80 \\
    LHCb~\cite{LHCb:2021huf} & 13000 GeV & $60$ GeV $ < M_{\ell\ell} < 120$ GeV, $p_{T_\ell} >$ 20 GeV, 2.0 $ < \eta_\ell < 4.5$ & 10 \\
    \hline
    \end{tabular}
    \caption{A list of all experiments used for the non-perturbative fit. The first column gives the name of the experiment, the second gives the center of mass energy, the third describes the experimental cuts applied for each set of events, and the last column gives the total number of data points for each experiment.}
    \label{tab:nonpert_datasets}
\end{table}

For the fits, we include the PDF uncertainty as an uncorrelated systematic uncertainty. The fit is performed through
the use of the Bayesian Analysis Toolkit (BAT) package~\cite{BAT}. BAT uses Markov Chain Monte Carlo (MCMC) to 
fit the optimal best fit, along with the correlation matrix, and uncertainties on all the parameters. The priors
for all the parameters are taken to be flat over the allowed range. The allowed ranges are $0 < g_1 < 3.5$,
$0 < g_2 < 1$, $0 < g_3 < 1$, $-0.5 < g_4 < 1$, $0 < g_5 < 10$, and $0.5 < g_6 < 5$.
The limits on the lower values of these parameters are chosen such that Eq.~\eqref{eq:IFY} remains positive.
We investigated the effects of choice of prior, and found no major difference when changing the priors.

The best fit was found by minimizing the log-likelihood
between the theory predictions and the experimental data.
The log-likelihood function used in this work
is the same as defined in 
Refs.~\cite{Pumplin:2002vw,Hou:2019efy} for an experiment $E$ and reproduced here
for convenience
\begin{equation}
\chi^2_E = \sum_{k=1}^{N_{pt}} \frac{1}{s_k^2}\left(
    D_k - T_k(g) - \sum_{\alpha=1}^{N_\lambda} \lambda_\alpha \beta_{k\alpha}
    \right)^2+\sum_{\alpha=1}^{N_\lambda}\lambda_\alpha^2\,.
\end{equation}
Each data point $k$ comes with the value for the data ($D$), a statistical uncertainty ($s_{k,{\rm stat}}$),
and a uncorrelated systematic uncertainty ($s_{k,{\rm uncorr. sys}}$). The total uncorrelated uncertainty
is then obtained as $s_k = \sqrt{s_{k,{\rm stat}}^2+s_{k,{\rm uncorr. sys}}^2}$ for a given data point.
Additionally, the theory prediction for each data point is given by $T_k(g)$, where $g$ is the vector
of all parameters in the non-perturbative function. Finally, each data point may come with $N_\lambda$ 
correlated systematic uncertainties given by $\beta_{k\alpha}$. Traditionally, these correlated
uncertainties are handled by introducing a nuisance parameter $\lambda_\alpha$ that is assumed to be
sampled from a standard normal distribution. As discussed in Appendix B of Ref.~\cite{Hou:2019efy},
the optimal values for $\lambda_\alpha$ can be directly calculated as a function of the fit parameters.
In this work, we implement this technique to handle the correlated systematic uncertainties.

In addition to the experimentally reported uncertainties, we also include the theory uncertainty arising
from the PDFs. In this work we use the CT18NNLO PDFs~\cite{Hou:2019efy}.
One can consider the uncertainty associated with the PDF extraction as an experimental uncertainty from the fitted data.
Therefore, it should be included in the $\chi^2$ calculation. In this work, we treat these uncertainties as uncorrelated, similar to Ref.~\cite{Bacchetta:2019sam}.
While this approximation is not completely accurate, separating out the correlated and uncorrelated
components between each data point is beyond the scope of this work. The PDF uncertainty is assessed by fixing the non-perturbative parameters and
calculating the spread from the central PDF value. This value is stored as a percentage shift of the theory prediction for each data point.

The ResBos2 calculation used in the fit is setup to use the scale choices of $4C_1=C_3=4 b_0$, $C_2=1$, $\mu_F=\mu_R=M_T$, and setting $b_{\rm max} = b_0 \simeq 1.123 \ {\rm GeV}^{-1} $, where $M_T=\sqrt{p_T^2+Q^2}$ is the transverse mass of the Drell-Yan pair, and $b_0=2 e^{-\gamma_E}$. 
The choice of $C_3=4b_0$ is to ensure that the ratio $C_3/b^*$ is always greater than the cutoff scale for the PDF to ensure that there is no extrapolation needed in the calculation.
Fixing this value before the fit is allowed since any difference caused by this scale choice will be absorbed in the non-perturbative fit.
Additionally, the choice of $b_{\rm max} = b_0$ is suggested by the disagreement between the BLNY fit and the lattice data seen in Fig. 15 of Ref.~\cite{Avkhadiev:2023poz}.
In the lattice results, the fall-off of the BLNY result occurs around $b=0.5\ {\rm GeV}^{-1}$ (i.e. the $b_{\rm max}$ value for the BLNY fit), suggesting that a larger $b_{\rm max}$ be used.
In addition to the scale choices,
certain experiments only provide data that includes experimental cuts and not unfolded back to an inclusive level. In our data set, this only consists of the data from the LHC.
Performing the full Monte-Carlo integration for each set of non-perturbative values would be computationally prohibitive. To address this issue, we calculate a cut efficiency for the
initial set of non-perturbative parameters and assume that this efficiency is not sensitive to the non-perturbative parameters chosen. After the fit is complete, we check again the cut
efficiency by performing a full calculation including the Monte-Carlo integration for the $\chi^2$. From these cross checks, we validate that the cut efficiency is insensitive to the non-perturbative
parameters. 
To increase the weight of the rapidity dependence data from the global $\chi^2$ fit, the PDF uncertainties of ATLAS, CMS, and LHCb data set are divided by a factor of $3$. 
For other data set, the PDF uncertainty remains the original value. 
After fitting, the $\chi^2$ are calculated again using the normal PDF uncertainty for every data set.

The fits are performed using 8 Markov chains that are run until they converge. This ensures that the
fit that we find is the global minimum instead of a local minimum. After the chains have converged,
we perform an additional 100,000 iterations to determine the optimal fit values, the uncertainty of each
parameter, and the correlations between the different parameters. Additional details on the procedure
can be found in Ref.~\cite{BAT}.

Over the course of the fits, we found that the best fit results had $g_3$ consistent with zero. Therefore, for the final fit result presented here
we set $g_3$ to zero.
For the best fit value of $g_5$, we find that simultaneous fitting of $g_5$ and $g_6$ leads to a very flat distribution. To address this, we perform an iterated approach until the results converge. First, we fix $g_5$ to 1 and fit all the other parameters. Then we fix the other parameters to their best fit results and only fit $g_5$. This procedure is repeated until the values for all the parameters remain at their previous best fit.
The total $\chi^2/dof$ from this fit is $380.8/384$, and the
break-down for each experiment is given in Tab.~\ref{tab:chi2}.
The best fit values are given in Tab.~\ref{tab:IY6_BestFit}. Furthermore, we perform an update to the
SIYY fit from Ref.~\cite{Sun:2014dqm} using the same data set and fitting techniques described above. The results
for the SIYY fit are also given in Tab.~\ref{tab:IY6_BestFit} and the $\chi^2/dof$ is given as $481.7/384$. This fit is slightly worse than the IFY fit.
For completeness, we summarize the result of the SIYY fit in Appendix~\ref{app:SIYY}.

\begin{table}[ht]
    \centering
    \begin{tabular}{|c|c|c|c|c|c|}
        \hline
         Experiment & $N_{pts}$ & $\chi^2$ (IFY) & $\chi^2$/dof (IFY) & $\chi^2$ (SIYY) & $\chi^2/dof$ (SIYY) \\
         \hline \hline
         CDF1~\cite{Affolder:1999jh} & 32 & 19.8 & 0.62 & 20.6 & 0.64 \\
         CDF2~\cite{Aaltonen:2012fi} & 41 & 48.5 & 1.18 & 62.6 & 1.53 \\
         D01~\cite{Abbott:1999wk} & 15 & 11.1 & 0.74 & 14.1 & 0.94 \\
         D02~\cite{Abazov:2007ac} & 8 & 19.4 & 2.43 & 21.9 & 2.74 \\
         E288 200~\cite{Ito:1980ev} & 28 & 31.4 & 1.12 & 60.0 & 2.14 \\
         E288 300~\cite{Ito:1980ev} & 35 & 34.6 & 0.99 & 42.0 & 1.20 \\
         E288 400~\cite{Ito:1980ev} & 42 & 89.9 & 2.14 & 104.0 & 2.48 \\
         E605~\cite{Moreno:1990sf} & 35 & 54.5 & 1.56 & 71.1 & 2.03 \\
         R209~\cite{Antreasyan:1981uv} & 10 & 8.5 & 0.85 & 11.3 & 1.13 \\
         ATLAS~\cite{ATLAS:2015iiu} & 48 & 9.2 & 0.19 & 11.4 & 0.24\\
         CMS~\cite{CMS:2019raw} & 80 & 35.7 & 0.45 & 39.7 & 0.50 \\
         LHCb~\cite{LHCb:2021huf} & 10 & 18.3 & 1.83 & 23.2 & 2.32 \\
         \hline
         Total & 384 & 380.8 & 0.992 & 481.7 & 1.25 \\
         \hline
    \end{tabular}
    \caption{The summary of the number of points, $\chi^2$, and $\chi^2$/dof for each experiment included in the fit. The reference to each experiment is given in the table. }
    \label{tab:chi2}
\end{table}

\begin{table}[ht]
    \centering
    \begin{tabular}{|c|c|}
       \hline
       Parameter  & Value \\
       \hline
       \hline
       $g_1$  & 1.034 $\pm$ 0.026 \\
       $g_2$  & 0.053 $\pm$ 0.025 \\
       $g_4$  & -0.143 $\pm$ 0.014 \\
       $g_5$  & 13.45 $\pm$ 2.0 \\
       $g_6$  & 1.468 $\pm$ 0.108 \\
       \hline
    \end{tabular}
    \caption{The best fit result to the experimental data included in the IFY fit.}
    \label{tab:IY6_BestFit}
\end{table}

\begin{figure}
    \centering
    \includegraphics[width=0.47\textwidth]{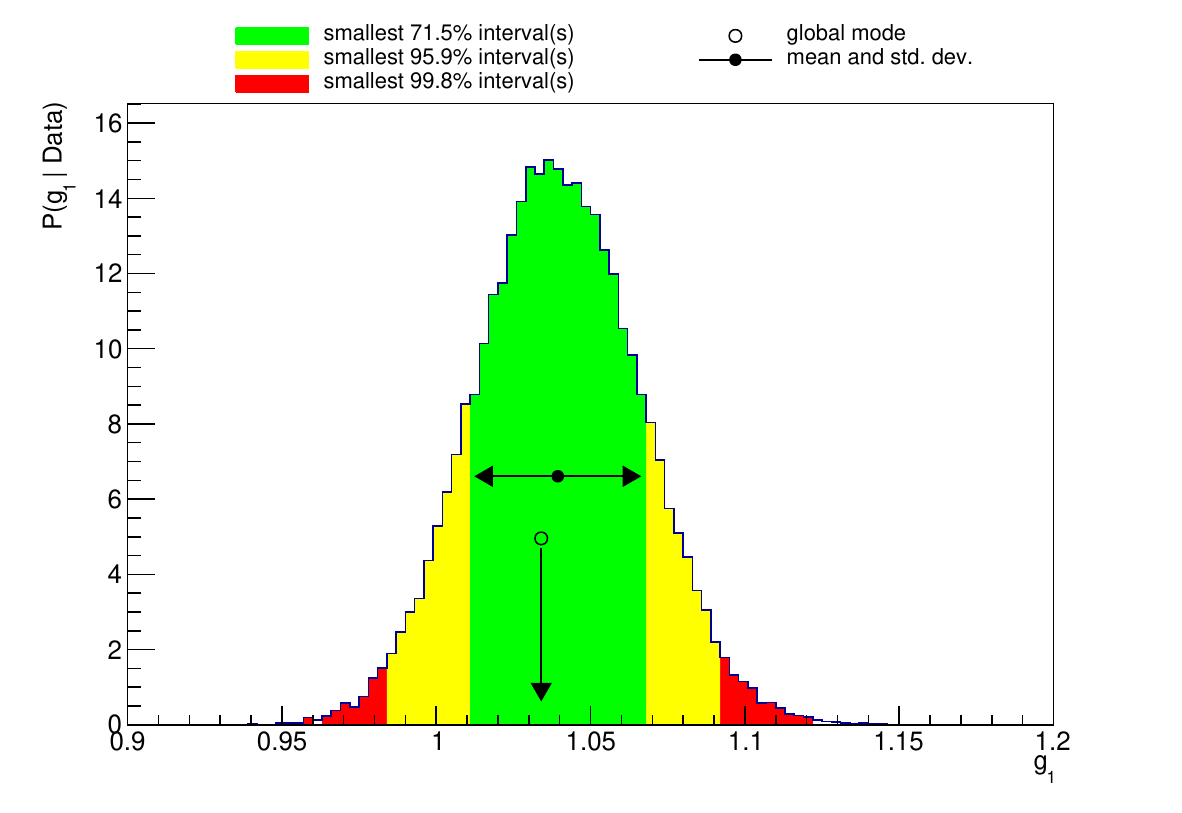}
    \hfill
    \includegraphics[width=0.47\textwidth]{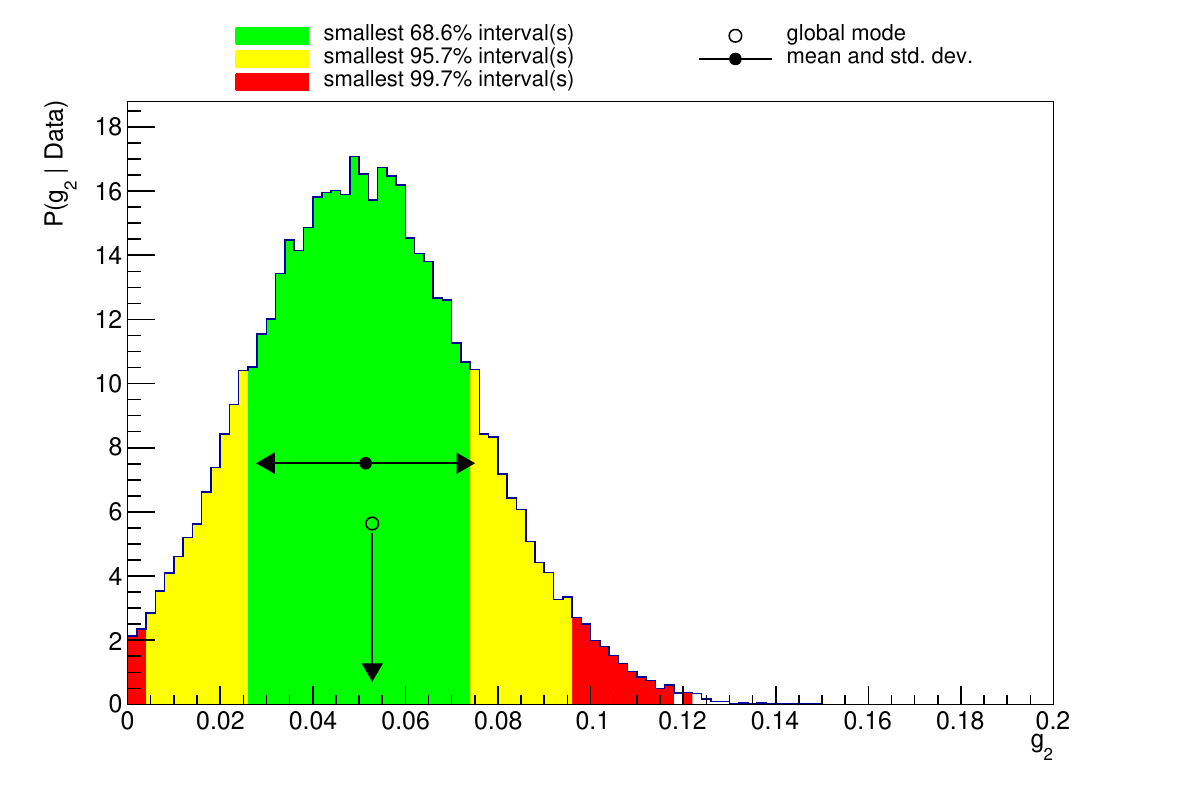} \\
    \includegraphics[width=0.47\textwidth]{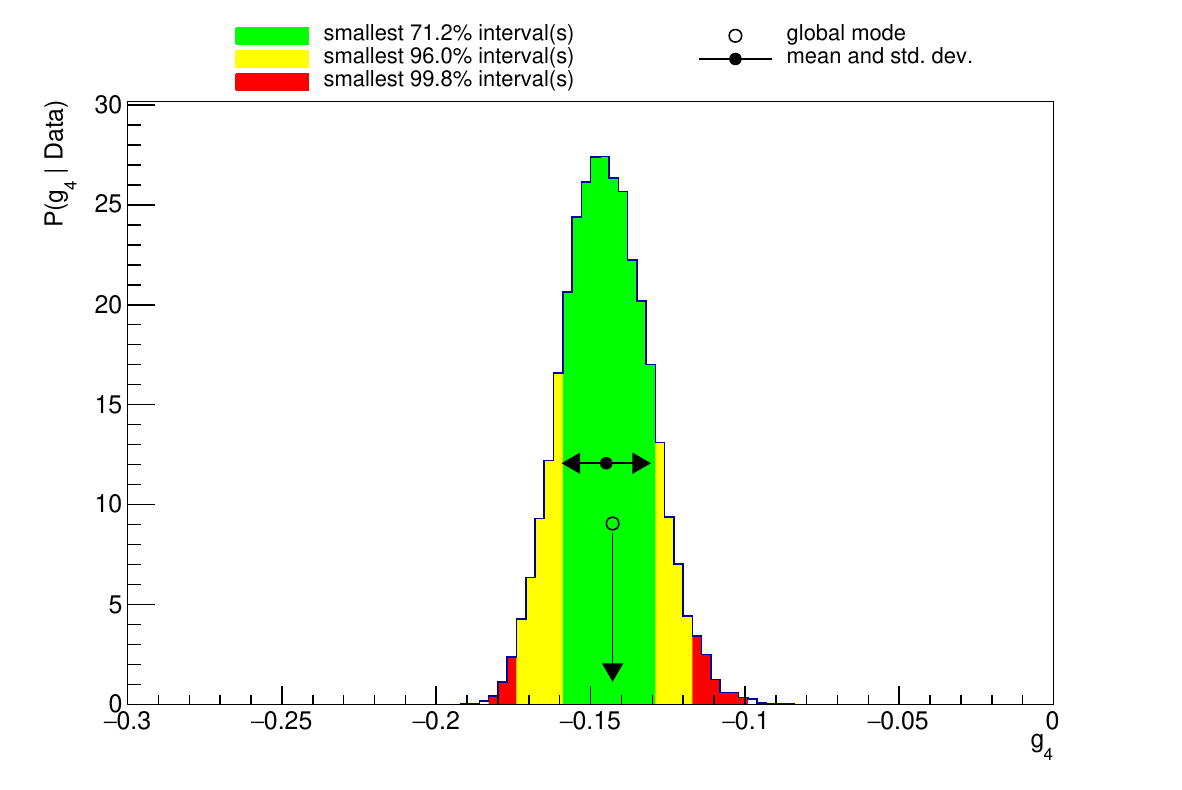}
    \hfill
    \includegraphics[width=0.47\textwidth]{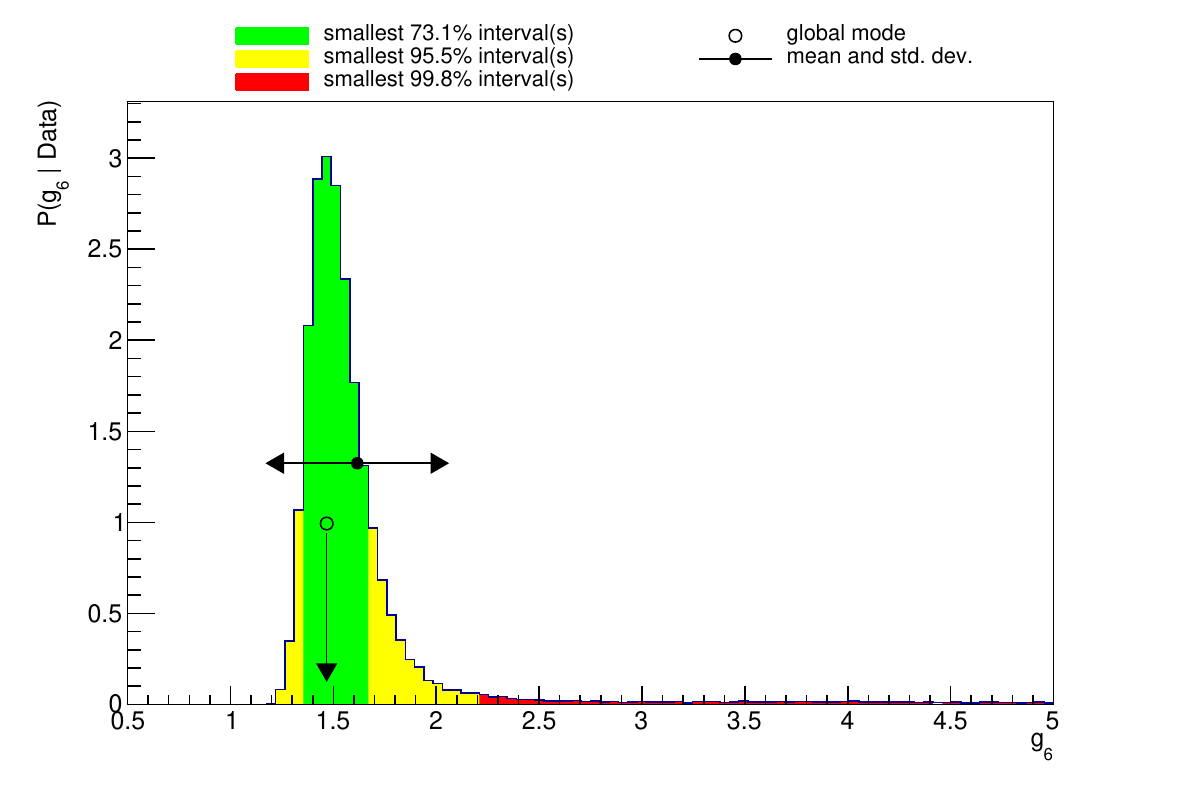}
    \caption{Best fit values with 1,2,3 $\sigma$ ranges in green, yellow, and red respectively. The value for $g_1$ is in the upper left plot, $g_2$ in the upper right, $g_4$ on the left in the second row, $g_5$ on the right of the second row, and $g_6$ on the bottom. The value for $g_3$ was fixed to be zero. Additional details in the text.}
    \label{fig:IY6_values}
\end{figure}

\begin{figure}
    \centering
    \includegraphics[width=0.47\textwidth]{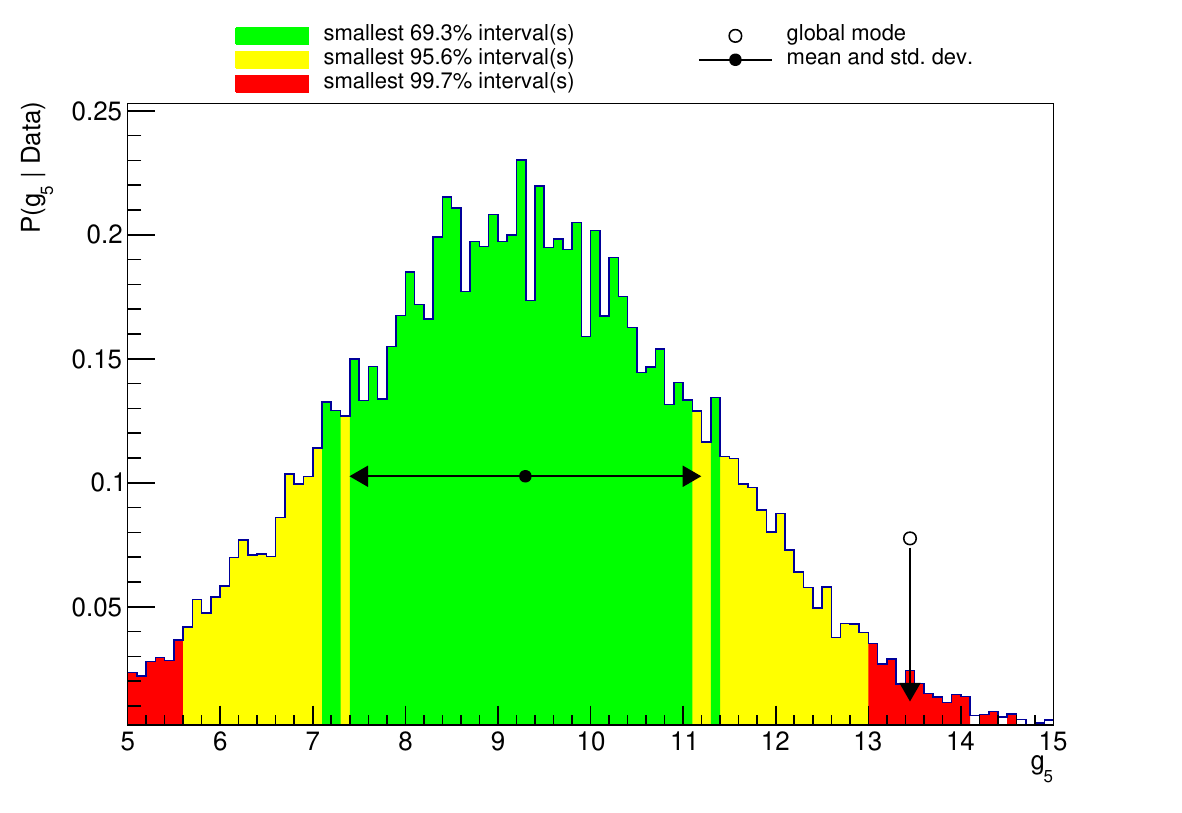}
    \caption{Best fit values on the marginalized $g_5$ with 1,2,3 $\sigma$ ranges in green, yellow, and red respectively.}
    \label{fig:IY6_g5}
\end{figure}

The correlation matrix is given as, for a set of parameters $(g_1, g_2, g_4, g_6)$,
\begin{equation}
    C = 
    \begin{pmatrix}
    1 & 0.288 & -0.139 & 0.350 \\
    0.288 & 1 & 0.906 & -0.174 \\
    -0.139 & 0.906 & 1 & -0.334 \\
    0.350 & -0.174 & -0.334 & 1 \\
    \end{pmatrix}\,.
\end{equation}
Here we see a strong correlation between $g_2$ and $g_4$, with minor correlations between $g_1$ and the
other coefficients and $g_4$ and $g_6$. The correlation between $g_1$ and the other coefficients is expected since the
$g_1$ is just an overall constant and shifting around any other coefficient will result in a need
to also shift $g_1$ to compensate. The posterior distributions marginalized over the other parameters for each of the fitted coefficients is shown
in Figs.~\ref{fig:IY6_values} and ~\ref{fig:IY6_g5}, along with denoting the one, two, and three $\sigma$ confidence intervals. The one $\sigma$  region is in green, the two $\sigma$
region is in yellow, and the three $\sigma$ region is in red. The filled circle denotes the mean value along with the standard deviation. This is obtained after
marginalizing over the other parameters. The open circle denotes the global mode value. This is the point that is sampled the most in the full parameter space by
the MCMC. The consistency between the global mode and the mean values shows that the MCMCs have converged well to the true minimum.

\begin{table}[ht]
    \centering
    \begin{tabular}{c|c|c}
       \hline
         & $\chi^2$ for $g_5=0$ & $\chi^2$ for $g_5=13.45$ \\
       \hline
       \hline
       $p_T=1.1$  & 1.422 & 0.042 \\
       $p_T=2.8$  & 0.864 & 0.122 \\
       $p_T=4.0$  & 2.666 & 2.086 \\
       $p_T=5.2$  & 0.028 & 0.065 \\
       \hline
    \end{tabular}
    \caption{The $\chi^2$ of LHCb data of the first four points for $g_5=0$ and $g_5=13.45$.}
    \label{tab:chi2_LHCb}
\end{table}

\begin{figure}
    \includegraphics[width=0.47\textwidth]{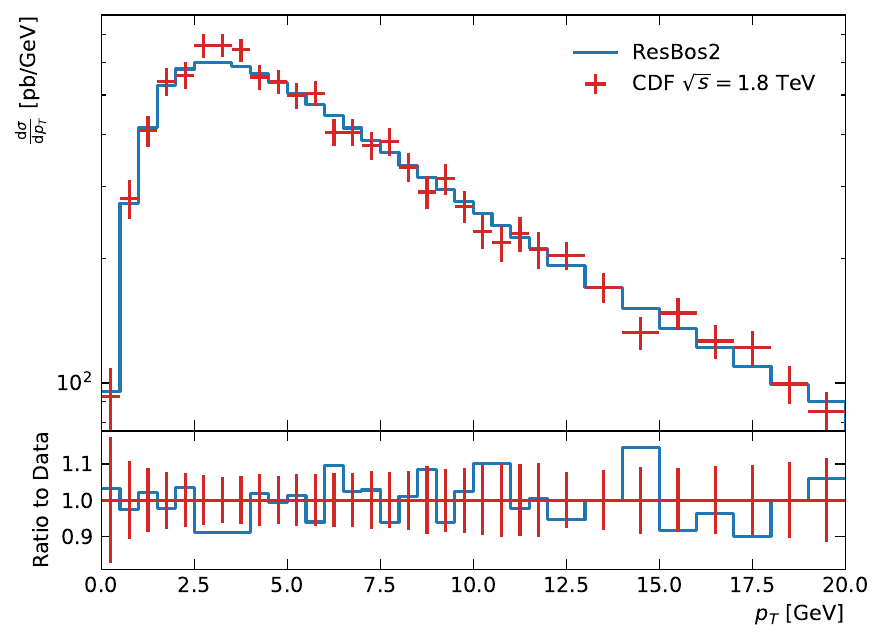}
    \hfill
    \includegraphics[width=0.47\textwidth]{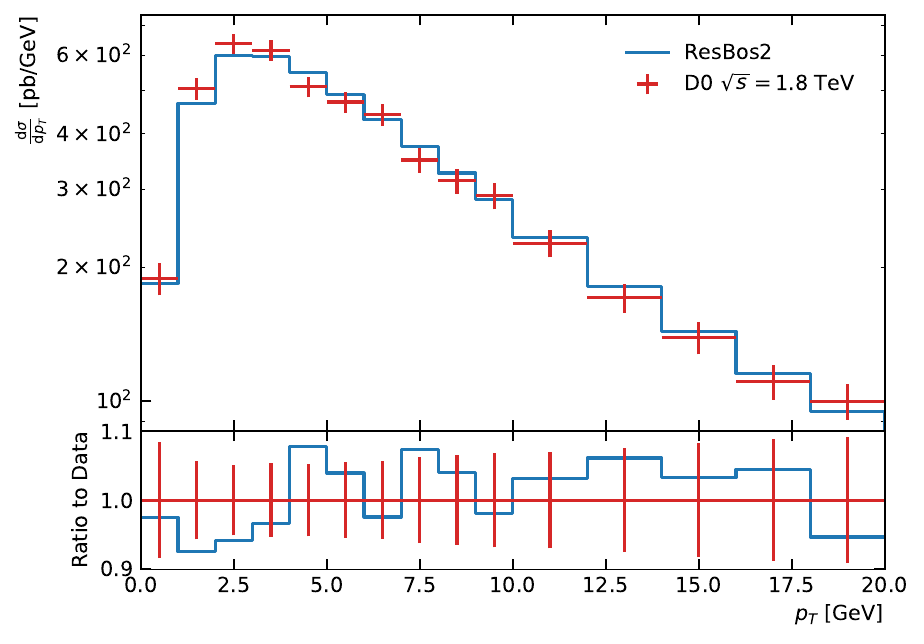} \\
    \includegraphics[width=0.47\textwidth]{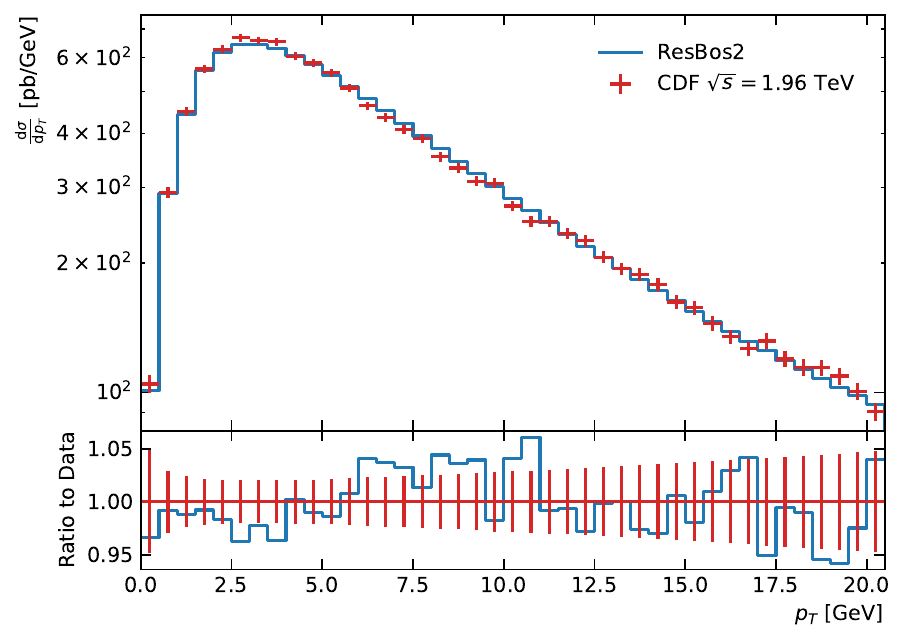}
    \hfill
    \includegraphics[width=0.47\textwidth]{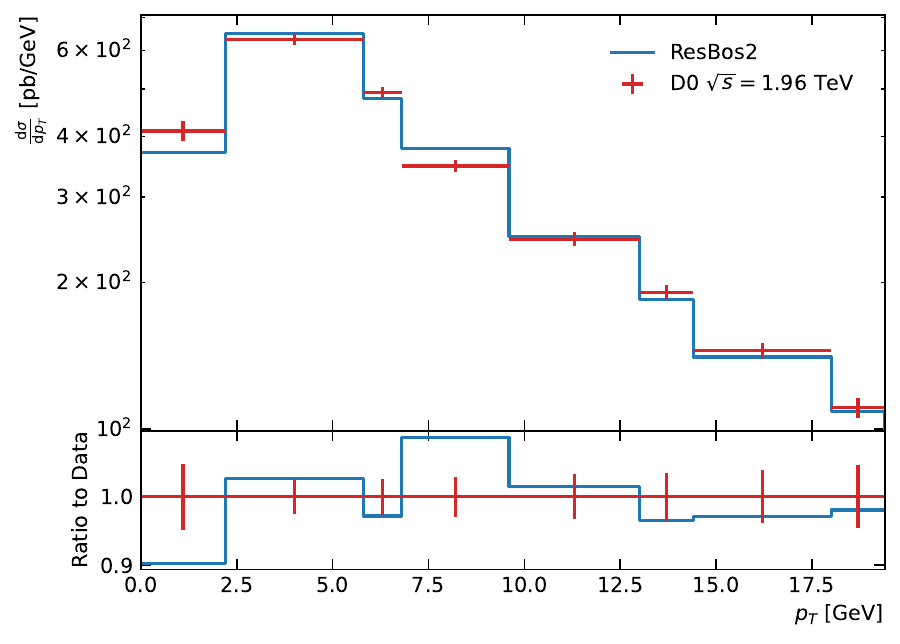} \\
    \caption{Comparison of the IFY Non-perturbative fit with the N${}^{3}$LL+NNLO accurate ResBos2 calculation to the Tevatron $Z$ boson transverse momentum data. The top row are the results from run I, with CDF~\cite{Affolder:1999jh} on the left and D0~\cite{Abbott:1999wk} on the right. The bottom row is the same as the top but with run II data~\cite{Aaltonen:2012fi,Abazov:2007ac}.}
    \label{fig:IY6_Tevatron}
\end{figure}

A comparison of the IFY non-perturbative fit described above to the Tevatron data from
CDF~\cite{Affolder:1999jh,Aaltonen:2012fi} and D0~\cite{Abbott:1999wk,Abazov:2007ac} are shown in
Fig.~\ref{fig:IY6_Tevatron}. The experimental data was collected with $\sqrt{s}=1800$ GeV for the Run I data (top row of the figure), and with $\sqrt{s}=1960$ GeV (bottom row) for the Run II data. The experimental data has been
unfolded to the inclusive $Z$ production level in the invariant mass window of 
66 GeV $< M_{\ell\ell} <$ 116 GeV. The cuts applied in the analysis are shown in 
Tab.~\ref{tab:nonpert_datasets}. The comparison is made using the N${}^{3}$LL+NNLO prediction within 
ResBos2 and comparing to the unshifted data. In these comparisons, the scale variations and PDF
uncertainties in the calculation are intentionally left out in the comparison to data to more easily demonstrate the quality of the fit.

\begin{figure}
    \centering
    \includegraphics[width=0.47\textwidth]{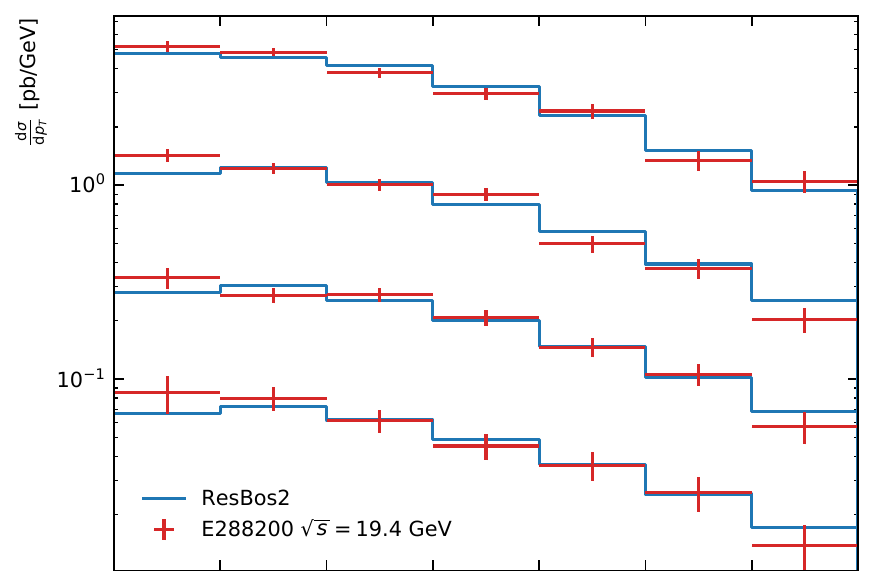}
    \hfill
    \includegraphics[width=0.47\textwidth]{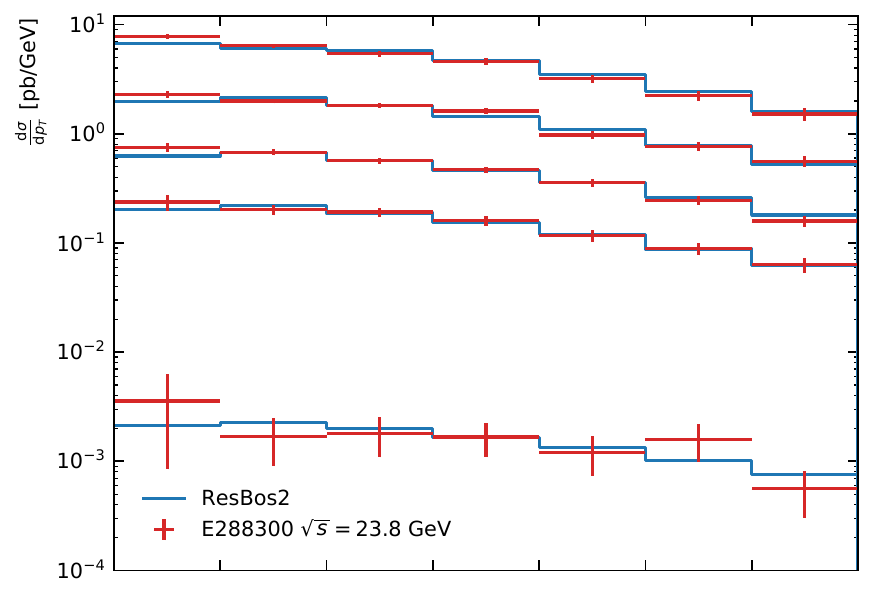} \\
    \includegraphics[width=0.47\textwidth]{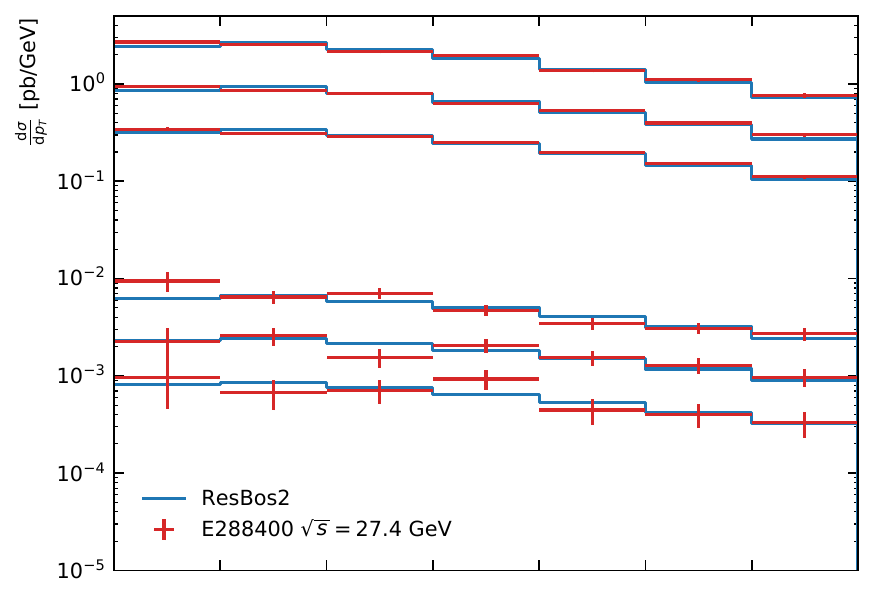}
    \hfill
    \includegraphics[width=0.47\textwidth]{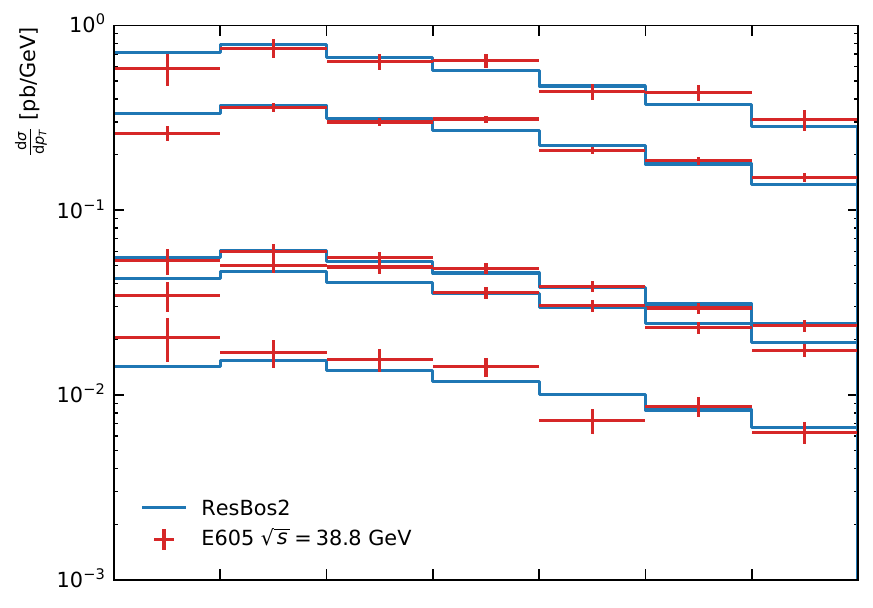} \\
    \includegraphics[width=0.47\textwidth]{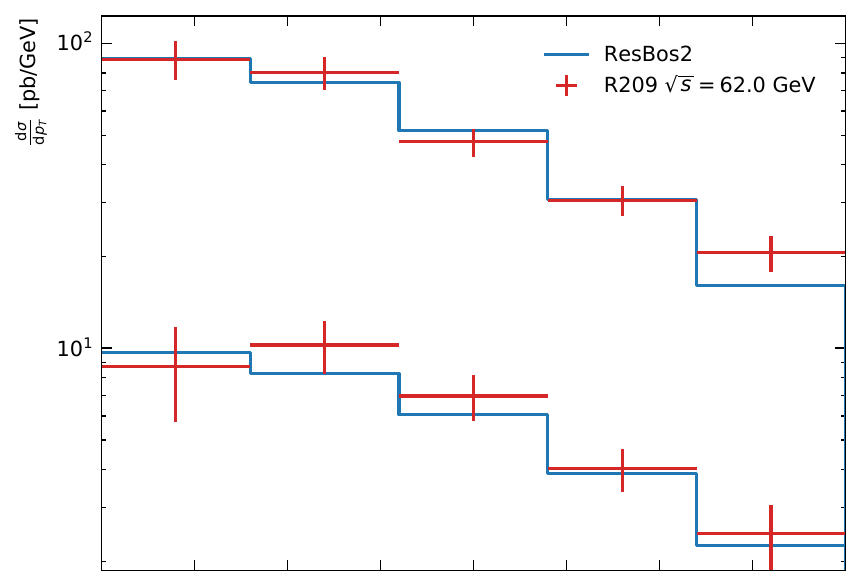}
    \caption{Comparison of the IFY Non-perturbative fit with the $N^{3}LL+NNLO$ accurate ResBos2 calculation to fixed target data. The top row and the plot on the second row on the left are the results from E288~\cite{Ito:1980ev}, the plot on the second row on the right is the comparison to E605 data~\cite{Moreno:1990sf}. The bottom row is a comparison to R209 data~\cite{Antreasyan:1981uv}.}
    \label{fig:IY6_FixedTarget}
\end{figure}

A similar comparison is made to various low energy experiments. In this work, we consider
the E288 experiment~\cite{Ito:1980ev}, the E605 experiment~\cite{Moreno:1990sf}, and the R209 experiment~\cite{Antreasyan:1981uv}.
The E288 experimental data set consists of three different incident
proton energies (200 GeV, 300 GeV, and 400 GeV). The different incident proton energies are separated
into three different sets labeled E288 200, E288 300, and E288 400, respectively. The E288 experiment
consisted of colliding a proton beam on a copper target. The E605 experimental data set was taken at
a center of mass energy of 38.8 GeV with a 800 GeV proton beam on a copper target. The experiment
measured the invariant dimuon cross section ($E\frac{{\rm d}^3\sigma}{{\rm d}^3p}$) for a fixed
longitudinal momentum ($x_F = 0.1$). Finally, the R209 experiment consisted of colliding two proton
beams at a center of mass energy of $\sqrt{s} = 62$ GeV. The data was separated into various invariant
mass windows and was inclusive on the rapidity of the dimuon pair. The comparison to the various low
energy experiments can be found in Fig.~\ref{fig:IY6_FixedTarget}. Overall, we see good agreement
between the ResBos2 IFY fit and the experimental data. Again, the scale uncertainties and PDF
uncertainties are not included to help make it clear the quality of the fit.

Overall, we see excellent agreement between the fitted IFY form and the experimental data.
The fact that the values for $g_5$ and $g_6$ are not consistent with zero supports that there is rapidity
dependence in the non-perturbative contributions to the transverse momentum resummation calculation. The rapidity dependence of the IFY form is most strongly influenced by the LHCb data, specifically the low transverse momentum bins as shown in Tab.~\ref{tab:chi2_LHCb}. 
This is supported by the observation that the numerical contribution of the $g_5$ (and $g_6$) term of Eq.~\eqref{eq:IFY} is negligible for $|y|$ less than about 2.5. Hence, the quality of the fits to the ATLAS and CMS $Z$ boson data is not noticeable altered by the inclusion of this rapidity-dependent term of the IFY form.
A detailed comparison to all the LHC data is left to the following section.

\section{\texorpdfstring{$Z$}{Z} Boson Observables at the LHC}\label{sec:z_obs}

In this work, we compare to the ResBos2 prediction to all available $Z$ boson transverse momentum and
$\phi^*_\eta$ distributions available at 7, 8, and 13 TeV.
The $\phi^*_\eta$ observable was proposed in Refs.~\cite{Vesterinen:2008hx,Banfi:2010cf}
and only depends on the angular distribution of the final state leptons, but has a direct correlation
to the transverse momentum of the $Z$ boson, and was first measured in D0~\cite{D0:2010qhp}. The observable is defined as
\begin{equation}
    \phi^*_\eta = \tan\left(\frac{\pi-\Delta\phi}{2}\right)\sin\left(\theta^*_\eta\right),
\end{equation}
where $\Delta\phi$ is the azimuthal separation of the two leptons and $\theta^*_\eta$ is the
measurement of the scattering angle with respect to the proton beam direction in the rest frame of
the $Z$ boson, i.e.
\begin{equation}
    \cos\left(\theta^*_\eta\right) = \tanh\left(\frac{\eta^- -\eta^+}{2}\right).
\end{equation}
In the above equation, $\eta^{\pm}$ corresponds to the rapidity of the positively or negatively
charged lepton, respectively. In the limit of small transverse momentum of the $Z$ boson, one can
approximate $\phi^*_\eta$ as
\begin{equation}
    \phi^*_\eta \approx \frac{p_T}{M_{\ell\ell}}\sin \phi_{\rm CS},
\end{equation}
where $\phi_{\rm CS}$ is the $\phi$ angle in the Collins-Soper frame~\cite{PhysRevD.16.2219}. This limit demonstrates
the correlation between the transverse momentum of the $Z$ boson and the $\phi^*_\eta$ observable.

Throughout the following sections, the resummation calculation is performed using CT18NNLO~\cite{Hou:2019efy},
with the central scale given by $C_1 = b_0$, $C_2 = 1$, $C_3=4b_0$, $\mu_F=M_T$, and $\mu_R=M_T$, and
using the IFY non-perturbative functional form described in the previous section. Additionally, we
vary all the scales by factors of two under the constraints that the ratio of all scaling factors
is in the range $1/2 < s_1/s_2 < 2$, where $s_1$ and $s_2$ are two of the scaling factors. 
Furthermore, we require that the scale factor for $C_1$, $C_3$, and $\mu_F$ are kept the same with
each other since all of these scales relate to the cut-off between the non-perturbative region
and the perturbative region. After all the above constraints, we are left with 15 scale variations.
To obtain the scale uncertainty, we take the maximum variation of all the scales in each bin forming
an envelope uncertainty. Additional details on the scale variations and the dependency
of the scales for the $A$, $B$, and $C$ coefficients are given in App.~\ref{app:ScaleVariation}.

\begin{figure}
    \centering
    \includegraphics[width=0.3\textwidth]{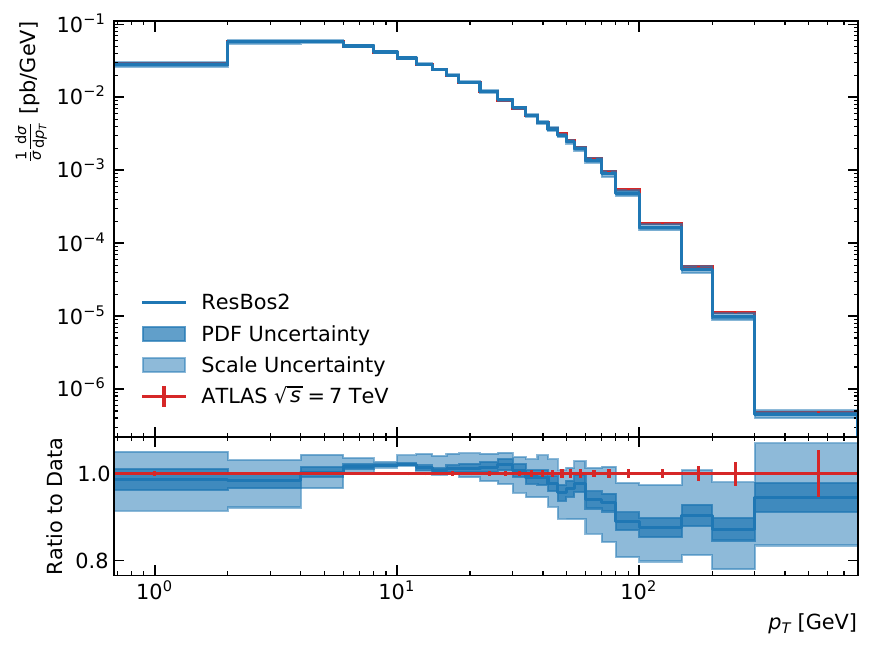} \hfill
    \includegraphics[width=0.3\textwidth]{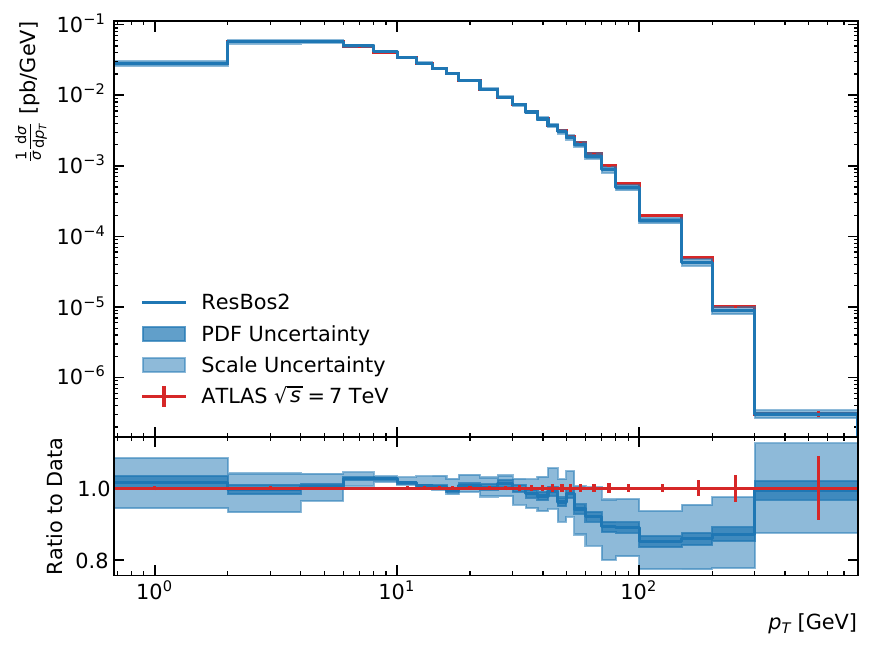} \hfill
    \includegraphics[width=0.3\textwidth]{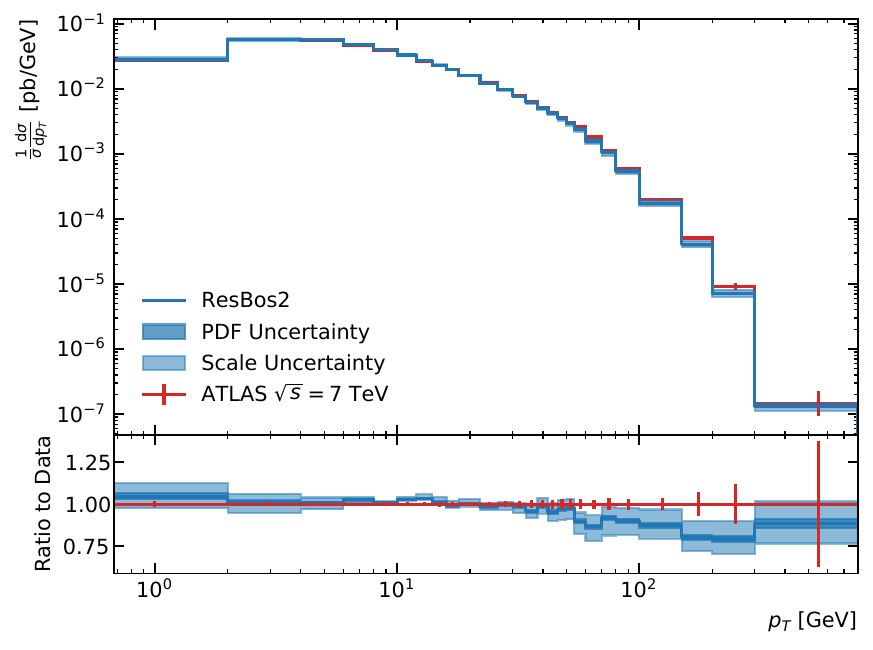} \\
    \caption{Comparison between the ResBos2 calculation and the ATLAS 7 TeV $p_T$ distributions from Ref.~\cite{ATLAS:2014alx}.}
    \label{fig:ATLAS7TeV_pT}
\end{figure}

\begin{figure}
    \centering
    \includegraphics[width=0.3\textwidth]{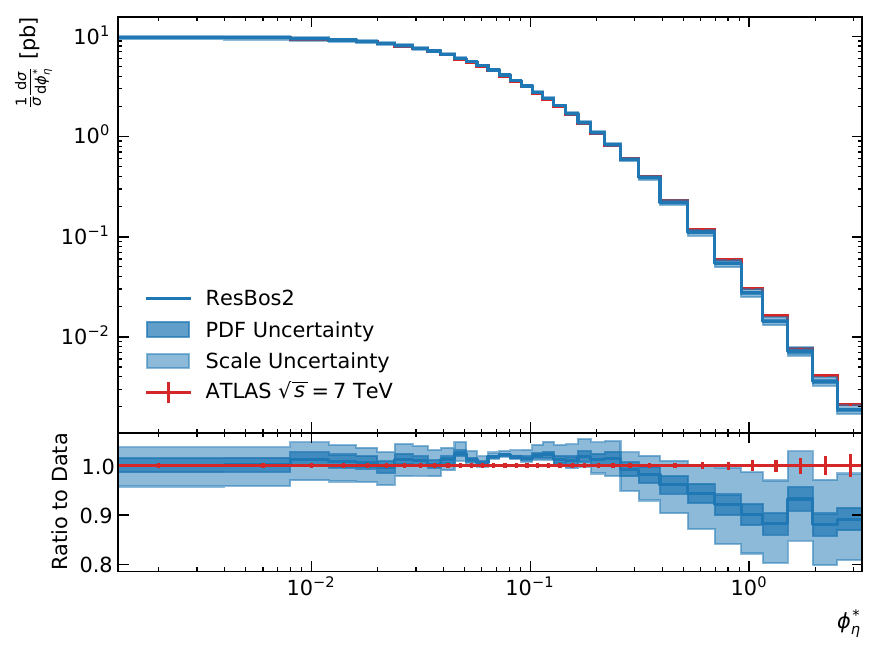} \hfill
    \includegraphics[width=0.3\textwidth]{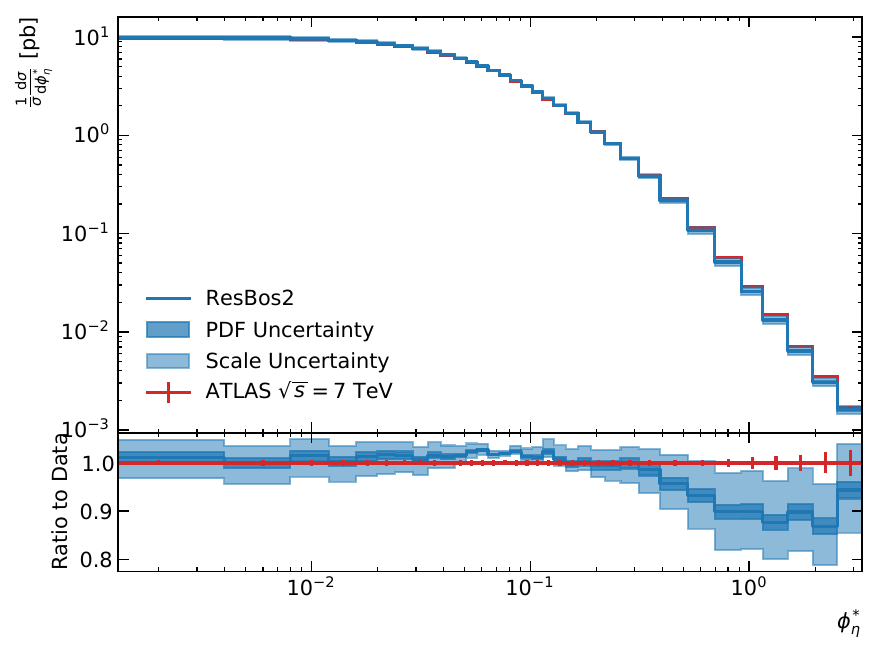} \hfill
    \includegraphics[width=0.3\textwidth]{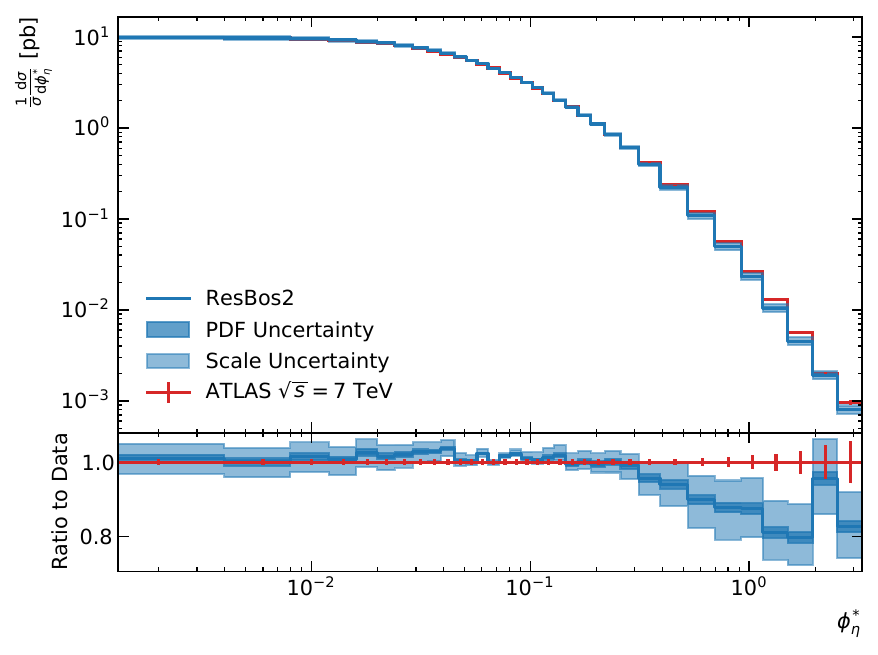} \\
    \caption{Comparison between the ResBos2 calculation and the ATLAS 7 TeV $\phi^*_\eta$ distributions from Ref.~\cite{ATLAS:2012ewf}.}
    \label{fig:ATLAS7TeV_Phi}
\end{figure}

\begin{figure}
    \centering
    \includegraphics[width=0.3\textwidth]{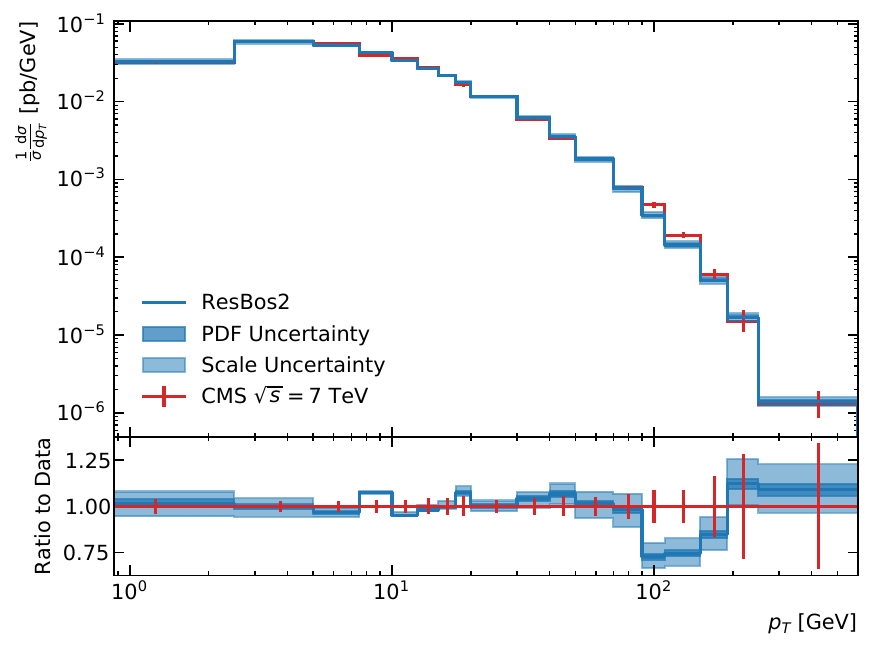}
    \caption{Comparison between the ResBos2 calculation and the CMS 7 TeV $p_T$ distributions from Ref.~\cite{CMS:2011wyd}.}
    \label{fig:CMS7TeV_pT}
\end{figure}

The 7 TeV data sets consists of a $p_T$ and $\phi^*_\eta$ measurement from ATLAS~\cite{ATLAS:2014alx,ATLAS:2012ewf} and a $p_T$ measurement from CMS~\cite{CMS:2011wyd}.
The comparisons can be see in Figs.~\ref{fig:ATLAS7TeV_pT},~\ref{fig:ATLAS7TeV_Phi}, and~\ref{fig:CMS7TeV_pT}. In each of these calculations, the data is given with
the error bars denoting the quadrature sum of the systematic and statistical uncertainties. The ResBos2 prediction with the IFY non-perturbative fit described above is given
with the PDF uncertainty in dark blue and the scale uncertainty combined with the PDF uncertainty in light blue. We can see that the PDF uncertainty is typically on the order of
a few percent, while the scale uncertainty is about 5\% for small $p_T$ and $\phi^*_\eta$ and growing to about 10\% at high $p_T$ and $\phi^*_\eta$. There is a disagreement
between the ResBos2 prediction and the experimental data at high transverse momentum and $\phi^*_\eta$. This discrepancy can be accounted for by the N${}^3$LO corrections
not included in the ResBos2 calculation as seen in Ref.~\cite{Gehrmann-DeRidder:2016cdi}. Overall, the ResBos2 prediction shows excellent agreement with the data in the small $p_T$ and $\phi^*_\eta$ regions,
in the intermediate region the agreement begins to degrade due to the break-down of the validity of the resummation calculation and corrections related to matching that are left to a future work.

\begin{figure}
    \centering
    \includegraphics[width=0.3\textwidth]{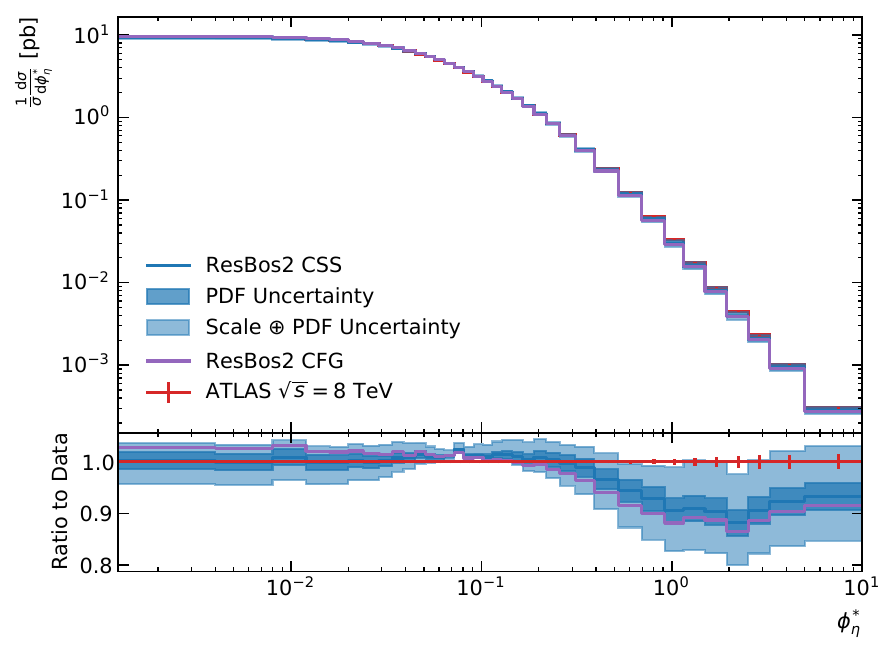} \hfill
    \includegraphics[width=0.3\textwidth]{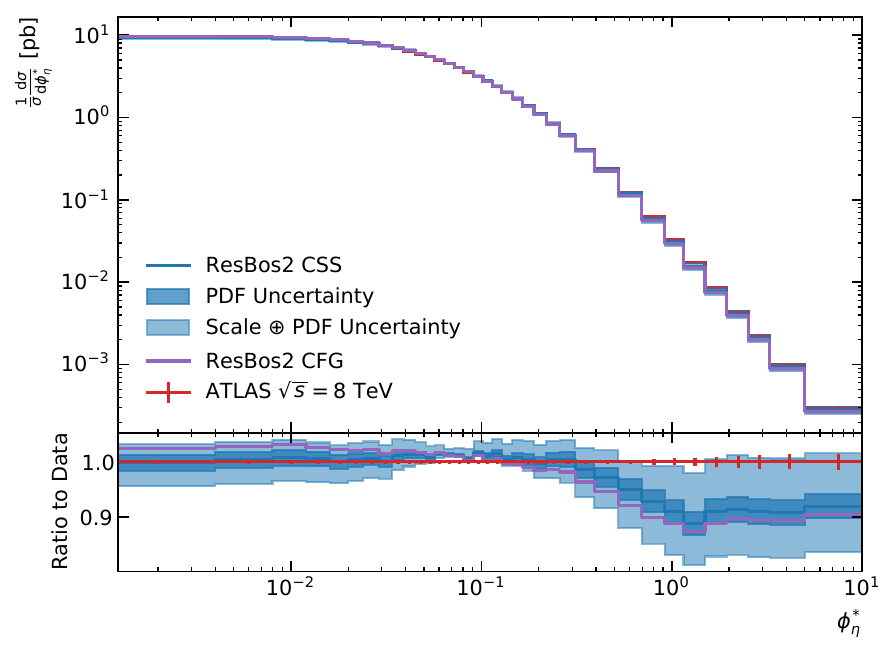} \hfill
    \includegraphics[width=0.3\textwidth]{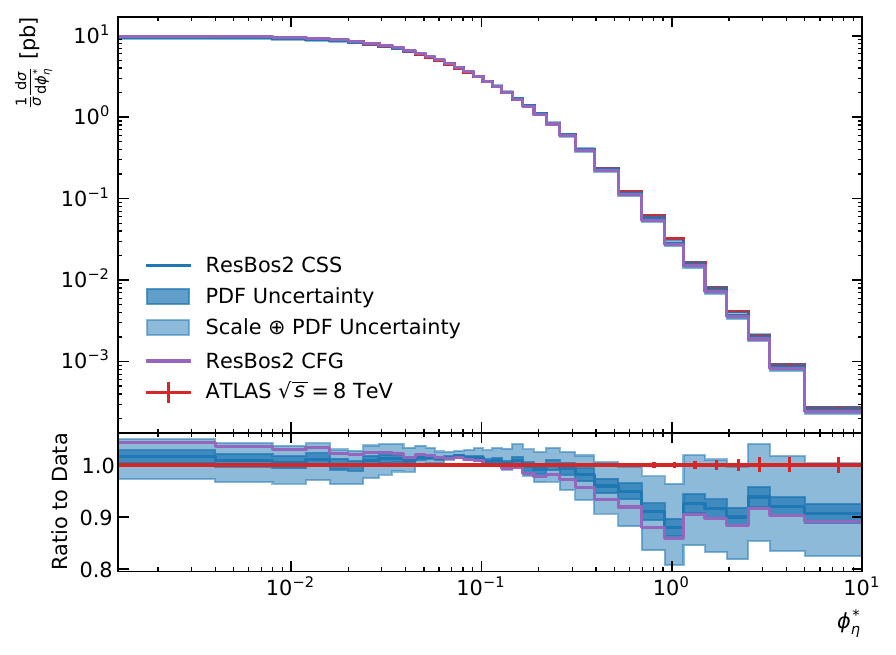} \\
    \includegraphics[width=0.3\textwidth]{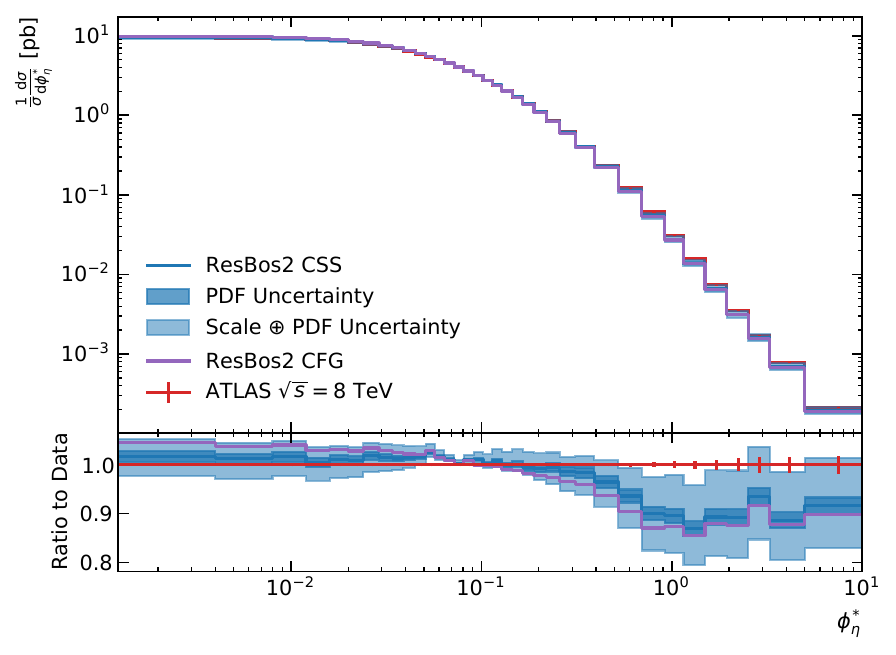} \hfill
    \includegraphics[width=0.3\textwidth]{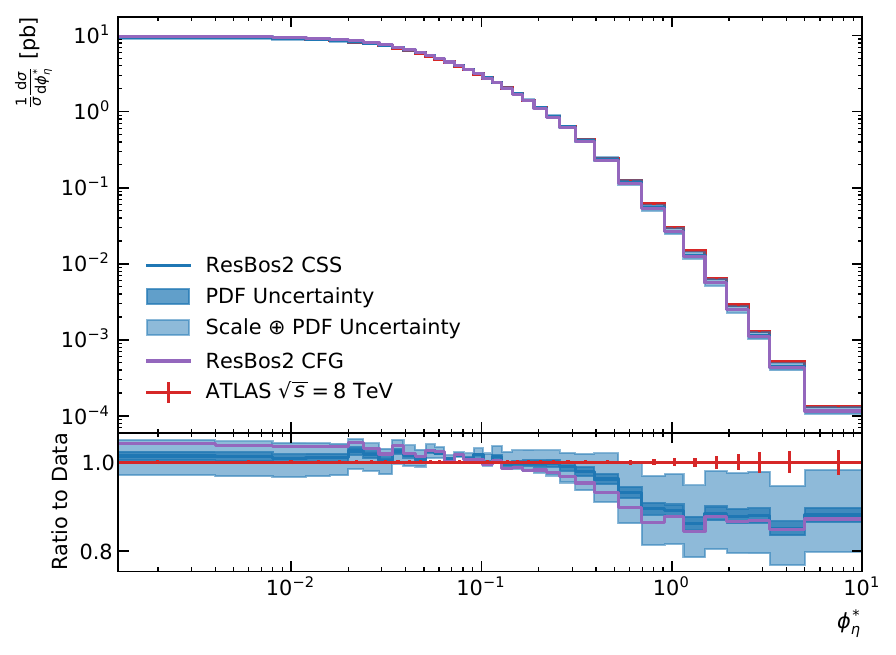} \hfill
    \includegraphics[width=0.3\textwidth]{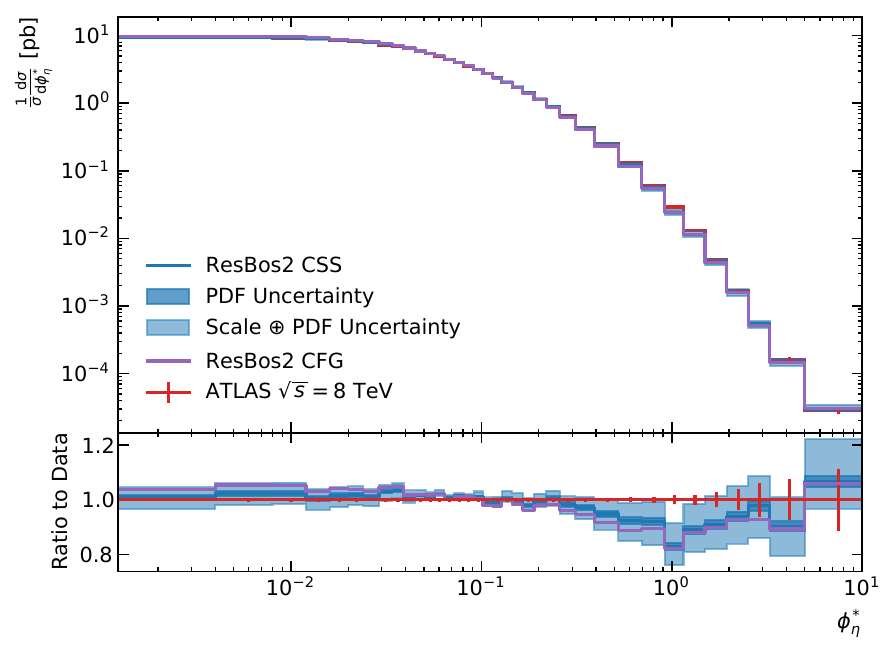}
    \caption{Comparison between the ResBos2 calculation and the ATLAS 8 TeV $\phi^*_\eta$ distributions from Ref.~\cite{ATLAS:2015iiu}.}
    \label{fig:ATLAS8TeV_Phi}
\end{figure}

The complete LHC 8 TeV dataset on $Z$ boson transverse momentum and $\phi^*_\eta$ comes from both ATLAS~\cite{ATLAS:2015iiu} and CMS~\cite{CMS:2016mwa}. The comparison between the ResBos2 calculation
and the experimental data shows very similar features as the comparison to the 7 TeV data, as expected. Again, we see excellent agreement in the small transverse momentum (or $\phi^*_\eta$) region, a disagreement between the
theory prediction and the data in the intermediate region due to needed improvements in the matching region, and a prediction below the data at high transverse momentum that can be improved by matching to the N${}^3$LO prediction.
As an example, we show in Fig.~\ref{fig:ATLAS8TeV_Phi} the 
comparison between the ResBos2 calculation and the ATLAS 8 TeV $\phi^*_\eta$ distributions from Ref.~\cite{ATLAS:2015iiu}.
In this figure, we also include the comparison to the prediction of CFG resummation formalism, calculated with the same choice of scales and the central set of CT18 NNLO PDFs. It shows that the predictions of CSS and CFG 
resummation formalisms agree well within the scale uncertainty of the CSS prediction. We have also confirmed that both formalisms yield almost identical predictions when using the canonical scales in the numerical calculations.

\begin{figure}
    \centering
    \includegraphics[width=0.3\textwidth]{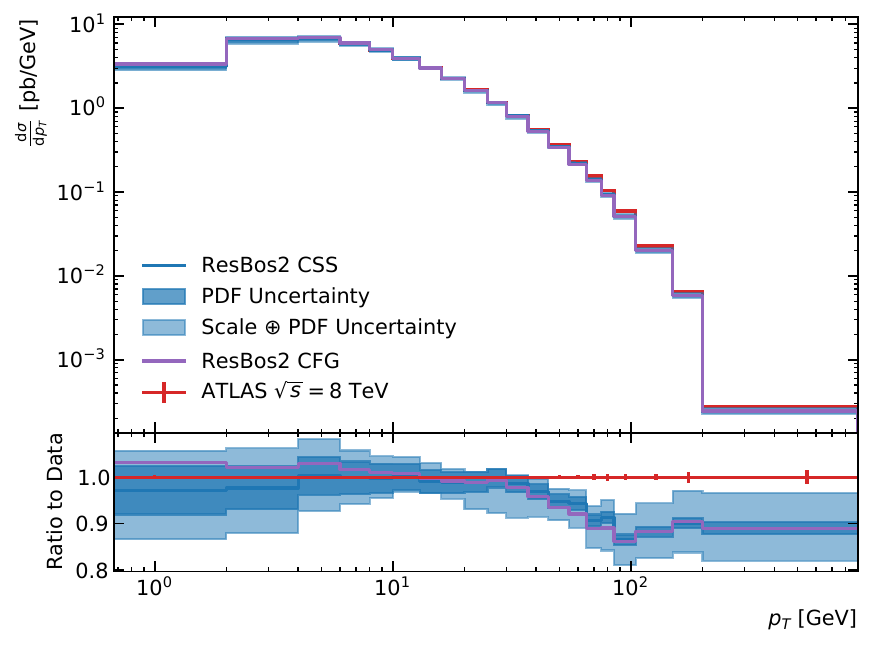} \hfill
    \includegraphics[width=0.3\textwidth]{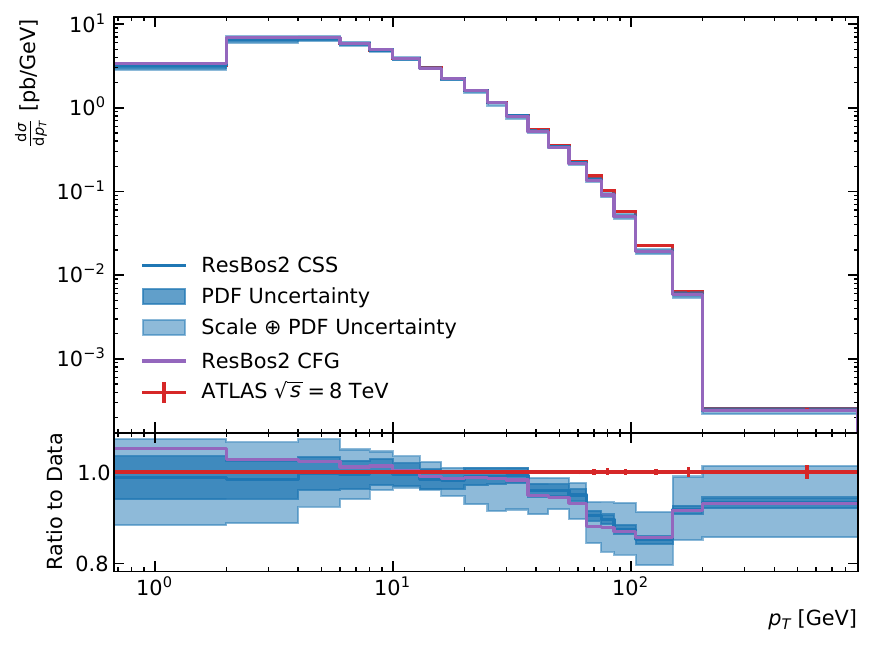} \hfill
    \includegraphics[width=0.3\textwidth]{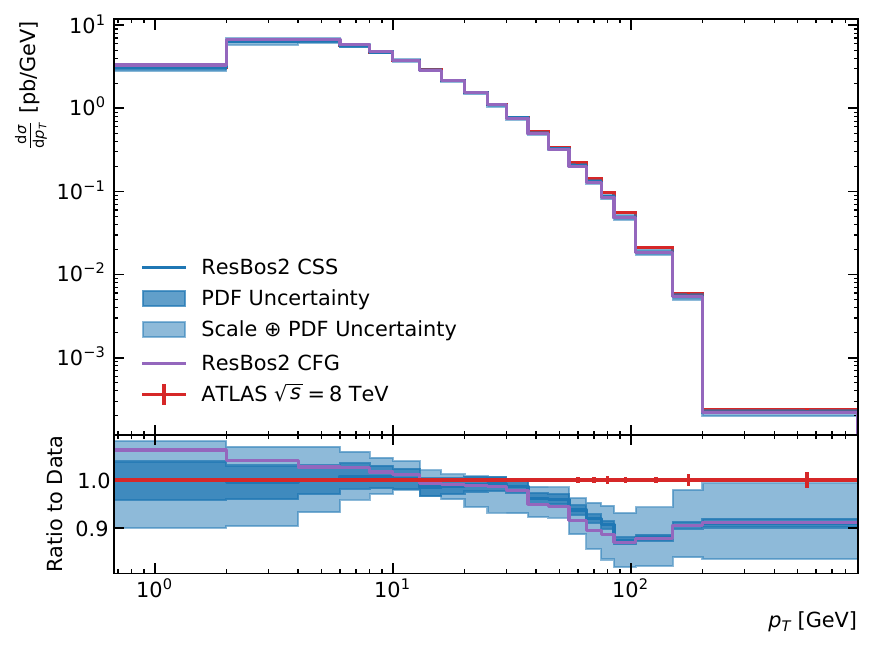} \\
    \includegraphics[width=0.3\textwidth]{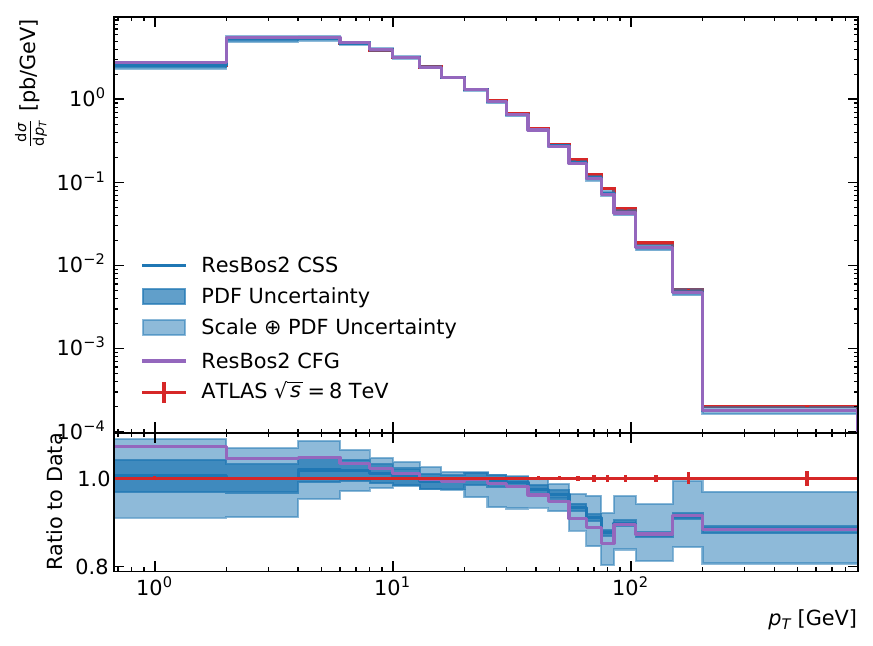} \hfill
    \includegraphics[width=0.3\textwidth]{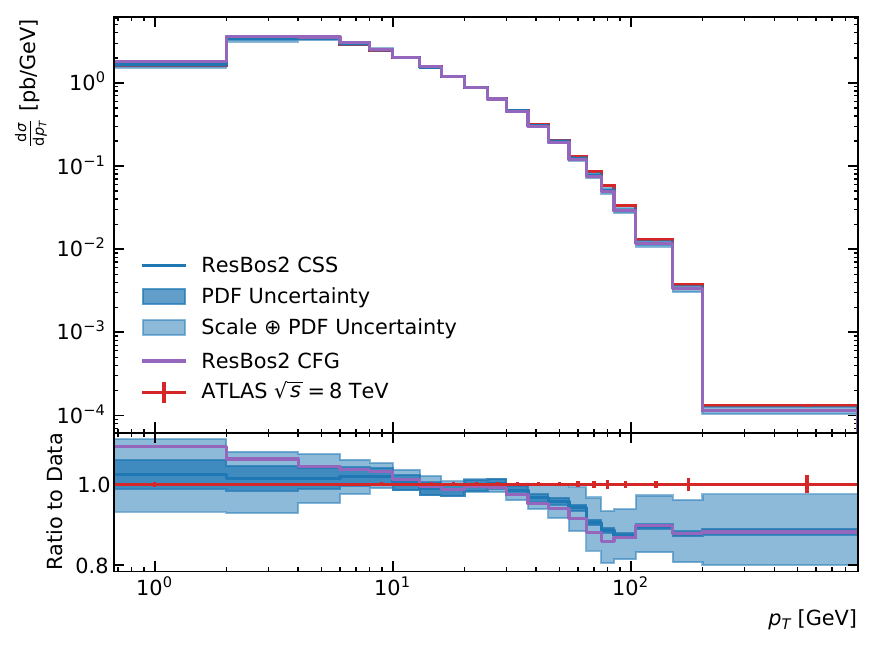} \hfill
    \includegraphics[width=0.3\textwidth]{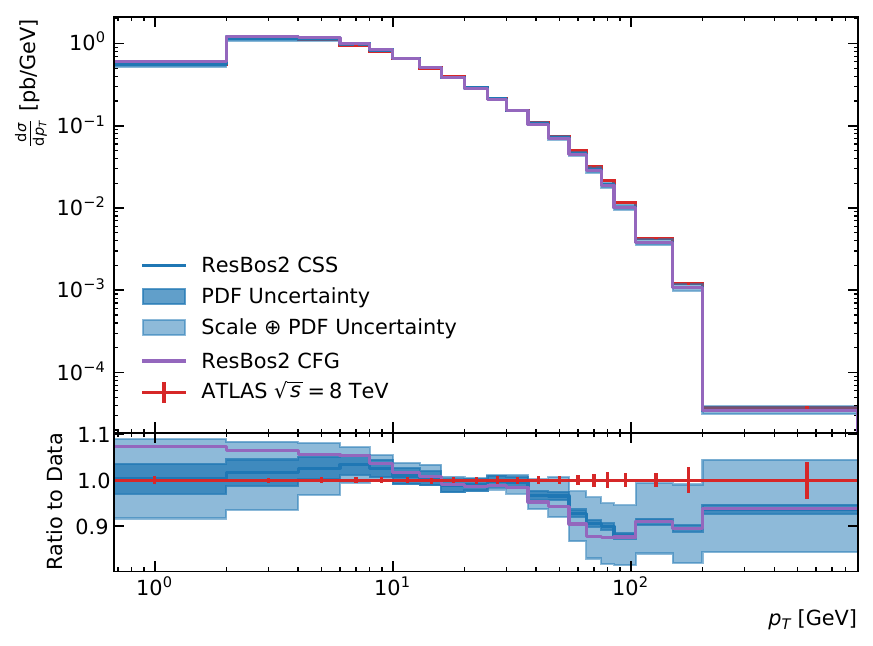} 
    \caption{Comparison between the ResBos2 calculation and the ATLAS 8 TeV $p_T$ distributions from Ref.~\cite{ATLAS:2015iiu}.}
    \label{fig:ATLAS8TeV_pT}
\end{figure}

\begin{figure}
    \centering
    \includegraphics[width=0.3\textwidth]{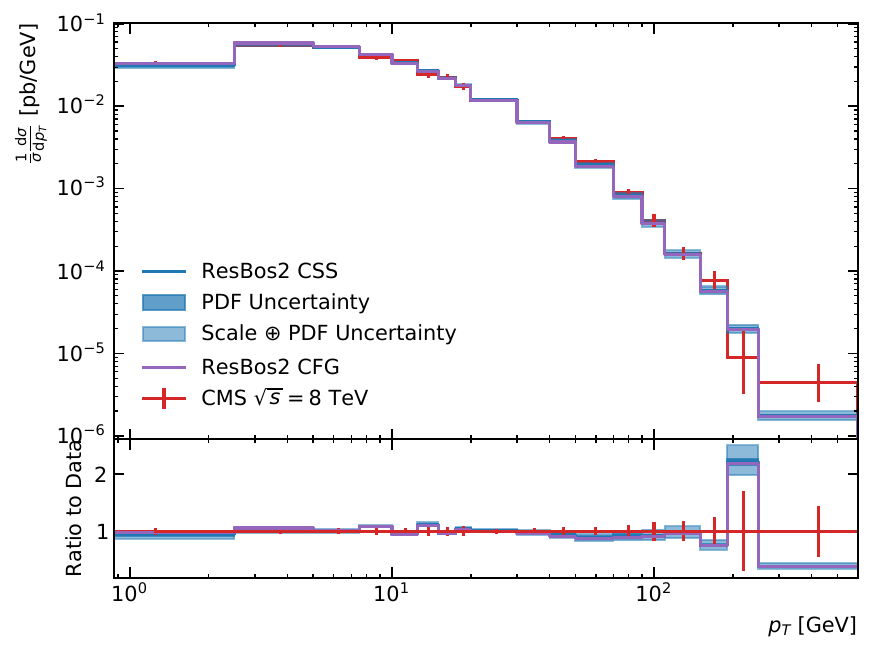}
    \caption{Comparison between the ResBos2 calculation and the CMS 8 TeV $p_T$ distributions from Ref.~\cite{CMS:2016mwa}.}
    \label{fig:CMS8TeV_pT}
\end{figure}

Finally, we compare the ResBos2 predictions to the LHC 13 TeV datasets on $Z$ boson transverse momentum and $\phi^*_\eta$ distributions. The measurements at 13 TeV come from ATLAS~\cite{ATLAS:2019zci},
CMS~\cite{CMS:2019raw}, and LHCb~\cite{LHCb:2021huf}. The overall agreement between the ResBos2 prediction and the experimental data is consistent with what is seen at 7 and 8 TeV. However,
there appears to be better agreement in the intermediate transverse momentum and $\phi^*_\eta$ region compared to 7 and 8 TeV. Again, the disagreement at large transverse momentum can be addressed
with the inclusion of higher order corrections.

\begin{figure}
    \centering
    \includegraphics[width=0.3\textwidth]{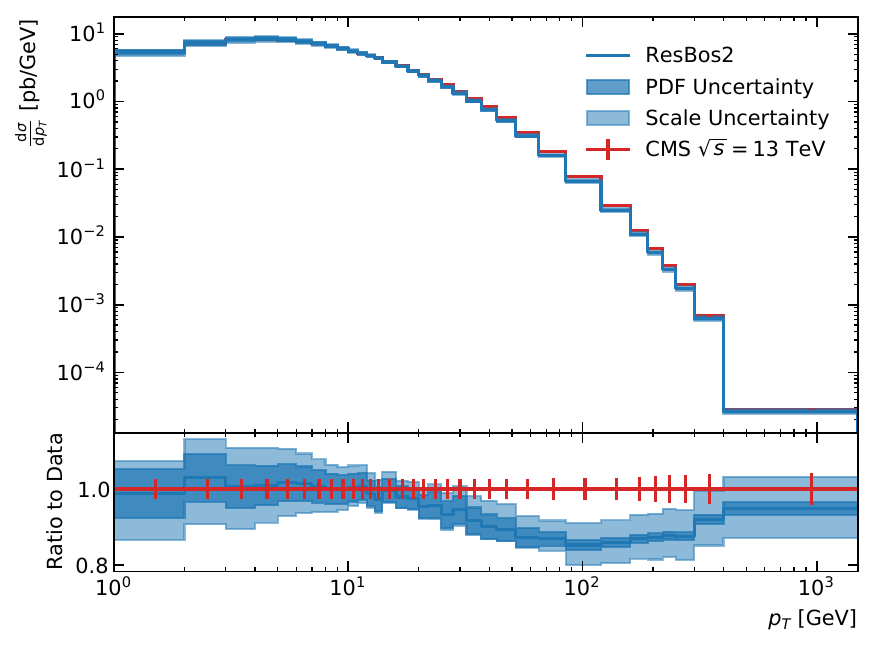} \hfill
    \includegraphics[width=0.3\textwidth]{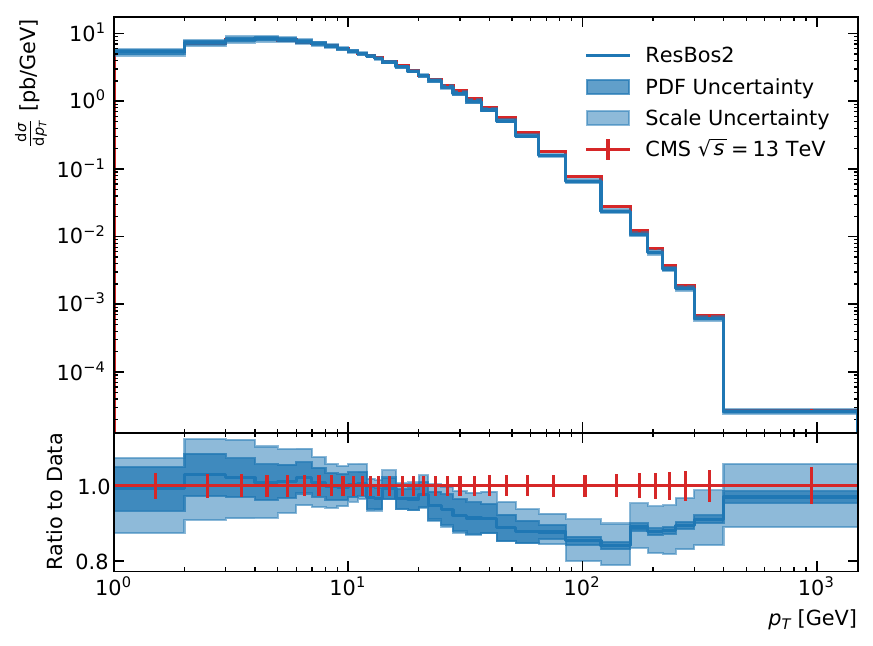} \hfill
    \includegraphics[width=0.3\textwidth]{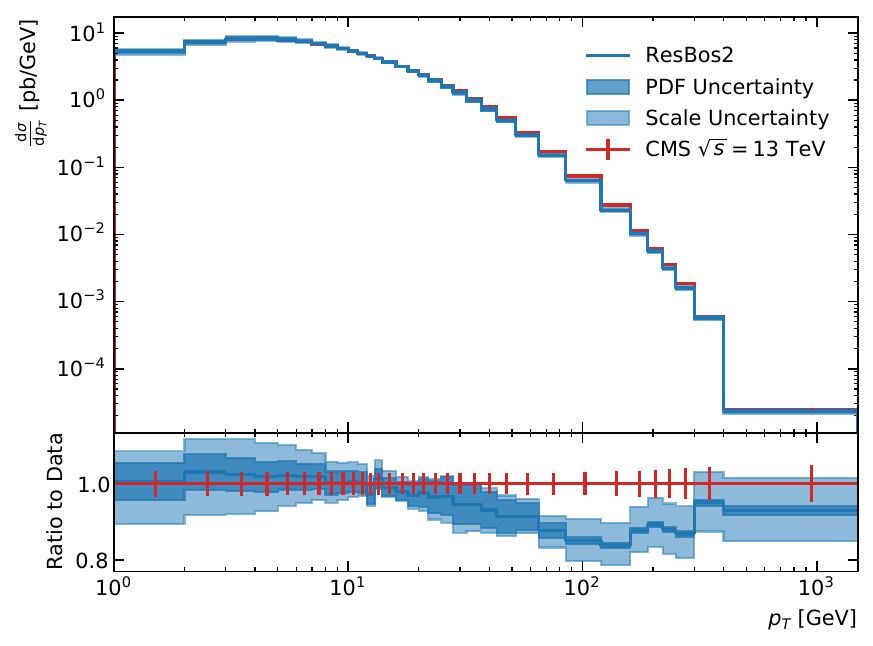} \\
    \includegraphics[width=0.3\textwidth]{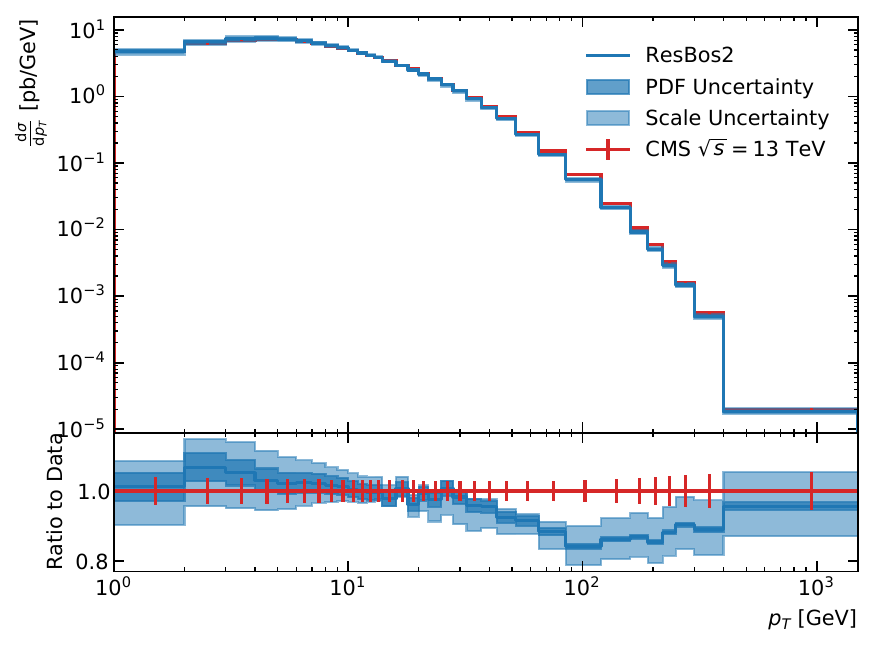} \hfill
    \includegraphics[width=0.3\textwidth]{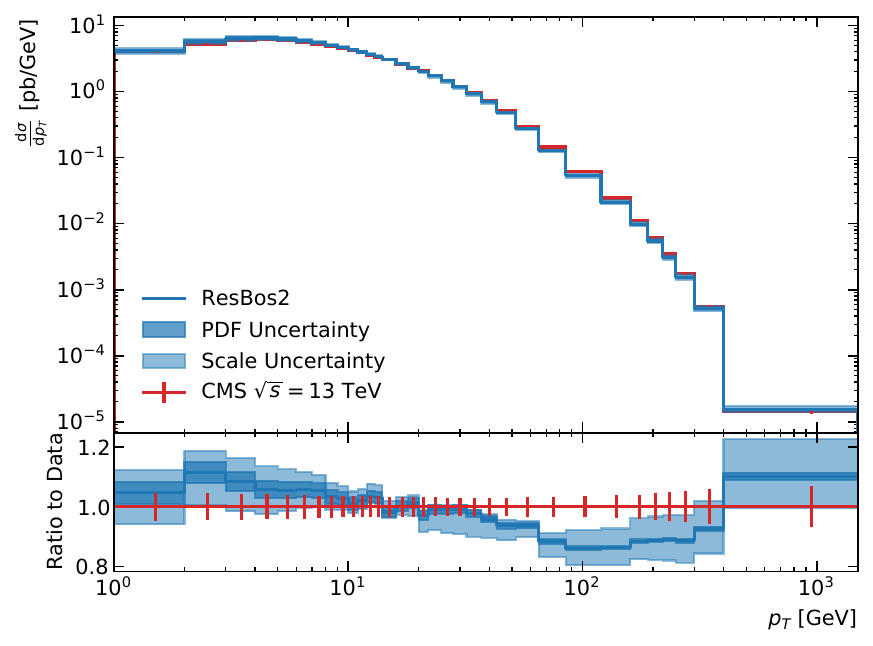} \hspace{0.338\textwidth} \hfill
    \caption{Comparison between the ResBos2 calculation and CMS~\cite{CMS:2019raw} for the $p_T$ distribution in different rapidity bin.}
    \label{fig:CMS13TeV_pT}
\end{figure}

\begin{figure}
    \centering
    \includegraphics[width=0.3\textwidth]{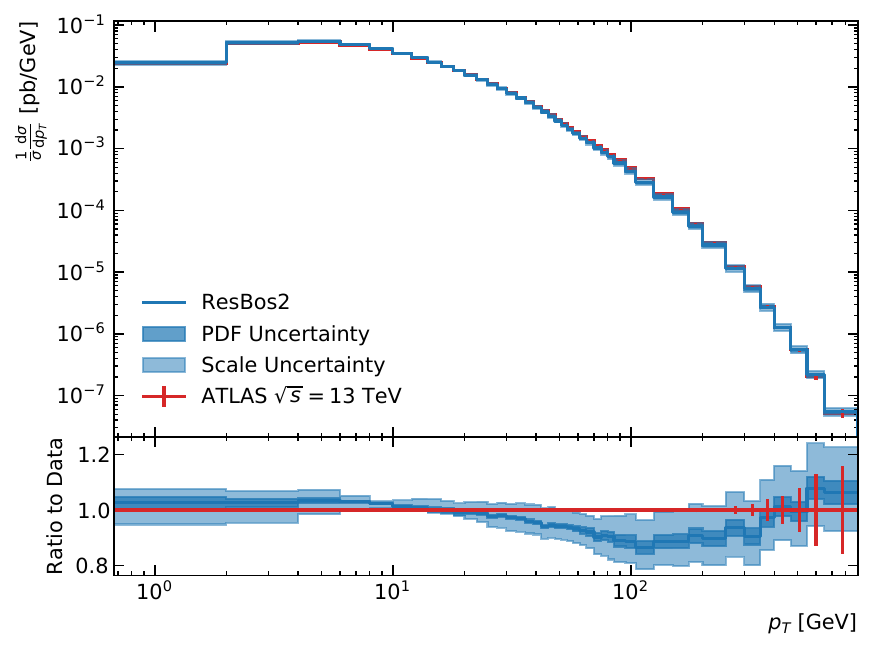} \hfill
    \includegraphics[width=0.3\textwidth]{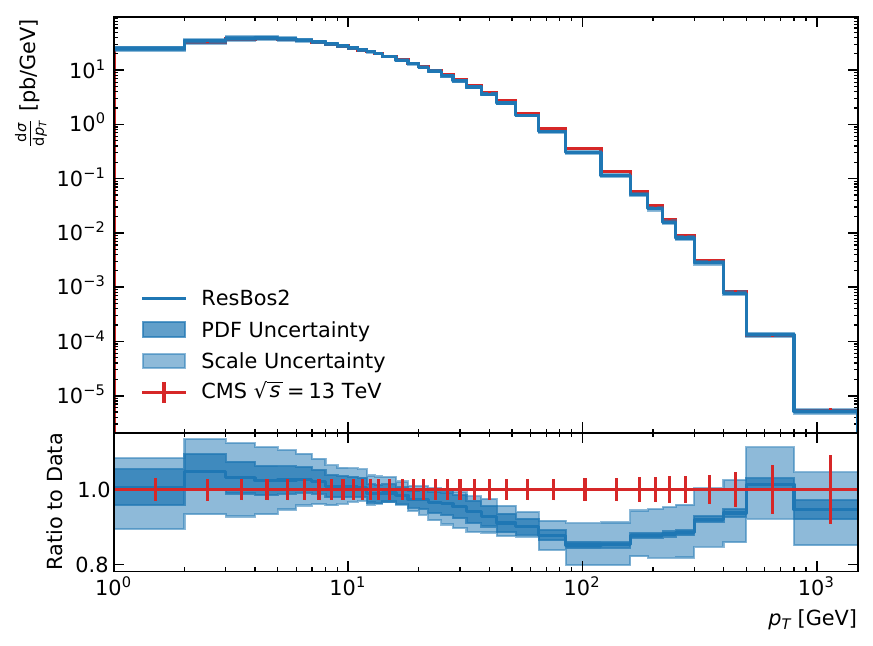} \hfill
    \includegraphics[width=0.3\textwidth]{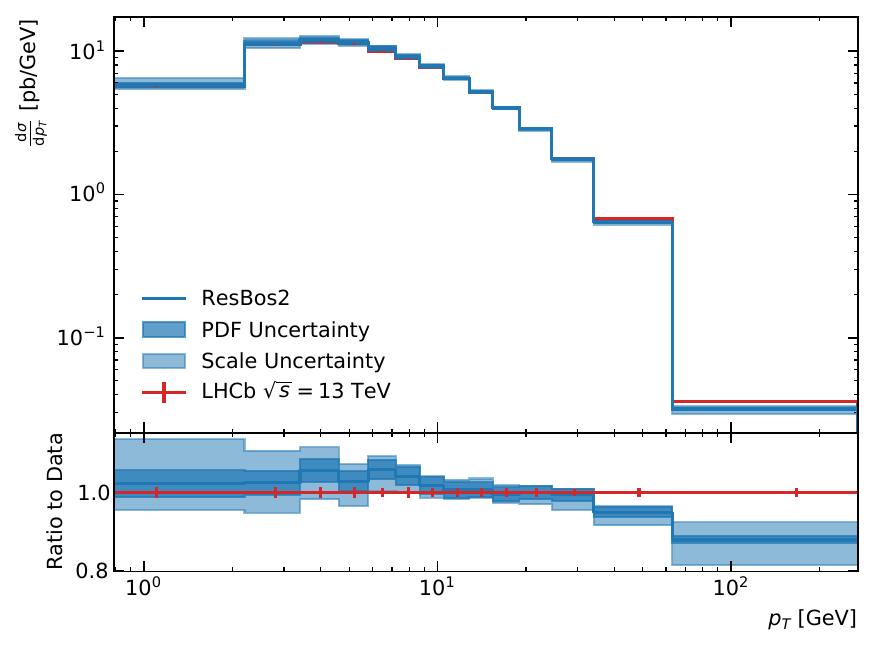}
    \caption{Comparison between the ResBos2 calculation and ATLAS~\cite{ATLAS:2019zci} on the left, CMS~\cite{CMS:2019raw} in the middle, and LHCb~\cite{LHCb:2021huf} on the right for the $p_T$ distribution.}
    \label{fig:LHC13TeV_pT}
\end{figure}

\begin{figure}
    \centering
    \includegraphics[width=0.3\textwidth]{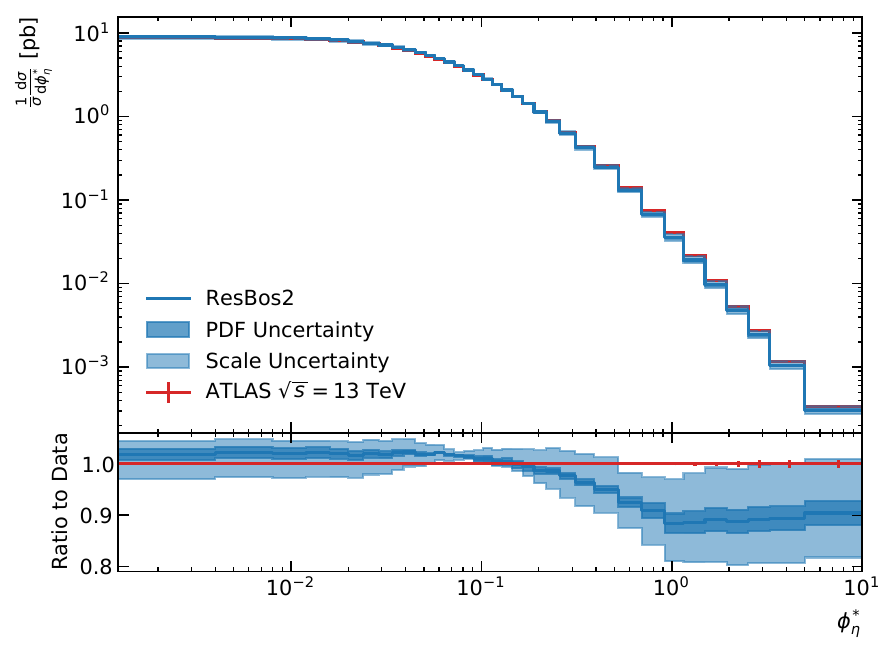} \hfill
    \includegraphics[width=0.3\textwidth]{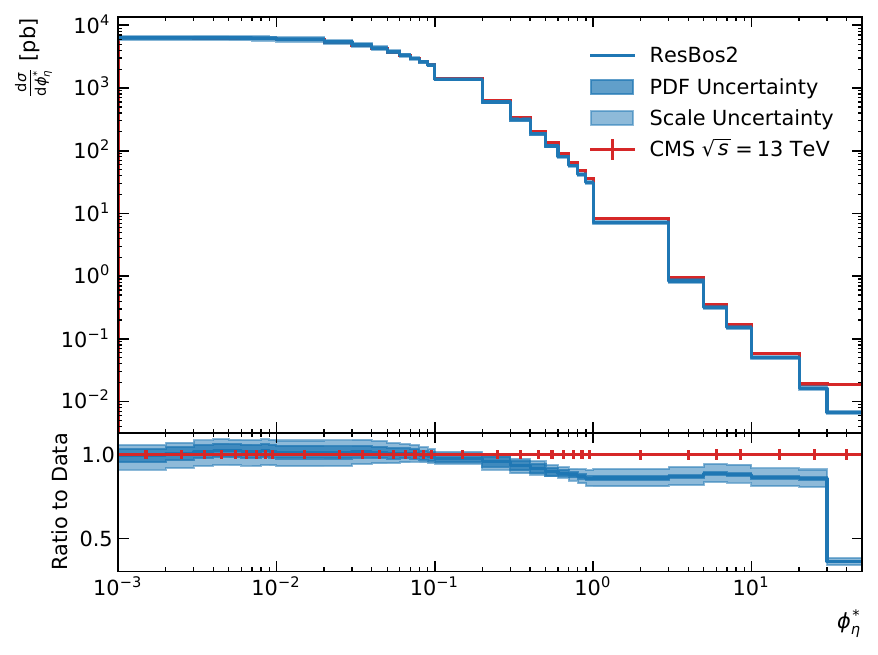} \hfill
    \includegraphics[width=0.3\textwidth]{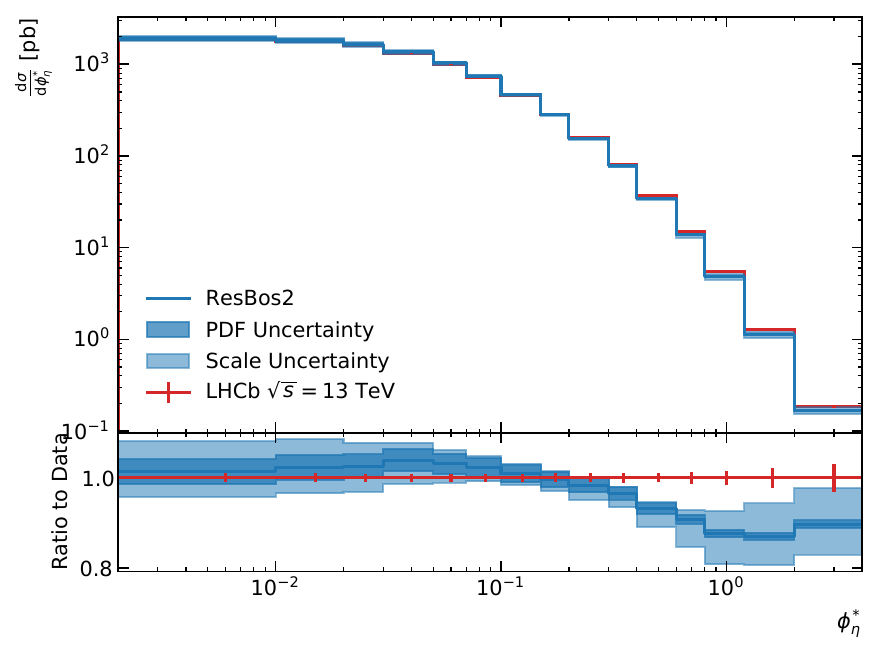}
    \caption{Comparison between the ResBos2 calculation and ATLAS~\cite{ATLAS:2019zci} on the left, CMS~\cite{CMS:2019raw} in the middle, and LHCb~\cite{LHCb:2021huf} on the right for the $\phi^*_\eta$ distribution.}
    \label{fig:LHC13TeV_Phi}
\end{figure}

Overall, ResBos2 can accurately describe the LHC data for $Z$ boson production in the small transverse momentum and $\phi^*_\eta$ regions, as expected. Additionally, since the ResBos2 prediction only matches to NNLO at large transverse momentum, there is missing strength compared to the experimental data that can be resolved by matching to N${}^3$LO. The matching to N${}^3$LO is left to a future work as well.
Here, we only show the detailed comparison to the ATLAS~\cite{ATLAS:2015iiu} and CMS~\cite{CMS:2016mwa} 8 TeV $p_T$ data (Figs.~\ref{fig:ATLAS8TeV_pT}) and ~\ref{fig:CMS8TeV_pT}, respectively), 
the CMS 13 TeV $p_T$ data~\cite{CMS:2019raw} (Fig.~\ref{fig:CMS13TeV_pT}), and the LHCb 13 TeV $p_T$ data~\cite{LHCb:2021huf} (Fig.~\ref{fig:LHC13TeV_pT} right panel). 
The non-perturbative function obtained above is still able to accurately describe the other LHC $p_T$ datasets (Figs.~\ref{fig:ATLAS7TeV_pT},~\ref{fig:CMS7TeV_pT},~\ref{fig:CMS8TeV_pT}, and~\ref{fig:LHC13TeV_pT}) and all of the $\phi^*_\eta$ datasets (Fig. ~\ref{fig:LHC13TeV_Phi}).

\section{\texorpdfstring{$W$}{W} Mass Details}\label{sec:w_mass}

Recently, CDF measured the $W$ mass to be 80,433 $\pm$ 9 MeV~\cite{CDF:2022hxs}.
This is the most precise direct measurement of the $W$ mass.
However, this result deviates from the Standard Model predicted mass of 80,359.1 $\pm$ 5.2 MeV~\cite{deBlas:2021wap} by 7$\sigma$.
The version of ResBos used by the CDF experiment was only accurate to NNLL+NLO. In Ref.~\cite{Isaacson:2022rts} and this work we investigate the impact of including higher order corrections into the study.
Additionally, in this work we delve into additional questions that were raised about the validity of the theory calculation used by CDF besides the formal accuracy of the calculation, including width effects, handling of the scales, detector smearing, and PDF effects.

\subsection{Width Effects}\label{app:width}

\begin{figure}[htpb]
	\centering
	\includegraphics[width=0.47\textwidth]{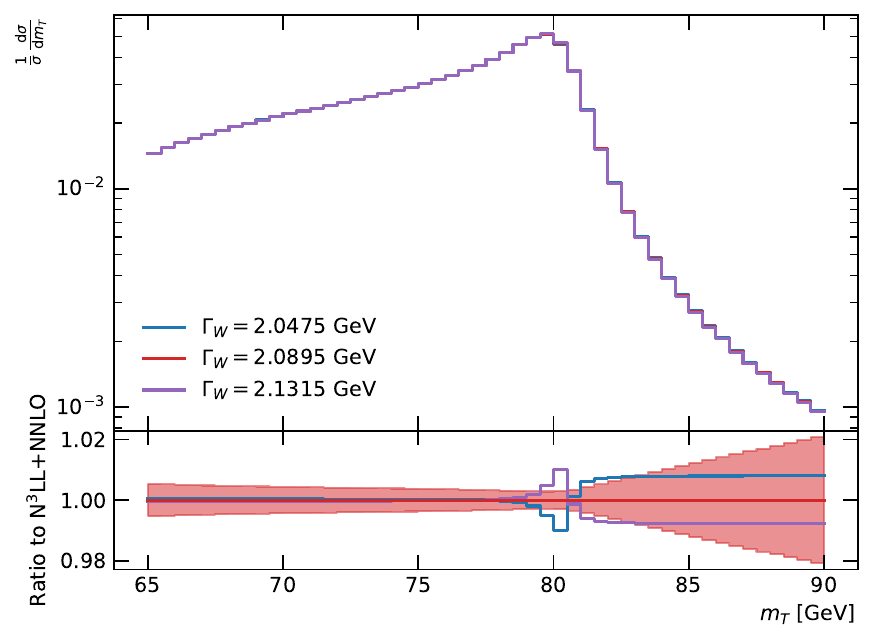}
	\caption{Comparison of the $m_T$ distribution for various different choices of $\Gamma_W$. The width used by CDF was 2.0895 GeV (red curve), and the blue and purple curve represent the shift in the width up and down by one standard deviation of the uncertainty quoted by the PDG~\cite{ParticleDataGroup:2020ssz}.}
	\label{fig:width}
\end{figure}

In the extraction of the $W$ mass, the CDF experiment kept fixed the width of the $W$ boson to 2.0895 GeV. We follow the approach taken by CDF and write the propagator of the $W$ boson as a Breit-Wigner shape with an energy-dependent width. The couplings of gauge bosons to fermions are defined in the $G_\mu$ scheme.
To estimate the impact of varying the width on the CDF result, we varied the width by 0.042 GeV based on the uncertainty from the global width measurement from the PDG~\cite{ParticleDataGroup:2020ssz}. Additionally, a fourth variation was used in which the width was fixed to the Standard Model prediction for the width at NLO, in which the width is proportional to $M_W^3$. The experimental observable most sensitive to the width is $m_T=\sqrt{2 \left(p_T(\ell) p_T(\nu) - \vec{p}_T(\ell) \cdot \vec{p}_T(\nu)\right)}$, and thus we only preformed the extraction for this observable. The effect of the width on $m_T$ can be found in Fig.~\ref{fig:width}, where the red uncertainty band gives the statistical uncertainty from CDF. It is clear that the effect of the width is important at high $m_T$ (\textit{i.e.} $m_T > 80$ GeV). Table~\ref{tab:width} shows the extracted mass shift for the different mass scenarios described above.

\begin{table}[htpb]
	\centering
	\begin{tabular}{|c|c|}
		\hline
		Width & Mass Shift [MeV] \\
		\hline
		2.0475 GeV & 2.0 $\pm$ 0.5 \\
		2.1315 GeV & 0.3 $\pm$ 0.5 \\
		NLO & 1.2 $\pm$ 0.5 \\
		\hline
	\end{tabular}
	\caption{The shift in $M_W$ due to changing the width. The width is varied by the uncertainty from the PDG~\cite{ParticleDataGroup:2020ssz}, with the central value set to 2.0895 GeV used by the CDF collaboration~\cite{CDF:2022hxs}. Additionally, the Standard Model prediction for the width at NLO is considered.}
	\label{tab:width}
\end{table}

\subsection{The ratio for the normalized \texorpdfstring{$p_T(W)$}{pT(W)} to the normalized \texorpdfstring{$p_T(Z)$}{pT(Z)}}
\label{app:ratio_pT}

\begin{figure}[htbp]
	\includegraphics[width=0.47\textwidth]{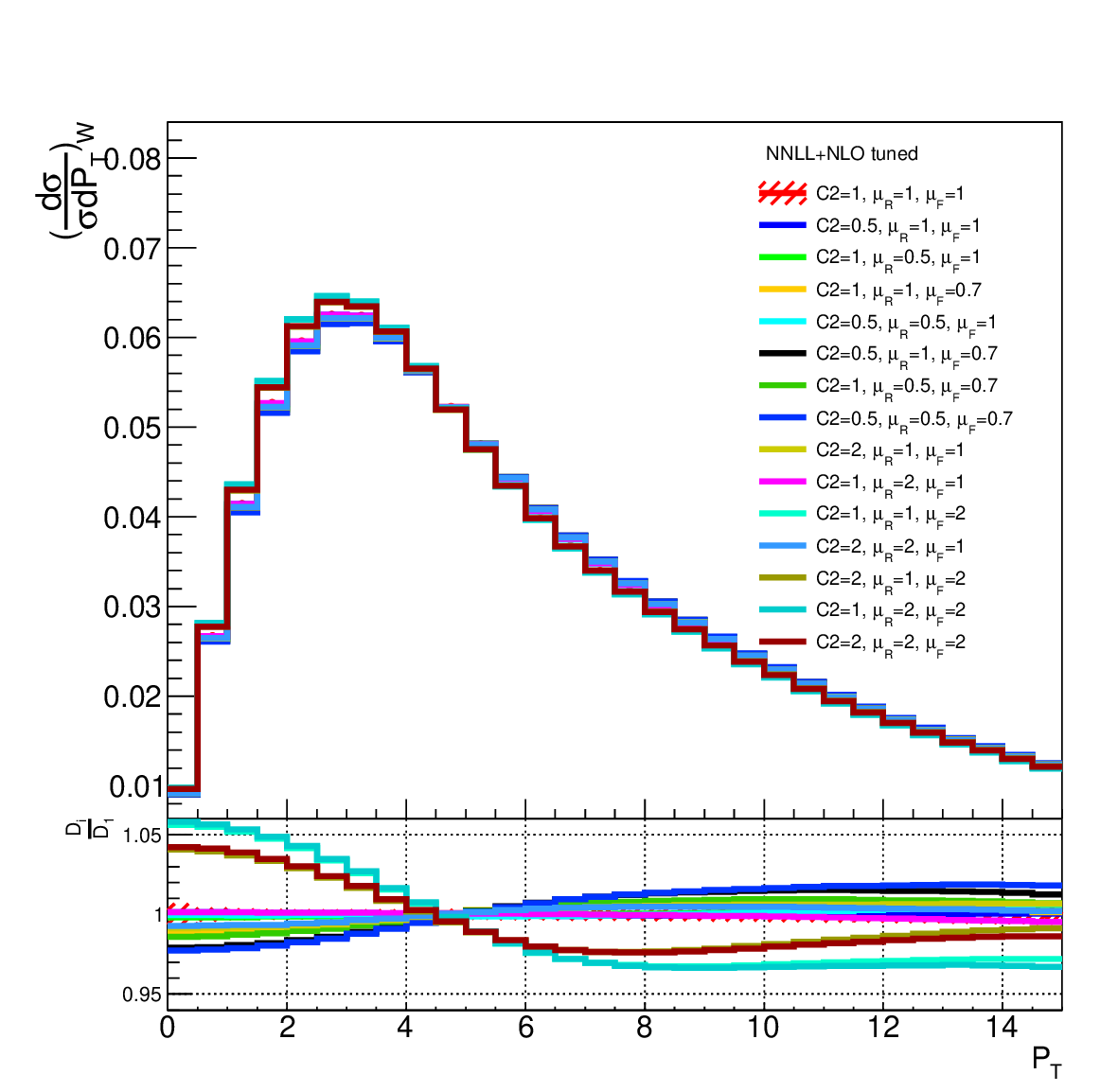}
	\caption{Normalized $P_T(W)$ distributions due to QCD scale variation.}
	\label{fig:pTW_15scale}
\end{figure}

A new feature in the recent CDF measurement was to use the ratio of the normalized $p_T(W)$ distribution to the normalized $p_T(Z)$ distribution to further constrain their systematic uncertainties. To estimate the theory uncertainty in this ratio, the CDF experiment used the DYQT code~\cite{Bozzi:2008bb,Bozzi:2010xn} instead of the ResBos code.
Here, we follow the procedure carried out by CDF. In this study, the scales are treated as fully correlated between $p_T(W)$  and $p_T(Z)$ 
in the ``tuned'' ResBos NNLL+NLO predictions (\textit{i.e.} the non-perturbative parameters in the ResBos calculation are modified to reproduce the $Z$ boson data as described in Ref.~\cite{Isaacson:2022rts}).
Figure~\ref{fig:pTW_15scale} shows in the upper panel the normalized $p_T(W)$ distribution, $D_i=(\frac{d \sigma}{\sigma d p_T})_W(i)$, for each of the 15 scale choices, and in the lower panel the ratio of $D_i/D_1$, where $D_1$ corresponds to the canonical scale choice. By applying the envelope method~\cite{CDF:2022hxs}, CDF further constrained the allowed QCD scale variation in the normalized $p_T(W)$ distribution by fitting to their measurement of the normalized $p_T(W)$ distribution with $\Delta \chi^2=1$. 
Given the allowed range of the normalized $p_T(W)$ distribution, one could extract the fitted value of $M_W$ from the corresponding $m_T$, $p_T(\ell)$ and $p_T(\nu)$ distributions, respectively. This result is shown in Table~\ref{tab:mw_envelope}.
We have checked that a similar conclusion also holds when using the N${}^3$LL+NNLO predictions.

\begin{table*}[ht]
	\centering
	\scalebox{0.9}{
		\begin{tabular}{|c|c|c|c|c|c|c|}
			\hline
			& \multicolumn{6}{c|}{Mass Shift [MeV]} \\
			\hline
			& \multicolumn{2}{c|}{$m_T$} & \multicolumn{2}{c|}{$p_T(\ell)$} & \multicolumn{2}{c|}{$p_T(\nu)$} \\
			\hline
			Scale & \textsc{ResBos2} & +Detector Effect+FSR & \textsc{ResBos2} & +Detector Effect+FSR & \textsc{ResBos2} & +Detector Effect+FSR \\
			\hline
			Upper & 1.2 $\pm$ 0.5 & 0.8 $\pm$ 1.8 $\pm$ 1.1 & 3.1 $\pm$ 2.1 & -6.5 $\pm$ 2.7 $\pm$ 1.3 & 1.4 $\pm$ 2.1 & -4.9 $\pm$ 3.4 $\pm$ 2.0 \\
			\hline
			Lower & 1.2 $\pm$ 0.5 & -0.7 $\pm$ 1.8 $\pm$ 01. & 1.8 $\pm$ 2.1 & 9.4 $\pm$ 2.6 $\pm$ 1.2 & 0.0 $\pm$ 2.1 & 4.8 $\pm$ 3.4 $\pm$ 1.9 \\
			\hline
		\end{tabular}
	}
	\caption{The shift in $M_W$ due to QCD scale variation in the ratio of the normalized $p_T(W)$ distribution to the normalized $p_T(Z)$ distribution, following the envelope method carried out by CDF.}
	\label{tab:mw_envelope}
\end{table*}

\subsection{Detector Smearing and Final State Radiation Effects}\label{app:detector}

\begin{figure}[htpb]
    \centering
    \includegraphics[width=0.47\textwidth]{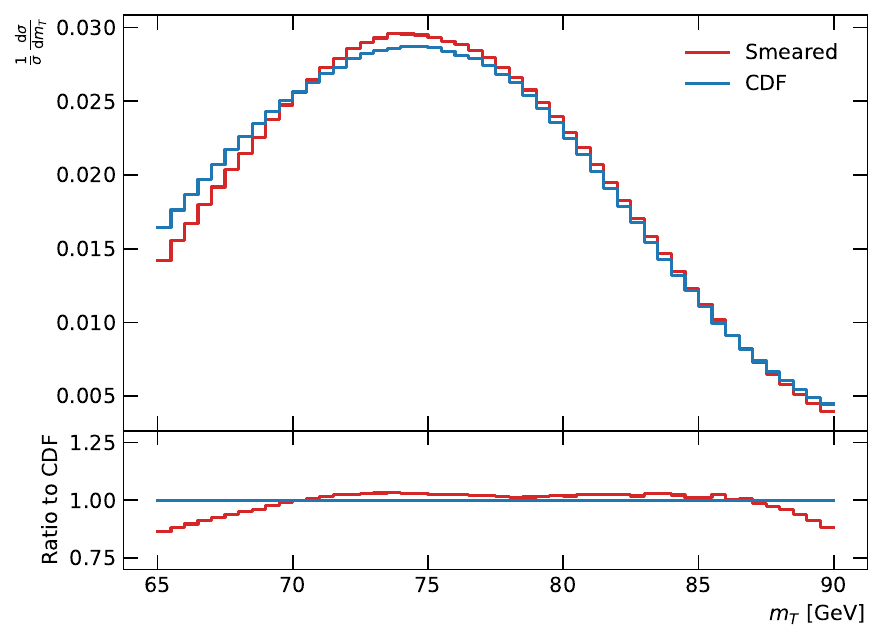}
    \caption{Comparison of the smeared $m_T$ distribution to the CDF data. The red curve is the result of the smearing, and the blue curve is the extracted CDF data in the electron channel.}
    \label{fig:detector_smear}
\end{figure}

The \textsc{ResBos} code does not contain final state QED radiation (FSR) nor detector effects in the calculation. These effects smear out the observables used to measure the $W$ mass, especially $m_T$. To investigate the impact of these effects, we parameterize the smearing effect and fit it to the data observed by CDF. In particular, we chose to study the electron channel since the impact of FSR and the background are both smaller than the muon channel. The comparison of the tuned smearing to the CDF data can be seen in Fig.~\ref{fig:detector_smear}. We used a three parameter fit to determine the width of the Gaussian based on the energy of the particle given by
$
\frac{\sigma}{E} = a \oplus \frac{b}{\sqrt{E}} \oplus \frac{c}{E},
$
where $a$, $b$, and $c$ are the parameters of the fit, and the terms are added in quadrature to obtain the width. Both the pseudodata and the mass templates are smeared using the same functional form, and the results are given in Tab.~II of Ref.~\cite{Isaacson:2022rts}. This form will not capture the detection efficiency nor all of the FSR effects, but can reproduce the general shape of the $m_T$ distribution shown in Fig.~\ref{fig:detector_smear}. We leave a detailed study of FSR to a future work, and require additional information from the CDF collaboration to accurately model the detector acceptance.

To estimate the impact of the approximation of the detector response on the extraction of $M_W$, we preform an additional check in which a simple Gaussian with a width of 5\% for electrons and 11\% for neutrinos is used to smear the results. The comparison between the two approaches are given in Tab.~\ref{tab:detector}. We find that the choice of smearing does not have an impact on the extracted result of $M_W$, so long as the smearing of the data and the templates are identical. The accuracy of the model used by CDF to smear the theory templates is beyond the scope of this work, but the inaccuracy of our model compared to that used by CDF does not change the conclusions drawn within this work.

\begin{table}[htpb]
    \centering
    \begin{tabular}{|c|c|c|}
        \hline
        & \multicolumn{2}{c|}{Mass Shift [MeV]} \\
        \hline
        Observable & Smearing 1 & Smearing 2 \\
        \hline
        $m_T$ & 0.2 $\pm$ 1.8 $\pm$ 1.0 & 1.0 $\pm$ 2.1 $\pm$ 1.3 \\
        $p_T(\ell)$ & 4.3 $\pm$ 2.7 $\pm$ 1.3 & 4.5 $\pm$ 2.6 $\pm$ 1.4  \\
        $p_T(\nu)$ & 3.0 $\pm$ 3.4 $\pm$ 2.2 & 3.8 $\pm$ 4 $\pm$ 2.7\\
        \hline
    \end{tabular}
    \caption{Summary of the shift in $M_W$ due to two different smearing methods. The first uncertainty denotes the statistical uncertainty, and the second uncertainty results from an approximate model simulating the detector effect and FSR, calculated from generating 100 different smearings on the data. ``Smearing 1'' refers to the fit result to the CDF data, and ``Smearing 2'' refers to the crude Gaussian smearing.}
    \label{tab:detector}
\end{table}

\subsection{PDF-induced Correlations}
\label{app:pdf_correlations}

\begin{figure}
    \centering
    \includegraphics[width=0.47\textwidth, clip, trim=10mm 25mm 0mm 15mm]{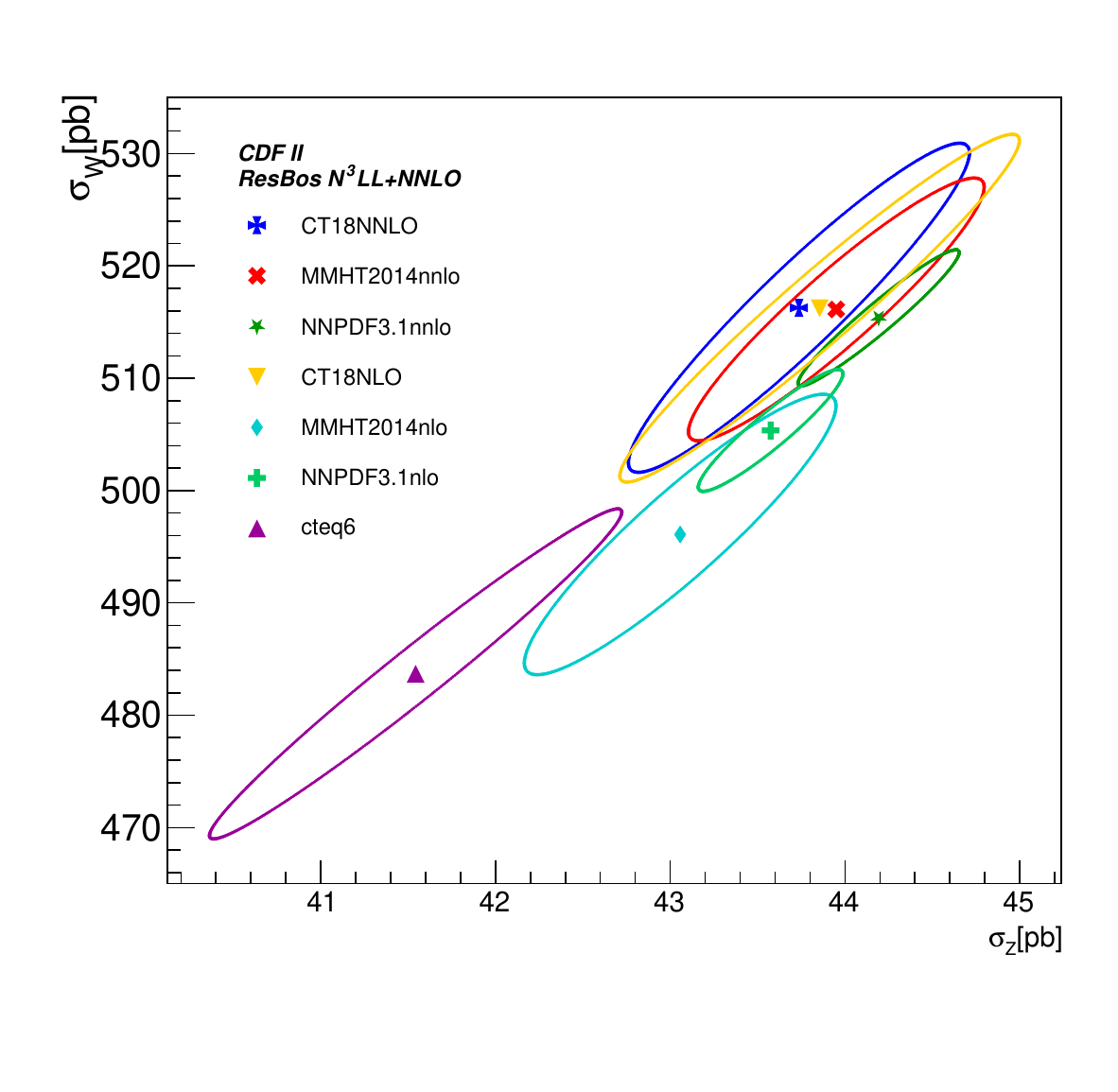}
    \caption{PDF-induced correlation ellipses, at the 68\% confidence level (C.L.), between the fiducial cross sections of $W$ and $Z$ boson  production at the Tevatron Run II.}
    \label{fig:Corr_ellipse}
\end{figure}

\begin{figure}
    \centering
    \includegraphics[width=0.47\textwidth, clip, trim=10mm 25mm 0mm 15mm]{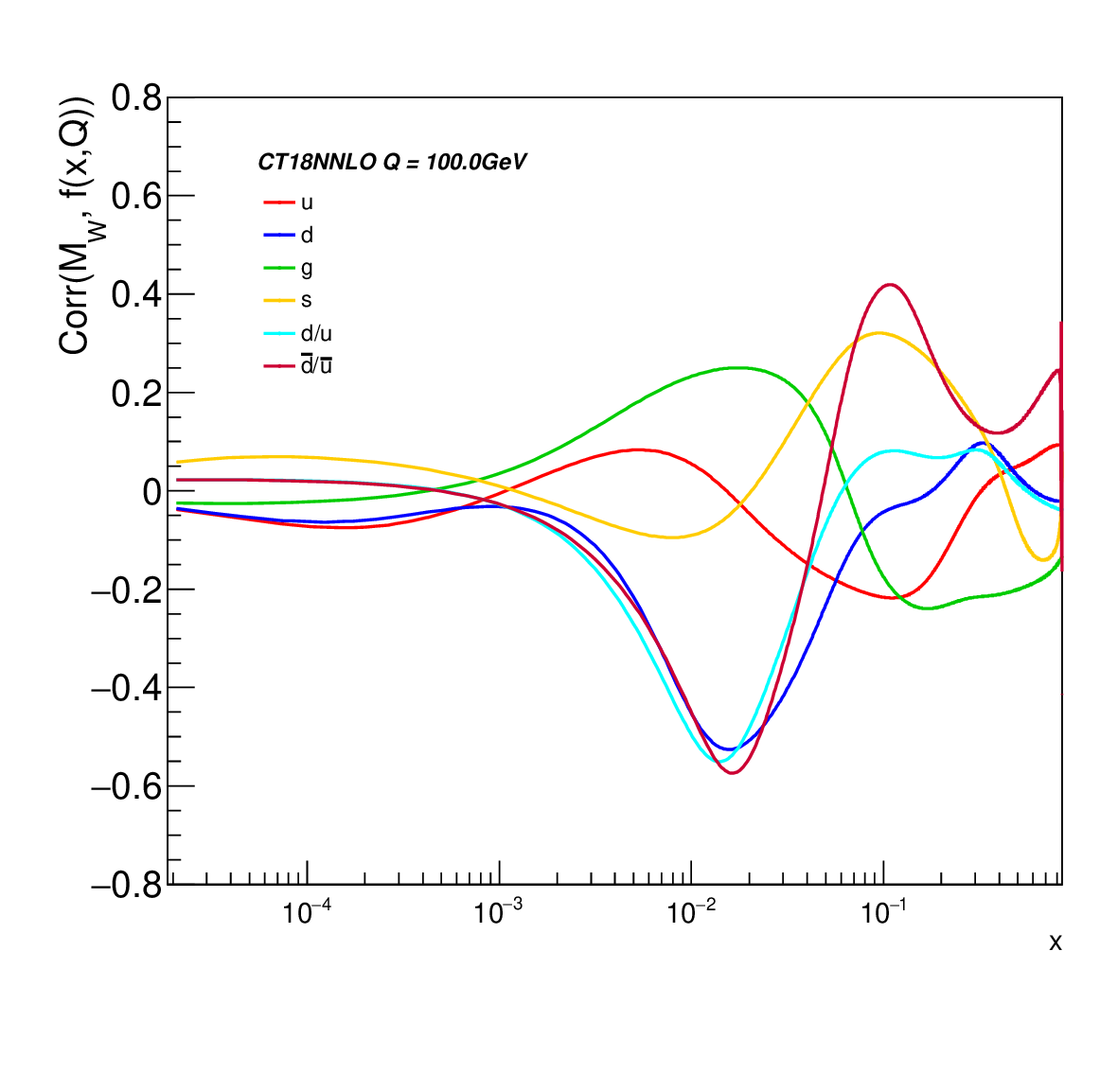}
    \caption{PDF-induced correlation cosine between the extracted $W$ boson mass (from $m_T$ distribution) and the CT18 NNLO PDFs at the specified $x$ value with $Q=100$ GeV. } 
   \label{fig:CorrCosine}
\end{figure}

In Table~\ref{tab:pdf_pt}, we compared the shift of $M_W$ for different PDF sets using $m_T$, $p_T(\ell)$, and $p_T(\nu)$. 
Again, the central prediction used was \texttt{CT18NNLO}~\cite{Hou:2019efy} with a mass of 80,385 MeV. The uncertainties quoted are the PDF uncertainties for the given PDF set. 
As discussed in Ref.~\cite{Isaacson:2022rts}, it is unclear to us how to appropriately propagate the uncertainties from individual observables to the final mass extraction done by CDF. Therefore,
we simply quote the individual values and make no attempt at a combination.

\begin{table*}[htbp]
    \centering
    \begin{tabular}{|c|c|c|c|c|c|c|}
        \hline
         &\multicolumn{2}{c|}{$m_T$} & \multicolumn{2}{c|}{$p_T(\ell)$} & \multicolumn{2}{c|}{$p_T(\nu)$}\\
        \hline
        PDF Set & NNLO & NLO  & NNLO & NLO & NNLO & NLO \\
        \hline
        \texttt{CT18} & 0.0 $\pm$ 1.3 & 1.8 $\pm$ 1.2 & 0.0 $\pm$ 15.9 & 2.0 $\pm$ 14.3 & 0.0 $\pm$ 15.5 & 2.9 $\pm$ 14.2 \\
        \texttt{MMHT2014} & 1.0 $\pm$ 0.6 & 2.6 $\pm$ 0.6 & 6.2 $\pm$ 7.8 & 36.7 $\pm$ 7.0 & 3.9 $\pm$ 7.5 & 36.0 $\pm$ 6.7 \\
        \texttt{NNPDF3.1} & 1.1 $\pm$ 0.3 & 2.1 $\pm$ 0.4 & 2.1 $\pm$ 3.8 & 13.5 $\pm$ 4.9 & 5.4 $\pm$ 3.7 & 10.0 $\pm$ 4.9 \\
        \texttt{CTEQ6M} & N/A  & 2.8 $\pm$ 0.9 & N/A & 19.0 $\pm$ 10.4 & N/A & 20.9 $\pm$ 10.2 \\
        \hline
    \end{tabular}  
	\caption{Comparison of the shift of $M_W$ for different PDF sets using the $m_T$, $p_T(\ell)$ and $p_T(\nu)$ observables, respectively. The central prediction used was \texttt{CT18NNLO} with a mass of 80,385 MeV. The uncertainties quoted are the PDF uncertainties for the given PDF set.
	}
	\label{tab:pdf_pt}
\end{table*}

Below, we briefly summarize a few PDF-induced correlations, predicted at N$^3$LL+NNLO, relevant to the CDF analysis. 

Figure~\ref{fig:Corr_ellipse} shows the PDF-induced correlation ellipses between the fiducial cross sections of $W$ and $Z$ boson productions ($\sigma_W$ vs. $\sigma_Z$) at the Tevatron Run II for various PDF sets. Here, the fiducial region is defined for $Z$ boson events as $p_T(Z) < 15$ GeV, $30 < p_T(\ell) < 55$ GeV, $\lvert\eta(\ell)\rvert < 1$, and $66 < M_{\ell\ell} < 116$ GeV. Similarly, the $W$ boson fiducial region is required to satisfy $p_T(W) < 15$ GeV, $30 < p_T(\ell) < 55$ GeV, $30 < p_T(\nu) < 55$ GeV, $\lvert\eta(\ell)\rvert < 1$, and $60 < m_T < 100$ GeV. Figure~\ref{fig:CorrCosine} displays the PDF-induced correlation of $M_W$ (extracted from the $m_T$ distribution) and CT18 NNLO error PDFs, for various flavors as a function of $x$ at $Q=100$ GeV. It shows that at the typical value of $x$, for the inclusive production of $W$ boson at the Tevatron, around $M_W/\sqrt{S} \simeq 80/1960 \simeq 0.04$, the PDF-induced error in $M_W$ is mainly correlated to that of the PDF-ratios ${\bar d} / {\bar u}$, $d/u$ and $d$-PDFs. 

\begin{figure}
    \centering
	\includegraphics[width=0.47\textwidth, clip, trim=10mm 25mm 0mm 10mm]{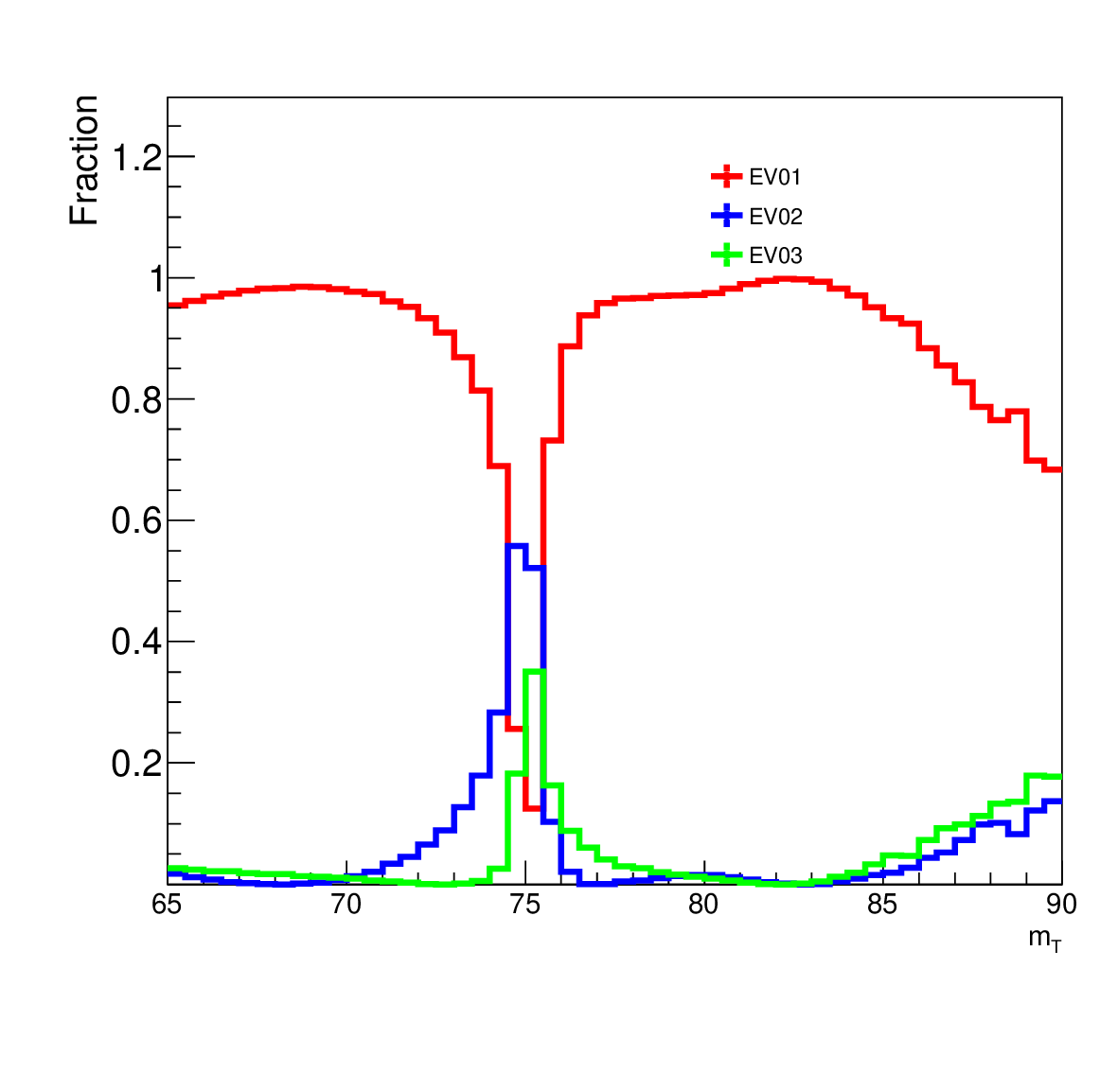}
	\caption{Fractional contribution of the three leading optimized eigenvector PDFs (EV01, EV02 and EV03) to the variance of the $m_T$ distribution, normalized to each bin, obtained from the ePump-optimization analysis.    
}
	\label{fig:frac_ev}
\end{figure}

\begin{figure}
    \centering
    \includegraphics[width=0.47\textwidth]{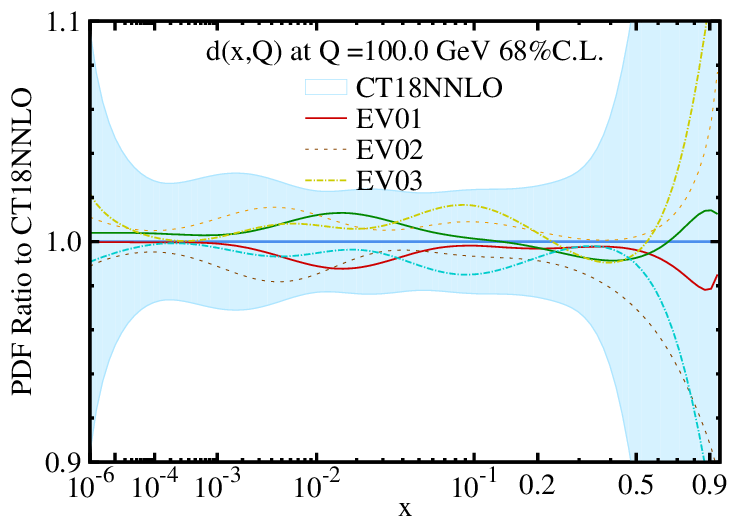}  \hfill
    \includegraphics[width=0.47\textwidth]{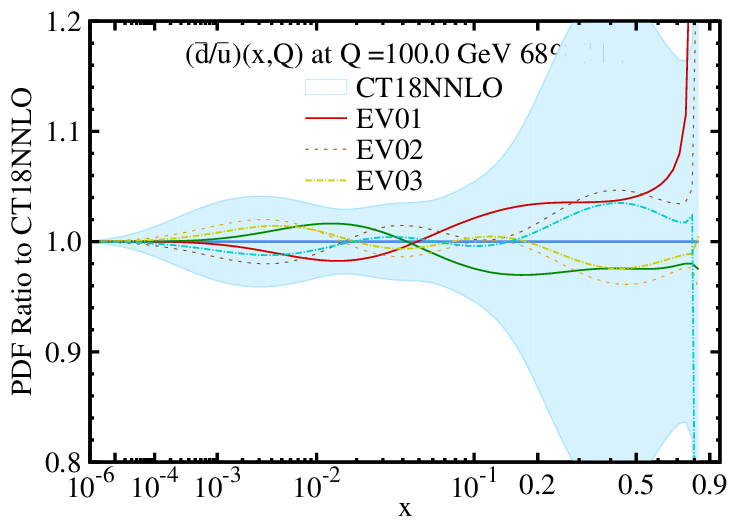} \\
    \includegraphics[width=0.47\textwidth]{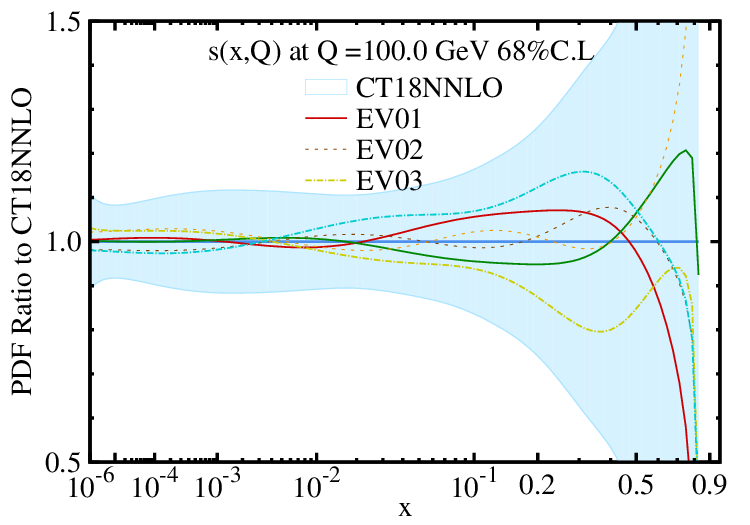} \hfill
    \includegraphics[width=0.47\textwidth]{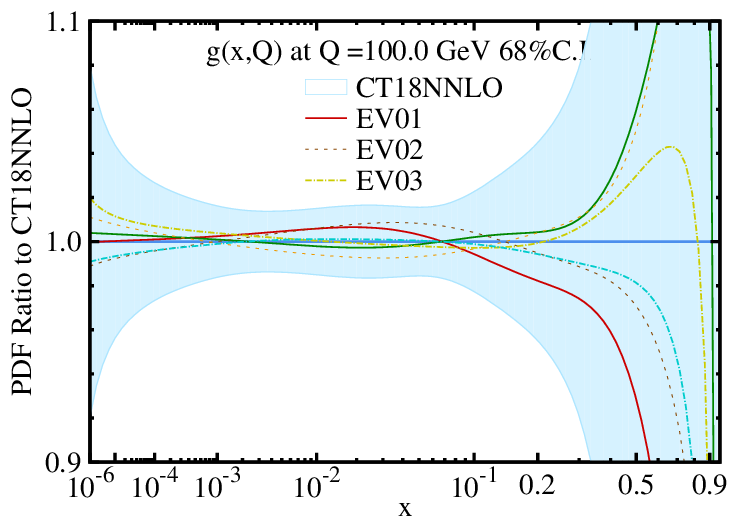}
    \caption{Ratios of the top three pairs of eigenvector PDFs and the original CT18 NNLO error PDFs, at $Q=100$ GeV, to the CT18 NNLO central value of $d$,  $\bar d/\bar u$, $s$ and $g$  PDFs. These eigenvector PDFs were obtained after applying the ePump-optimization to the original CT18 NNLO PDFs with respect to the $m_T$ distribution.}
    \label{fig:PDFFlavor}
\end{figure}

As discussed in Refs.~\cite{Schmidt:2018hvu,Hou:2019gfw}, one can easily find the first few leading eigenvector (EV) sets of error PDFs relevant to a particular experimental observable, such as the $m_T$ distribution, by applying the ePump-optimization procedure. This application of  \textsc{ePump} (error PDF Updating Method
Package) is based on ideas similar to that used in the data set diagonalization
method developed by Pumplin~\cite{Pumplin:2009nm}. It takes a set of Hessian error PDFs and constructs an equivalent set
of error PDFs that exactly reproduces the Hessian symmetric PDF uncertainties, but in addition each new eigenvector pair has an eigenvalue that quantitatively describes its contribution to the PDF uncertainty of a given data set or sets. 
The new eigenvectors can be considered as projecting the original error PDFs to the given data set, and be optimized or re-ordered so that it is easy to choose a reduced set that covers the PDF uncertainty for the input data set to any desired accuracy~\cite{Schmidt:2018hvu,Hou:2019gfw}.
The result is shown in Figs.~\ref{fig:frac_ev} and \ref{fig:PDFFlavor}. The eigenvalues of the three leading EV sets, after applying the ePump-optimization, are $44.5$, $3.0$, $2.4$, respectively. The combination of those top three optimized error PDFs contributes up to $99.6\%$ in the total PDF variance of the 50 given data points, {\it i.e.,} with 50 bins in $m_T$ distribution. This ePump-optimization allows us to conveniently use these three leading new eigenvectors (with a total of six error sets), in contrast to applying the full 58 error sets of the CT18 NNLO PDFs, to study the PDF-induced uncertainty of the  $m_T$ observable. Among them, the leading set EV01 dominates the Jacobian region, for $m_T$ around $M_W$. Hence, one could use those three leading pairs of EV sets to perform Monte Carlo study, such as studying the detector effect on the determination of $M_W$. The contributions provided by those three pairs of eigenvector PDFs to the PDF-induced uncertainty of $m_T$ distribution, for various parton flavors and $x$-ranges, are depicted in Fig.~\ref{fig:PDFFlavor}. The first eigenvector pair (EV01) gives the largest PDF contribution to the $m_T$ uncertainty, dominates the $d$ and ${\bar d} / {\bar u}$  uncertainties in the $x$ region of the $W$ boson production at the Tevatron. In the same figure, we also show the other two noticeable contributions of those three leading EV sets, which are found in the strangeness and gluon PDFs. 

\begin{figure}
    \centering
    \includegraphics[width=0.47\textwidth, clip, trim=10mm 25mm 0mm 15mm]{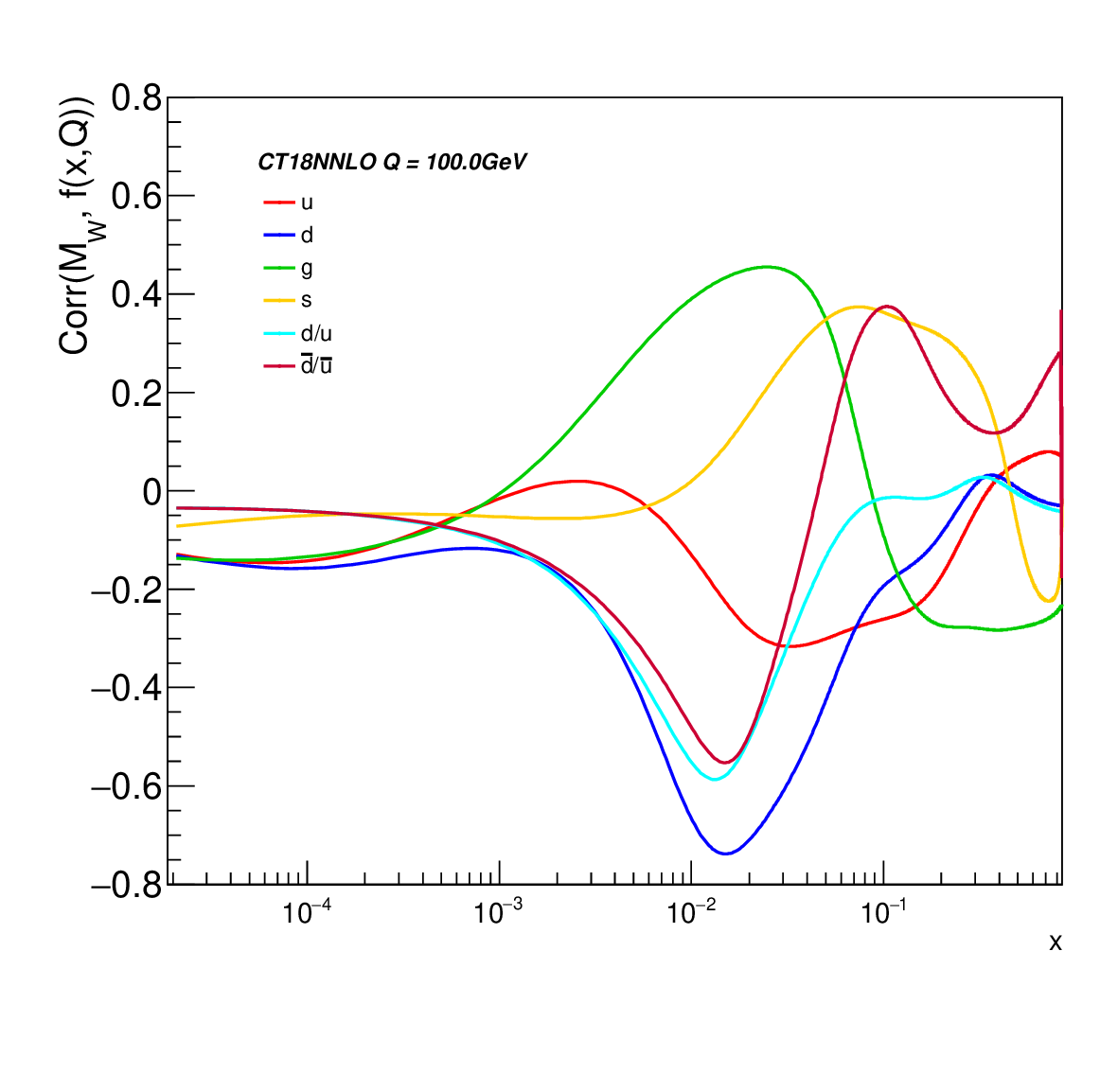}
    \caption{Similar to Fig.~\ref{fig:CorrCosine}, but from $p_T(\ell)$ distribution.}
   \label{fig:CorrCosine_Lep}
\end{figure}

For completeness, we also show in Fig.~\ref{fig:CorrCosine_Lep} the correlation cosine plot similar to Fig.~\ref{fig:CorrCosine}, but for the $W$ boson mass extracted from $p_T(\ell)$ distribution. (The same conclusion also holds for using the normalized $p_T(\nu)$ distribution.)
We note that the most relevant PDF flavors in these two cases are $d$ and $g$ PDFs. Due to the larger uncertainty in $g$-PDF, the PDF-induced uncertainty on the extracted $W$ boson mass is larger than the one extracted from the $m_T$ distribution, as shown in Table~\ref{tab:pdf_pt}.


\begin{figure}
    \centering
    \includegraphics[width=0.47\textwidth]{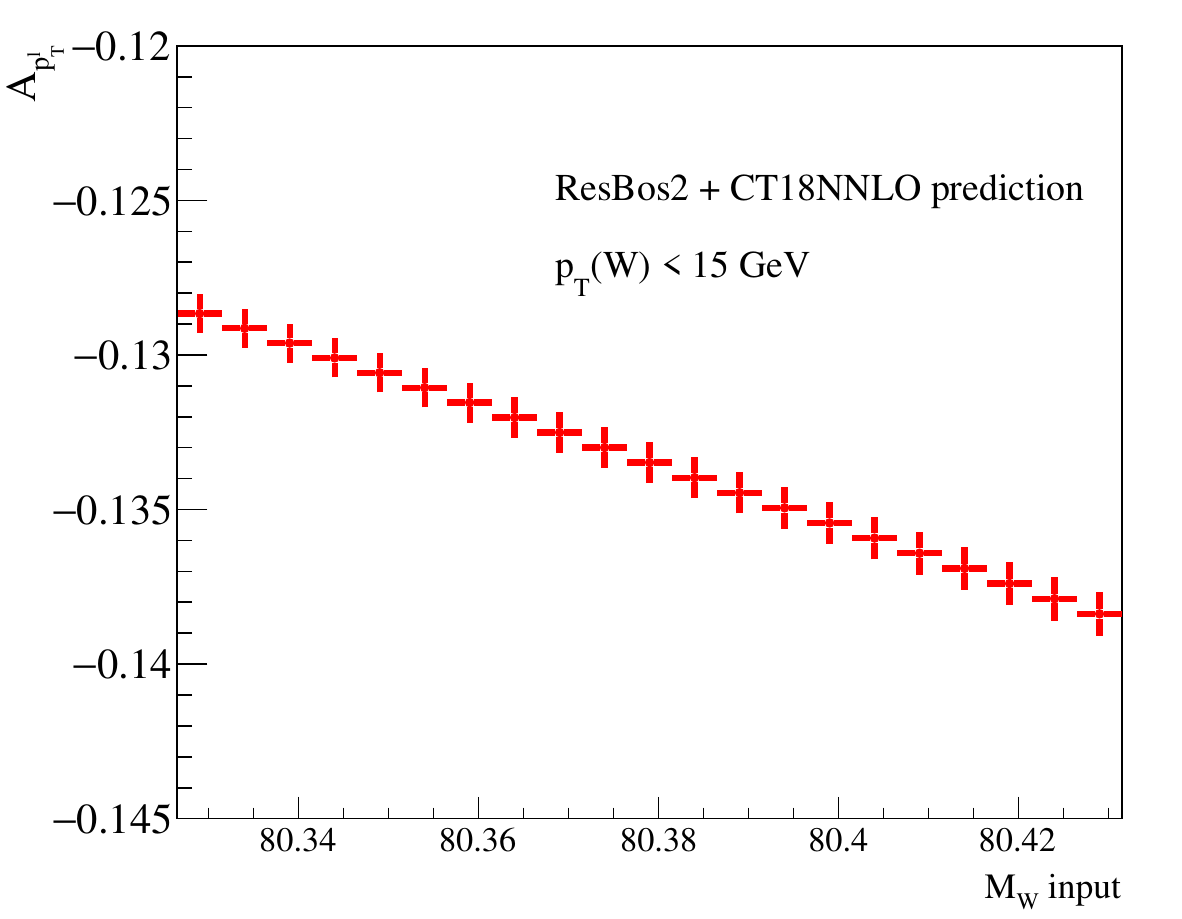}
    \caption{Jacobian asymmetry as a function of $M_W$ input. The phase space is the same as what reference~\cite{Rottoli:2023xdc} describes, in addition of the selection cut of $p_T(W)<15$~GeV. The PDF input is CT18NNLO. The error band represents the scale uncertainty of calculation.}
    \label{fig:JacobianAsymmetry}
\end{figure}

\begin{figure}
    \centering
    \includegraphics[width=0.3\textwidth]{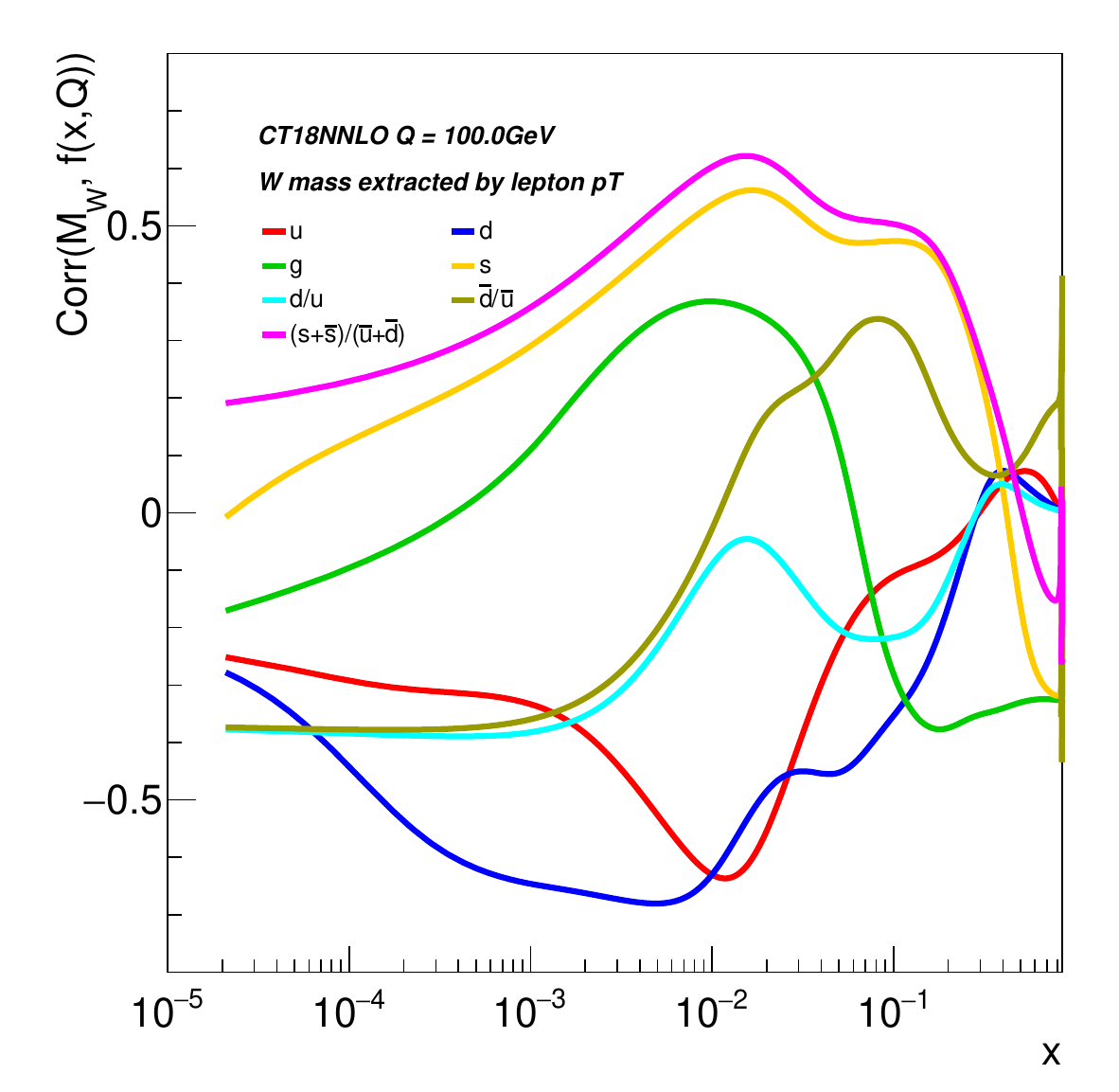}
    \includegraphics[width=0.3\textwidth]{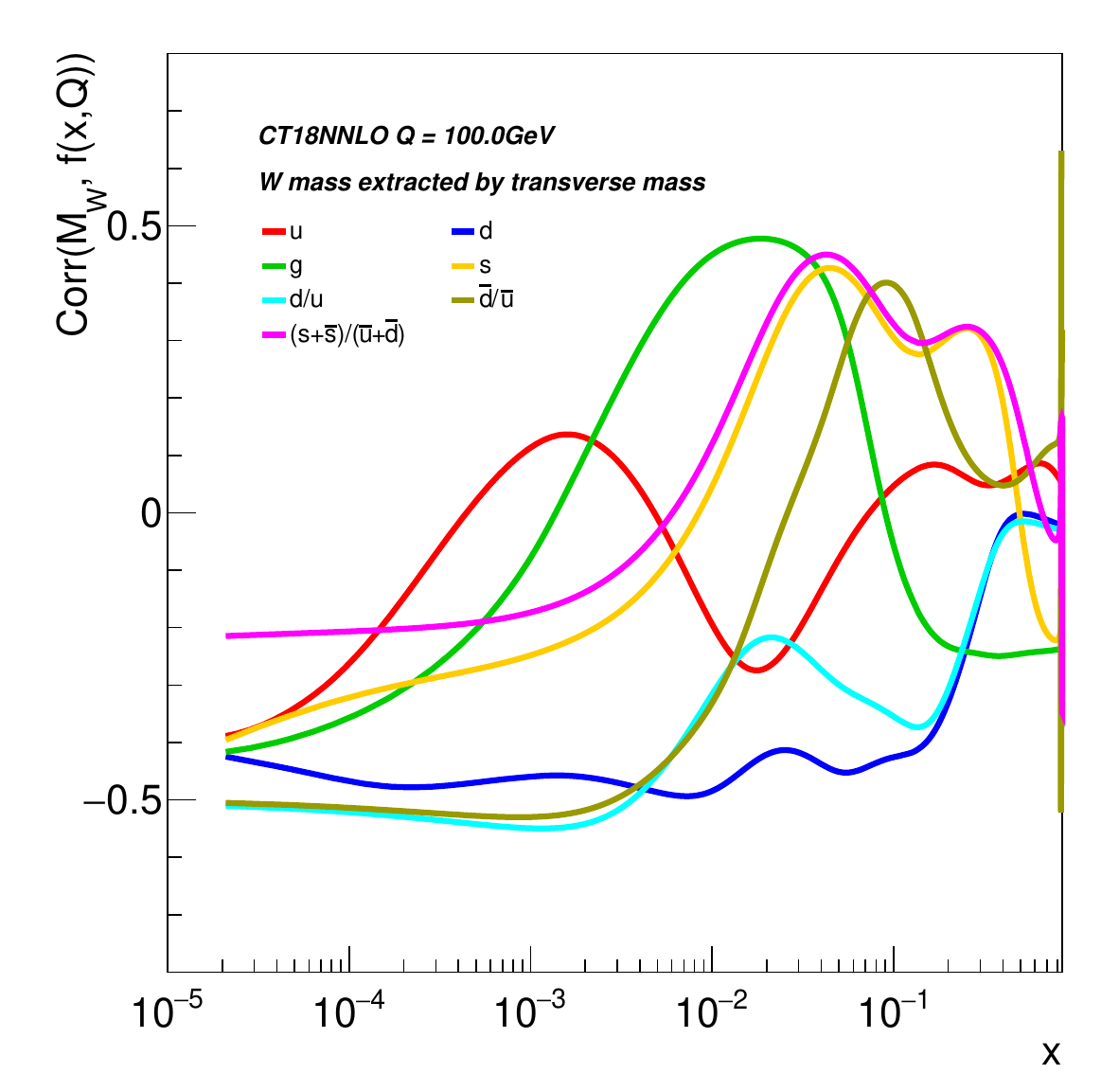}
    \includegraphics[width=0.3\textwidth]{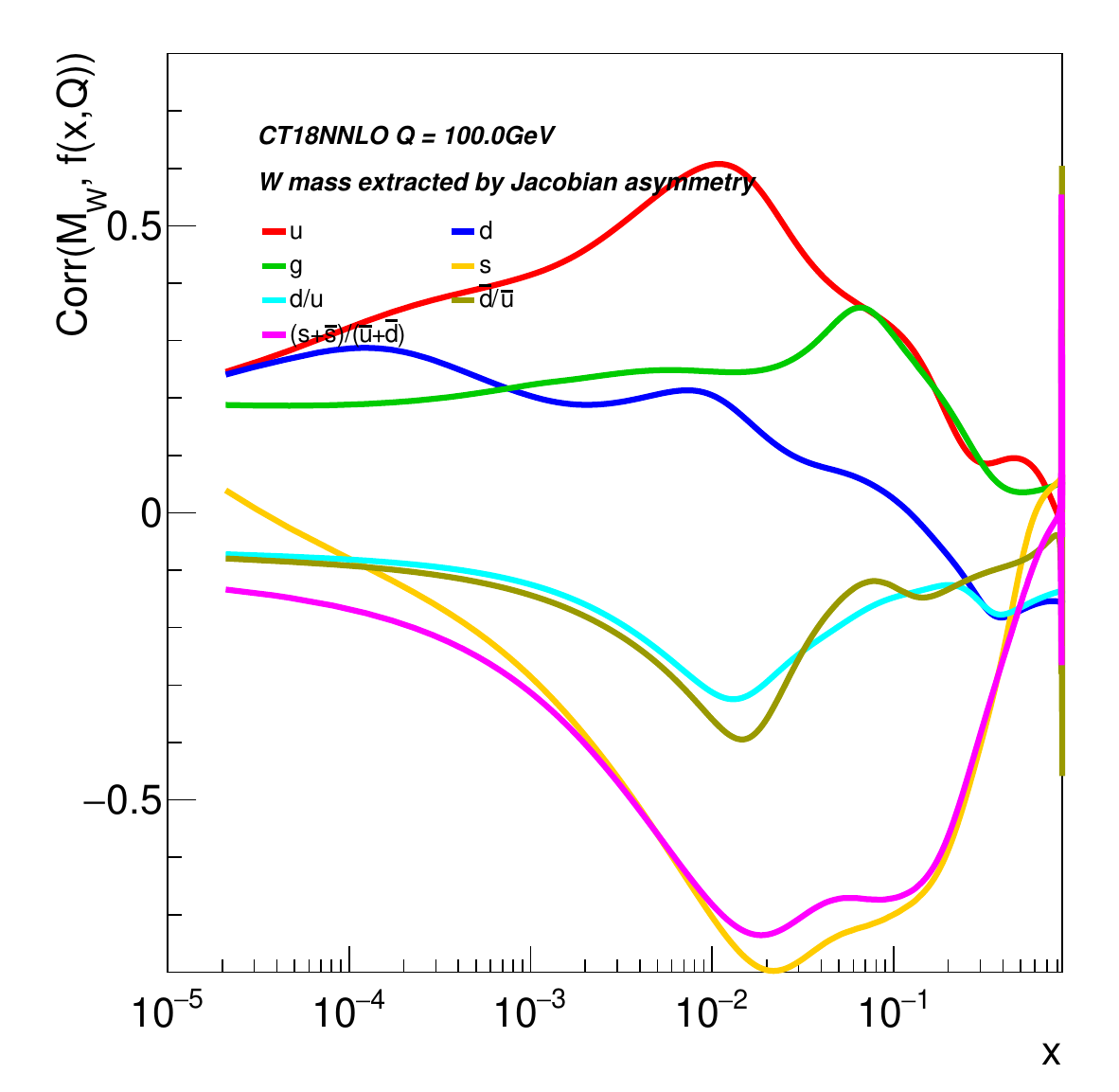}
    \caption{Similar to Fig.~\ref{fig:CorrCosine}, but in $13$ TeV LHC case. From left to right, the $M_W$ is extracted from $p_T(\ell)$ distribution, $m_T$ distribution, and the Jacobian asymmetry.}
   \label{fig:CorrCosine_LHC}
\end{figure}

We also investigate the PDF-induced correlations in $M_W$ measurement at the $13$ TeV LHC. In this study, we apply an acceptance cut of $p_T(\ell) > 30$ GeV, $p_T(\nu) > 30$ GeV, $p_T(W) < 30$ GeV, $m_T > 60$ GeV, $|\eta_{\ell}| < 2.5$ to mimic the ATLAS detector. 
We repeat the procedure described in Ref.~\cite{Isaacson:2022rts} to extract the $M_W$ using the $m_T$ distribution and the $p_T(\ell)$ distribution.
Additionally, we consider the Jacobian asymmetry introduced in Ref.~\cite{Rottoli:2023xdc}.
The Jacobian asymmetry is defined as
\begin{equation}
    \mathcal{A}_{p_T^\ell}(p_T^{\ell,{\rm min}}, p_T^{\ell, {\rm mid}}, p_T^{\ell,{\rm max}})=\frac{L_{p^\ell_T}-U_{p^\ell_T}}{L_{p^\ell_T}+U_{p^\ell_T}}\,
\end{equation}
where $L_{p_T^\ell}$ and $U_{p_T^\ell}$ are given as
\begin{equation}
L_{p_T^\ell} = \int_{p_T^{\ell,{\rm min}}}^{p_T^{\ell,{\rm mid}}} {\rm d}p_T^{\ell} \frac{{\rm d}\sigma}{{\rm d}p_T^{\ell}}\,,\quad\quad U_{p_T^\ell} = \int_{p_T^{\ell,{\rm mid}}}^{p_T^{\ell,{\rm max}}} {\rm d}p_T^{\ell} \frac{{\rm d}\sigma}{{\rm d}p_T^{\ell}}\,.
\end{equation}
Reference~\cite{Rottoli:2023xdc} show that this variable may be more sensitive to $M_W$ than traditional variables at the LHC. Fig.~\ref{fig:JacobianAsymmetry} shows the plot of asymmetry variable as a function of $M_W$ input. As the plot shown, the Jacobian asymmetry has linear relation with the $M_W$ input.
Therefore, we also choose to extract $M_W$ using this asymmetry for PDF-induced correlation studies.
By using the $M_W$ measured, we also show in Fig.~\ref{fig:CorrCosine_LHC} the correlation cosine plot similar to Fig.~\ref{fig:CorrCosine} and Fig.~\ref{fig:CorrCosine_Lep}. Different to the Tevatron case, the PDF-induced error in $M_W$ is mainly correlated to the $s$-PDFs if the Jacobian asymmetry of the $p_T(\ell)$ distribution is used to extract the $M_W$. Instead, if the $m_T$ distribution is used to extract the $M_W$, the PDF-induced error in $M_W$ is not obvious correlated to specific parton flavors.

\section{Conclusions}\label{sec:conclusions}

In this work, we discussed the recent improvements in the ResBos resummation program known as ResBos2. One significant improvement was in increasing the logarithmic accuracy from NNLL to N${}^3$LL in the resummation calculation. Additionally, the resummation formalism was expanded to include both the CSS and CFG formalisms into the same program. This enables a comparison between the two difference formalisms
while easily controlling for other implementation detail effects. Finally, the ResBos2 program improves upon the ResBos framework by including a complete NNLO calculation to be matched to. This is of vital
importance to ensure that the angular distributions for $W$ and $Z$ boson production are simulated correctly.

With the improvement of logarithmic accuracy, a new more accurate non-perturbative function of the Sudakov factor was also required to meet the precision needs of the LHC. In this work, we proposed the IFY functional form that includes rapidity
dependence. We find that the non-perturbative function has mild rapidity dependence based on the results of the fit. The IFY functional form was fit to data from fixed target experiments, the Tevatron, and the LHC. Overall,
there were 384 points included in the fit, and the best fit result had a $\chi^2$ of 380.8. 

The ResBos2 prediction with the IFY functional form was then validated against the LHC data sets which are intentional excluded from the fit.
In the comparison to this set of LHC data, we find excellent agreement in the small transverse momentum (or $\phi^*_\eta$) region as
expected. However, in the intermediate matching region and the large transverse momentum region the agreement
is not as good. The large transverse momentum (or $\phi^*_\eta$)  disagreement is expected due to the large contributions from
N${}^3$LO. The rough size of these corrections is of the same size as the disagreement between the ResBos2
prediction at N${}^3$LL+NNLO. We plan to interface with a N${}^3$LO prediction in a future work~\cite{Gehrmann-DeRidder:2017mvr,Neumann:2022lft}. The
disagreement in the intermediate region requires further investigation of the matching scheme and is left
to a future work.

Finally, we provide additional details of related to extracting the $W$ mass using ResBos2, expanding upon
Ref.~\cite{Isaacson:2022rts}. We include investigations into various explanations of the $W$ mass anomaly from CDF such as: different approaches for including the $W$ width, a discussion on using the $p_T(W)$ data to constrain the scale variations as described by the CDF collaboration in Ref.~\cite{CDF:2022hxs}, and an approximate modeling of the detector effects and QED final state radiation to improve our final conclusions in
Ref.~\cite{Isaacson:2022rts}. We conclude the $W$ mass discussion by investigating the impacts of PDFs for
both the LHC and Tevatron, and determine the most important contributions needed to reduce the PDF uncertainty
on any $W$ mass measurement. We find that the up and strange quark uncertainty is the dominate uncertainty for
extracting the $W$ mass at the LHC when using the Jacobian asymmetry of the $p_T(\ell)$ distribution.

\section*{Acknowledgments}
We thank Feng Yuan for useful discussions and comments on the manuscript. This manuscript has been authored by Fermi Research Alliance, LLC under Contract No. DE-AC02-07CH11359 with the U.S. Department of Energy, Office of Science, Office of High Energy Physics. It is also supported in part by the U.S.~National Science Foundation
under Grant No.~PHY-2013791 and No.~PHY-2310291. C.-P.~Yuan is also grateful for the support from the Wu-Ki Tung endowed chair in particle physics. The work of J.I. was supported by the U.S. Department of Energy,
Office of Science, Office of Advanced Scientific Computing Research, Scientific Discovery through Advanced Computing
(SciDAC-5) program, grant “NeuCol”.

\appendix

\pagebreak

\section{Scale Variations}
\label{app:ScaleVariation}

In the ResBos2 calculation, there exist 5 scales that can be varied to obtain the theoretical scale uncertainty. These 5 scales are the typical factorization and renormalization scales that appear in a fixed order calculation, and then three resummation scales that arise from the solution of the renormalization group equations (denoted $C_1$, $C_2$, and $C_3$). The prescription that we propose to be used with the ResBos2 code is to fix the factorization scale, $C_1$, and $C_3$ to be scaled by the same prefactor, and vary $C_2$ and $\mu_R$ independently. This combination gives a total of 15 scales to be calculated for the scale uncertainty.

Additionally, the values of $A$, $B$, and $C$ in the resummation formalism depend on the choice of $C_1$, $C_2$, and $C_3$.
To obtain the scale dependence, the resummation formalism is expanded for arbitrary scales and is compared to the canonical choice, ($C_1=C_3=b_0$, and $C_2=1$). In other words:
\begin{equation}
\label{eq: scaleDependence}
W\left(b,Q,C_1,C_2,C_3\right)|_{\order{n}}=W\left(b,Q,C_1=b_0,C_2=1,C_3=b_0\right)|_{\order{n}},
\end{equation}
where the definitions of $C_1, C_2$, and $C_3$ can be found in the scale dependent resummation formalism given as:
\begin{align}
W&=\exp\left(-\int_{C_1^2/b^2}^{C_2^2 Q^2} \frac{d\mu^2}{\mu^2}A(\mu;C_1)\log\left(\frac{C_2^2Q^2}{\mu^2}\right)+B(\mu;C_1,C_2)\right) \nonumber \\
&C\otimes f_a\left(x_1,\frac{C_1}{C_2},\frac{C_3}{b}\right)C\otimes f_b\left(x_2,\frac{C_1}{C_2},\frac{C_3}{b}\right).
\end{align}
performing the series expansion of the previous equation, and using Eq.~\ref{eq: scaleDependence}, to $\order{3}$, the scale dependence is given by:
\begin{align}
A^{(1)}&=A^{(1,c)} \\
A^{(2)}&=A^{(2,c)}-\beta _0 A^{(1,c)} \log \left(\frac{b_0^2}{C_1^2}\right)\\
A^{(3)}&=A^{(3,c)}+4 \beta _0^2 A^{(1,c)} \log ^2\left(\frac{b_0}{C_1}\right)-2 \log \left(\frac{b_0}{C_1}\right) \left(\beta _1 A^{(1,c)}+2 \beta _0 A^{(2,c)}\right) \\
B^{(1)}&= B^{(1,c)}-A^{(1,c)} \log \left(\frac{b_0^2 C_2^2}{C_1^2}\right)   \\
B^{(2)}&= B^{(2,c)}-A^{(2,c)} \log \left(\frac{b_0^2
   C_2^2}{C_1^2}\right) \nonumber \\ 
   &+\beta _0 \left(2 A^{(1,c)} \log ^2\left(\frac{b_0}{C_1}\right)-2 A^{(1,c)} \log ^2\left(C_2\right)+2 B^{(1,c)} \log \left(C_2\right)\right) \\
B^{(3)}&=B^{(3,c)}-A^{(3,c)} \log \left(\frac{b_0^2 C_2^2}{C_1^2}\right)\nonumber \\ 
&+2 \beta _1 \left(A^{(1,c)} \log ^2\left(\frac{b_0}{C_1}\right)+\log \left(C_2\right) \left(B^{(1,c)}-A^{(1,c)} \log \left(C_2\right)\right)\right) \nonumber \\
&-\frac{4}{3} \beta _0^2 \left(2 A^{(1,c)} \log
   ^3\left(\frac{b_0}{C_1}\right)+\log ^2\left(C_2\right) \left(2 A^{(1,c)} \log \left(C_2\right)-3 B^{(1,c)}\right)\right)\nonumber \\
   &+4 \beta _0 \left(A^{(2,c)} \log ^2\left(\frac{b_0}{C_1}\right)+\log
   \left(C_2\right) \left(B^{(2,c)}-A^{(2,c)} \log \left(C_2\right)\right)\right)
\end{align}
\begin{align}
C^{(1)}&=C_{ja}^{(1,c)}(\xi)+\delta_{ja}\delta(1-\xi)\left(-\frac{1}{4} A^{(1,c)} \log ^2\left(\frac{b_0^2 C_2^2}{C_1^2}\right)+\frac{1}{2} B^{(1,c)} \log \left(\frac{b_0^2 C_2^2}{C_1^2}\right)\right) \nonumber \\
 &- \frac{1}{2} P^{(1)}_{ja} \log\frac{C_3^2}{b_0^2} \\
C^{(2)}&=C_{ja}^{(2,c)}(\xi)+\delta_{ja}\delta(1-\xi)\left(-\frac{1}{4} \beta _0 A^{(1,c)} \log ^2\left(\frac{b_0^2 C_2^2}{C_1^2}\right) \log \left(\frac{b^2 \mu _F^2}{b_0^2}\right)\nonumber \right. \\
&\left.+\frac{1}{2} \beta _0 B^{(1,c)} \log \left(\frac{b_0^2
   C_2^2}{C_1^2}\right) \log \left(\frac{b^2 \mu _F^2}{b_0^2}\right)+\frac{1}{32} \left(A^{(1,c)}\right)^2 \log ^4\left(\frac{b_0^2 C_2^2}{C_1^2}\right)\nonumber\right. \\ 
   &\left.-\frac{1}{12} \beta _0 A^{(1,c)} \log
   ^3\left(\frac{b_0^2 C_2^2}{C_1^2}\right)-\frac{1}{8} A^{(1,c)} B^{(1,c)} \log ^3\left(\frac{b_0^2 C_2^2}{C_1^2}\right)+\frac{1}{2} B^{(2,c)} \log \left(\frac{b_0^2
   C_2^2}{C_1^2}\right)\nonumber\right. \\
   &\left.-\frac{1}{4} A^{(2,c)} \log ^2\left(\frac{b_0^2 C_2^2}{C_1^2}\right)+\frac{1}{8} \left(B^{(1,c)}\right)^2 \log ^2\left(\frac{b_0^2 C_2^2}{C_1^2}\right)+\frac{1}{4} \beta _0 B^{(1,c)}
   \log ^2\left(\frac{b_0^2 C_2^2}{C_1^2}\right) \nonumber \right)\\
   &+\frac{1}{2}B^{(1,c)}C^{(1,c)}_{ja}\log\left(\frac{b_0^2 C_2^2}{C_1^2}\right)-\frac{1}{4}A^{(1,c)}C^{(1,c)}_{ja}\log^2\left(\frac{b_0^2 C_2^2}{C_1^2}\right) \nonumber \\
   &+\left(\beta_0 C^{(1,c)}_{ja}-\frac{1}{2}C^{(1,c)}_{jb}\otimes P^{(1)}_{ba}-\frac{1}{4}P^{(2)}_{ja}\right)\log\frac{C_3^2}{b_0^2} \nonumber \\
   &+\frac{1}{8}A^{(1,c)}P^{(1)}_{jb}\otimes P^{(1)}_{ba}\log^2\frac{C_3^2}{b_0^2}-\frac{1}{4}B^{(1,c)}P^{(1)}_{ja}\log\left(\frac{b_0^2 C_2^2}{C_1^2}\right)\log\frac{C_3^2}{b_0^2} \nonumber \\
   &+\frac{1}{8}A^{(1,c)}P^{(1)}_{ja}\log^2\left(\frac{b_0^2 C_2^2}{C_1^2}\right)\log\frac{C_3^2}{b_0^2}-\frac{\beta_0}{4} P^{(1)}_{ja}\log^2\frac{C_3^2}{b_0^2},
\end{align}
Comparing these results to that from Ref.~\cite{Guzzi:2013aja}, it is important to note the differences in the definition of $\beta_0$ and $\beta_1$. In Ref.~\cite{Guzzi:2013aja}, the $\beta$ functions are $\beta_0=(11C_A-2n_f)/6$ and $\beta_1=(17C_A^2-5C_An_f-3C_Fn_f)/6$, while here $\beta_0=(11C_A-2n_f)/12$ and $\beta_1=(17C_A^2-5C_An_f-3C_Fn_f)/24$. Note that this result is consistent with Ref.~\cite{Guzzi:2013aja}, except for the scale dependence in $C^{(2)}$. Additionally, the calculation is extended to include $A^{(3)}$ and $B^{(3)}$. The maximum uncertainty for the Sudakov factor arises for the choice $C_1=b_0/2$ and $C_2=2$ and $C_1=2b_0$ and $C_2=1/2$, which can be understood from the fact that this has the largest impact on the value of the Sudakov integral. The dependence of $C_3$ for the uncertainty is more complicated, because it deals with the complex energy and $x$-dependence of the PDFs.

When using the canonical choice of the scales, the scale variation in the CFG formalism can be readily obtained from the conversion relations presented in Sec.~\ref{subsubsec:CSSvCFG}. However, when using the non-canonical scales, due to the different treatments of the hard factor $H$, cf. Sec.~\ref{subsubsec:CFG}, the numerical difference in the predictions of the CSS and CFG formalisms can increase as compared to the case of using the canonical scales.
The difference between CSS and CFG when using the canonical scales should differ starting at the next-to-next-to-next-to-next subleading log (\textit{i.e.} $\alpha_s^4 \log(q_T^2/Q^2)$).
When using the non-canonical scales, the difference arises first in the $B^{(2)}$ and $C^{(1)}$ terms due to the different
scales the hard function is now evaluated at (see Eqs.~\eqref{eq: CSStoCFG}).

\section{Computational Improvements}
\label{app:computational}

As the accuracy of the ResBos code increases, the computational needs also drastically increase. 
To address these needs, two approaches can be taken.
First, improved algorithms can be developed to improve the overall performance.
Second, different computing architectures can be leveraged to provide more computing resources.
Both of these approaches result in a smaller amount of walltime used to obtain predictions.
The ResBos2 code takes advantage of both methods to speed up calculations and take advantage of highly parallel computers.

\subsection{Algorithmic Improvements}

There are two major algorithmic improvements in the ResBos2 code that enable a significant speed improvement in overall compute requirements.
The first and most important algorithmic improvement is the use of Ogata quadrature~\cite{Ogata2005}.
The second improvement is in the handling of the convolution kernels required in the impact parameter calculation.

\subsubsection{Ogata Quadrature}

Ogata in Ref.~\cite{Ogata2005} proposes a novel quadrature formula with the zeros of the Bessel function as nodes, \textit{i.e.}
\begin{equation}
    \int_{-\infty}^{\infty} |x|^{2\nu+1} f(x) dx \approx h \sum_{k=-\infty, k\neq 0}^\infty w_{\nu k} |h \xi_{\nu k}|^{2\nu + 1} f(h\xi_{\nu k}),
\end{equation}
where 
\begin{equation}
    w_{\nu k} = \frac{Y_\nu(\pi \xi_{\nu |k|}}{J_{\nu+1}(\pi \xi_{\nu|k|}}=\frac{2}{\pi^2 \xi_{\nu|k|}j_{\nu+1}(\pi\xi_{\nu|k|})}\,, k=\pm 1, \pm 2, \ldots,
\end{equation}
with $\nu$ is a real constant greater than -1, $h$ is a positive step size, $\xi_{\nu k}$ are the zeros for the Bessel function $J_\nu(\pi x)$ of the first kind with of order $\nu$ ordered in an increasing order for increasing $k$ with $\xi_{\nu -k}=-\xi_{\nu k}$, and $Y_\nu$ is the Bessel function of the second kind of order $\nu$.
The approach takes influence from the double exponential (DE) quadrature formula~\cite{Takahasi1973}.
The DE formula is known to be optimal for a large class of integrals.
However, this approach fails for oscillatory functions over infinite intervals.
The use of DE for oscillatory functions of the Fourier transform type was first addressed in Ref.~\cite{OOURA1991353}.
The authors of Ref.~\cite{OOURA1991353} chose a DE transform such that the nodes of the quadrature rapidly approached the zeros of the function $\sin(x)$, allowing for the integral to be computed with a small number of function evaluations.

The work of Ogata~\cite{Ogata2005} was to extend the DE approach for Fourier type integrals to integrals of the Hankel transform, \textit{i.e.} 
\begin{equation}
    \int_0^{\infty} f(x) J_\nu(x) dx\,.
\end{equation}
These integrals show up in the resummation calculation, and thus finding a DE quadrature rule would significantly reduce the number of function evaluations needed.
To achieve a DE-type formula, the variables are transformed as
\begin{equation}
    x = \frac{\pi}{h}\psi(t)\quad\quad {\rm with}\quad\quad \psi(t) = t \tanh(\frac{\pi}{2} \sinh t)\,.
\end{equation}
This transforms the integral for the Hankel transform as
\begin{equation}
    \int_0^\infty f(x) J_\nu(x) dx \approx \pi \sum_{k=1}^{\infty} w_{\nu k}f(\frac{\pi}{h}\psi(h\xi_{\nu k}))J_{\nu}(\frac{\pi}{h}\psi(h\xi_{\nu k})\psi'(h\xi_{\nu k})\,.
\end{equation}
The above infinite sum can be truncated with a small number of function evaluations, as the value $\frac{\pi}{h}\psi(h\xi_{\nu k})$ approaches $\pi \xi_{\nu k}$ double exponentially as $k\rightarrow \infty$.

This approach still requires an optimization for the choice of $h$ and the upper bound on the sum ($N$). A method to achieve this was proposed in Ref.~\cite{Kang:2019ctl}. 
In this work, the authors investigate the uncertainty estimates to determine an optimal value for $h$ and $N$.
The uncertainty after the above transformation arises from two pieces.
The first arises from the approximation of the quadrature at a finite value of $h$ and is estimated as $\mathcal{O}(e^{-c/h})$, where $c$ is some positive constant which depends on the function being integrated.
The second term arises from the truncation of the sum at some finite value $N$ and is given as
\begin{equation}
    \mathcal{I}_{\nu N+1} = h\sum_{j=N+1}^\infty w_{\nu j} |h\xi_{\nu j}|^{2\nu + 1} f(h\xi_{\nu j}).
\end{equation}
In Ref.~\cite{Kang:2019ctl}, the authors found that the optimal choice of parameters can be obtained by maximizing the contribution of the first node, \textit{i.e.}
\begin{equation}
    \frac{\partial}{\partial h}(h(h\xi_{\nu 1}^{2\nu+1}f(h\xi_{\nu 1})) = 0.
\end{equation}
The equation is then solved for $h$ and the resulting value is defined as $h_u$.
However, it is important to note that the value for $h_u$ tends to be large, thus making the value of $\mathcal{O}(e^{-c/h})$ also large.
This can be resolved by noticing that for a fixed $N$, the final nodes can be placed in the same position for some new $h$ by enforcing
\begin{equation}
    h_u \xi_{\nu N} = \frac{\pi}{h}\psi(h\xi_{\nu N}),
\end{equation}
which has a solution defined as $h_t$ at
\begin{equation}
    h_t = \frac{1}{\xi_{\nu N}}\sinh^{-1}\left(\frac{2}{\pi}\tanh^{-1}\left(\frac{h_u}{\pi}\right)\right)\,.
\end{equation}
The above equation only has a solution for $h_u < \pi$. To address this issue, an upper bound on $h_u$ can be set to ensure that the calculated value of $h_t$ is always well defined. A demonstration of the effectiveness of this optimization technique can be seen in Fig.~6 from Ref.~\cite{Kang:2019ctl}.

This algorithm with the optimization is implemented in the ResBos code and significantly reduces the number of function evaluations needed in impact parameter space to achieve an accurate conversion back to momentum space. Overall, this introduces a speed up of 10-100 depending on the value of transverse momentum considered.

\subsubsection{Convolution Improvements}

A major use of computational resources in the ResBos calculation is the convolution of the hard collinear kernels with the PDF. This becomes especially expensive for calculating $C^{(2)}$.
However, it is straightforward to pre-tabulate these values similar to how PDFs are handled in tools like LHAPDF~\cite{Buckley:2014ana}.
The convolution can be written as
\begin{equation}
    C\otimes f(x, Q^2) = \int_x^1 C(z) f(x/z, Q^2) dz\,.
\end{equation}
Additionally, the convolution can be carried out using a fixed quadrature algorithm like Double Exponential Quadrature~\cite{Takahasi1973}.
When using a fixed quadrature algorithm, the values of $z$ evaluated for all $x$ and $Q^2$ are the same.
Thus the values for $C(z)$ can be tabulated in the first iteration and then just read off for each subsequent iteration.
This allows for a significant time improvement in generating convolution grids, which can then be interpolated to create a significant time improvement when generating the resummation and asymptotic grids.

\subsection{Architecture Improvements}

The ResBos calculation consists of two components.
Firstly, resummation, perturbation, and asymptotic grids in $Q$, $p_T$, and $y$ are generated over the required parameter space.
This piece only includes the production of the vector boson being considered.
Secondly, the grids are then interpolated and combined with the decay of the vector boson through the use of integration over the entire phase space.
The integration over the phase space is optimized using adaptive importance sampling via the VEGAS algorithm~\cite{Lepage:1977sw,Lepage:2020tgj}.
Both of these components are trivially parallelizable.
Therefore, it is possible to see a significant improvement in the performance through the use of a message passing interface (MPI) to enable communication between a set of CPUs.
This results in a overall walltime speed-up directly proportional to the number of CPUs used.
Investigations into using GPUs for the parallelization are currently underway and are left to a future work.

\pagebreak

\section{\texorpdfstring{$B^{(3)}$ and $C^{(2)}$}{B3 and C2} coefficients}
\label{app:B2C3}

The coefficients up to NNLL can be found in Section~\ref{subsubsec:CSS} and~\ref{subsubsec:CFG} for the CSS and CFG formalisms respectively. For the additional terms that appear at N${}^3$LL, they can be found in~\cite{Li:2016ctv,Catani:2012qa}, and are reproduced here for ease.

The $B$ anomalous dimension in CSS at $\order{3}$ is given as:
\begin{align}
B_{3}^{DY} &= \gamma^{DY}_2-\gamma_2^r+\beta_1 c_1^{DY} + 2\beta_0\left(c_2^{DY}-\frac{1}{2}\left(c_1^{DY}\right)^2\right),
\end{align}
substituting the numbers into the equation above, the numerical result is given as:
\begin{align}
B_{3}^{DY} &= 114.98-11.27 n_f +0.32 n_f^2,
\end{align}
where $n_f$ is the number of active flavors. Note the above equation differs from that in~\cite{Li:2016ctv}, since the expansion in~\cite{Li:2016ctv} is for $\frac{\alpha_s}{4\pi}$, while this work uses $\frac{\alpha_s}{\pi}$. 

The hard-collinear coefficients $C_{ij}^{(2)}$ at the NNLO for vector boson production are given by five different initial states: $q\bar{q}$, $q\bar{q}'$, $qq$, $qq'$, and $qg$. These coefficients can be respectively obtained from (reproduced from Ref.~\cite{Catani:2012qa} for convenience):
\begin{align}
2C_{q\bar{q}}^{(2)}\left(z\right)&+\delta\left(1-z\right)\left[H_q^{DY(2)}-\frac{3}{4}\left(H_q^{DY(1)}\right)^2+\frac{C_F}{4}\left(\pi^2-8\right)H_q^{DY(1)}\right]\nn\\
&+\frac{1}{2}C_F H_q^{DY(1)}\left(1-z\right)\nn\\
&= \mathcal{H}_{q\bar{q}\leftarrow q\bar{q}}^{DY(2)}\left(z\right)-\frac{C_F^2}{4}\left[\delta\left(1-z\right)\frac{\left(\pi^2-8\right)^2}{4}+\left(\pi^2-10\right)\left(1-z\right)-\left(1+z\right)\ln z\right], \\
C_{qg}^{(2)}\left(z\right)&+\frac{1}{4}H_q^{DY(1)}z\left(1-z\right)=\mathcal{H}_{q\bar{q}\leftarrow qg}^{DY(2)}\left(z\right)\nn\\
&-\frac{C_F}{4}\left[z\ln z + \frac{1}{2}\left(1-z^2\right)+\left(\frac{\pi^2}{4}-4\right) z\left(1-z\right) \right], \\
C_{qq}^{(2)}\left(z\right)&=\mathcal{H}_{q\bar{q}\leftarrow qq}^{DY(2)}\left(z\right), \\
C_{qq'}^{(2)}\left(z\right)&=\mathcal{H}_{q\bar{q}\leftarrow qq'}^{DY(2)}\left(z\right), \\
C_{q\bar{q}'}^{(2)}\left(z\right)&=\mathcal{H}_{q\bar{q}\leftarrow q\bar{q}'},
\end{align}
where the scheme independent hard-collinear coefficient functions ($\mathcal{H}$) are given by:
\begin{align}
\mathcal{H}^{DY(2)}_{q\bar q\leftarrow q \bar q}(z)&=
C_A C_F 
\bigg\{
\left(\frac{7 \zeta_3}{2}-\frac{101}{27}\right)\left(\frac{1}{1-z}\right)_+ %\text{D0}(z)
+\left(\frac{59 \zeta_3}{18}-\frac{1535}{192}+\frac{215 \pi ^2}{216}-\frac{\pi ^4}{240}\right) \delta(1-z)
\nn\\
&~~~~~~
+\frac{1+z^2}{1-z}
 \bigg(-\frac{\text{Li}_3(1-z)}{2}+\text{Li}_3(z)-\frac{\text{Li}_2(z) \log (z)}{2}
-\frac{1}{2} \text{Li}_2(z) \log (1-z)
\nn\\
&~~~~~~
-\frac{1}{24} \log ^3(z)
-\frac{1}{2} \log ^2(1-z) \log (z)+\frac{1}{12} \pi ^2 \log (1-z)-\frac{\pi ^2}{8}\bigg)
\nn\\
&~~~~~~
+\frac{1}{1-z} \bigg(
-\frac{1}{4} \left(11-3 z^2\right) \zeta_3
-\frac{1}{48} \left(-z^2+12 z+11\right) \log ^2(z)
\nn\\
&~~~~~~
-\frac{1}{36} \left(83 z^2-36 z+29\right) \log (z)+\frac{\pi ^2 z}{4}\bigg)
\nn\\
&~~~~~~
+(1-z) \bigg(\frac{\text{Li}_2(z)}{2}+\frac{1}{2} \log (1-z) \log (z)\bigg)
+\frac{z+100}{27}+\frac{1}{4} z \log (1-z)\bigg\}
\nn\\
&~~~~~~
+C_F n_F \bigg\{\frac{14}{27}\left(\frac{1}{1-z}\right)_+ %\text{D0}(z)
+\frac{1}{864} \left(192 \zeta_3+1143-152 \pi ^2\right) \delta(1-z)
\nn\\
&~~~~~~
+\frac{\left(1+z^2\right)}{72 (1-z)} \log (z) (3 \log (z)+10)+\frac{1}{108} (-19 z-37)\bigg\}
\nn\\
&~~~~~~
+C_F^2 \bigg\{\frac{1}{4} \left(-15 \zeta_3+\frac{511}{16}
-\frac{67 \pi ^2}{12}+\frac{17 \pi ^4}{45}\right) \delta(1-z)
\nn\\
&~~~~~~
+\frac{1+z^2}{1-z}
 \bigg(\frac{\text{Li}_3(1-z)}{2}-\frac{5 \text{Li}_3(z)}{2}+\frac{1}{2} \text{Li}_2(z) \log (1-z)
+\frac{3 \text{Li}_2(z) \log (z)}{2}
\nn\\
&~~~~~~
+\frac{3}{4} \log (z) \log ^2(1-z)+\frac{1}{4} \log ^2(z) \log (1-z)-\frac{1}{12} \pi ^2 \log (1-z)+\frac{5 \zeta_3}{2}\bigg)
\nn\\
&~~~~~~
+(1-z) \left(-\text{Li}_2(z)-\frac{3}{2} \log (1-z) \log (z)+\frac{2 \pi ^2}{3}-\frac{29}{4}\right)
+\frac{1}{24} \left(1+z\right) \log ^3(z)
\nn\\
&~~~~~~
+\frac{1}{1-z}\bigg(
\frac{1}{8} \left(-2 z^2+2 z+3\right) \log ^2(z)+\frac{1}{4} \left(17 z^2-13 z+4\right) \log (z)\bigg)
\nn\\
&~~~~~~
-\frac{z}{4}\log (1-z)
\bigg\}
\nn\\
&~~~~~~
+C_F \bigg\{\frac{1}{z}(1-z) \left(2 z^2-z+2\right) \left(\frac{\text{Li}_2(z)}{6}+\frac{1}{6} \log (1-z) \log (z)-\frac{\pi ^2}{36}\right)
\nn\\
&~~~~~~
+\frac{1}{216 z}(1-z) \left(136 z^2-143 z+172\right)-\frac{1}{48} \left(8 z^2+3 z+3\right) \log ^2(z)
\nn\\
&~~~~~~
+\frac{1}{36} \left(32 z^2-30 z+21\right) \log (z)+\frac{1}{24} (1+z) \log ^3(z)
\bigg\}\;, \\ \nn \\ \nn \\ \nn\\ \nn \\
{\cal H}^{DY(2)}_{q\bar q\ito q \bar{q}'}(z)&=
C_F \bigg\{
\frac{1}{12 z}(1-z) \left(2 z^2-z+2\right) \bigg(
\text{Li}_2(z)
+\log (1-z) \log (z)
-\frac{\pi^2}{6}
\bigg)
\nn\\
&~~~~~~
+\frac{1}{432 z}(1-z) \left(136 z^2-143 z+172\right)
+\frac{1}{48} (1+z) \log^3(z)
\nn\\
&~~~~~~
-\frac{1}{96} \left(8 z^2+3 z+3\right) \log^2(z)
+\frac{1}{72} \left(32 z^2-30 z+21\right) \log (z)
\bigg\}\;,
\label{h2qq}
\end{align}
\begin{align}
{\cal H}^{DY(2)}_{q\bar q\ito q q}(z)&=
C_F\left(C_F-\frac{1}{2}C_A\right)
 \bigg\{ \frac{1+z^2}{1+z}
\bigg(\frac{3 \text{Li}_3(-z)}{2}+\text{Li}_3(z)
+\text{Li}_3\left(\frac{1}{1+z}\right)-\frac{\text{Li}_2(-z) \log (z)}{2}
\nn\\
&~~~~~~
-\frac{\text{Li}_2(z) \log (z)}{2}-\frac{1}{24} \log ^3(z)
-\frac{1}{6} \log ^3(1+z)+\frac{1}{4} \log (1+z) \log ^2(z)
\nn\\
&~~~~~~
+\frac{\pi^2}{12}  \log (1+z)-\frac{3 \zeta_3}{4}\bigg)
+\left(1-z\right) \left(\frac{\text{Li}_2(z)}{2}+\frac{1}{2} \log (1-z) \log (z)+\frac{15}{8}\right)
\nn\\
&~~~~~~
-\frac{1}{2}(1+z) \big(\text{Li}_2(-z)+ \log (z) \log (1+z)\big)
+ \frac{\pi^2}{24}  (z-3)+\frac{1}{8} (11 z+3) \log (z)\bigg\}
\nn\\
&~~~~~~
+C_F \bigg\{\frac{1}{12 z}(1-z) \left(2 z^2-z+2\right) \left(\text{Li}_2(z)+\log (1-z) \log (z)
-\frac{\pi ^2}{6}\right)
\nn\\
&~~~~~~
+\frac{1}{432 z}(1-z) \left(136 z^2-143 z+172\right)-\frac{1}{96} \left(8 z^2+3 z+3\right) \log ^2(z)
\nn\\
&~~~~~~
+\frac{1}{72} \left(32 z^2-30 z+21\right) \log (z)+\frac{1}{48} (1+z) \log ^3(z)\bigg\}
\;,
\end{align}
%\newpage
%
\begin{align}
\label{h2qqp}
\mathcal{H}^{DY(2)}_{q\bar q\ito q q'}(z)=\mathcal{ H}^{DY(2)}_{q\bar q\ito q \bar{q}'}(z)\;,
\end{align}
\begin{align}
\label{h2qg}
{\cal H}^{DY(2)}_{q \bar q\ito qg}(z)
&=
C_A \bigg\{-\frac{1}{12 z}(1-z) \left(11 z^2-z+2\right) \text{Li}_2(1-z)
\nn\\
&~~~~~~
+\left(2 z^2-2 z+1\right) \bigg(\frac{\text{Li}_3(1-z)}{8}
-\frac{1}{8} \text{Li}_2(1-z) \log (1-z)+\frac{1}{48} \log^3(1-z)\bigg)
\nn\\
&~~~~~~
+\left(2 z^2+2 z+1\right) \bigg(\frac{3 \text{Li}_3(-z)}{8}
+\frac{\text{Li}_3\left(\frac{1}{1+z}\right)}{4}-\frac{\text{Li}_2(-z) \log(z)}{8}
-\frac{1}{24} \log^3(1+z)
\nn\\
&~~~~~~
+\frac{1}{16} \log^2(z) \log (1+z)
+\frac{1}{48} \pi^2 \log (1+z)\bigg)
+\frac{1}{4} z (1+z) \text{Li}_2(-z)+z \text{Li}_3(z)
\nn\\
&~~~~~~
-\frac{1}{2} z \text{Li}_2(1-z) \log(z)-z \text{Li}_2(z) \log(z)
-\frac{3}{8} \left(2 z^2+1\right) \zeta_3-\frac{149 z^2}{216}
\nn\\
&~~~~~~
-\frac{1}{96} \left(44 z^2-12 z+3\right) \log^2(z)
+\frac{1}{72} \left(68 z^2+6 \pi^2 z-30 z+21\right) \log(z)
\nn\\
&~~~~~~
+\frac{\pi^2 z}{24}+\frac{43 z}{48}
+\frac{43}{108 z}
+\frac{1}{48} (2 z+1) \log^3(z)
-\frac{1}{2} z \log (1-z) \log^2(z)
\nn\\
&~~~~~~
-\frac{1}{8} (1-z) z \log^2(1-z)
+\frac{1}{4} z (1+z) \log (1+z) \log(z)
\nn\\
&~~~~~~
+\frac{1}{16} (3-4 z) z \log (1-z)-\frac{35}{48}\bigg\}
\nn\\
&~~~~~~
+C_F \bigg\{\left(2 z^2-2 z+1\right) 
\bigg(\zeta_3-\frac{\text{Li}_3(1-z)}{8}
-\frac{\text{Li}_3(z)}{8}+\frac{1}{8} \text{Li}_2(1-z) \log (1-z)
\nn\\
&~~~~~~
+\frac{\text{Li}_2(z) \log(z)}{8}-\frac{1}{48} \log^3(1-z)
+\frac{1}{16} \log(z) \log^2(1-z)
\nn\\
&~~~~~~
+\frac{1}{16} \log^2(z) \log (1-z)
\bigg)
\nn\\
&~~~~~~
-\frac{3 z^2}{8}-\frac{1}{96} \left(4 z^2-2 z+1\right) \log^3(z)
+\frac{1}{64} \left(-8 z^2+12 z+1\right) \log^2(z)
\nn\\
&~~~~~~
+\frac{1}{32} \left(-8 z^2+23 z+8\right) \log(z)+\frac{5}{24} \pi^2 (1-z) z
+\frac{11 z}{32}+\frac{1}{8} (1-z) z \log^2(1-z)
\nn\\
&~~~~~~
-\frac{1}{4} (1-z) z \log (1-z) \log(z)
-\frac{1}{16} (3-4 z) z \log (1-z)-\frac{9}{32}\bigg\}\;,
\end{align}
\begin{align}
\label{h2gg}
\mathcal{H}^{DY(2)}_{q\bar q\ito gg}(z)&=
-\,\frac{z}{2}\,\left(\,1-z+\frac{1}{2}\,(1+z)\,\log (z)\, \right)\;,
\end{align}
where 
${\rm Li}_k(z)$ $(k=2,3)$ are the
polylogarithm functions,
\begin{equation}
{\rm Li}_2(z)= - \int_0^z \frac{dt}{t} \;\ln(1-t) \;\;,
\quad \quad {\rm Li}_3(z)=  \int_0^1 \frac{dt}{t} \;\ln(t)\;\ln(1-zt) \;\;,
\end{equation}
and the $H$ factors are the scheme dependent resummation factors. For CSS, $H$ is 1 to all orders, while for CFG, $H$ has $\alpha_s$ dependence.

\section{Bessel Integrals}
\label{app:bessel_int}

In the calculation of the asymptotic expansion, integrals with varying powers of the impact parameter ($b$).
The main integration needed can be expressed of the form:
\begin{equation}
\int_0^\infty d\eta \eta J_0\left(\eta\right) F\left(\eta\right)\,,
\end{equation}
 This integral can be performed by using the following integration by parts identity:
\begin{equation}
\label{eq: BesselIBP}
\int_0^\infty d\eta \eta J_0\left(\eta\right) F\left(\eta\right) = - \int_0^\infty d\eta \eta J_1\left(\eta\right) \frac{dF\left(\eta\right)}{d\eta},
\end{equation}
which is true given that the boundary term vanishes, $\left(\eta J_1\left(\eta\right)F\left(\eta\right)_{\eta=0}^\infty=0\right)$. Additionally, the following integral results will be important in obtaining both the asymptotic and singular piece up to $\order{3}$:
\begin{equation}
\label{eq: BesselInt}
\int_0^\infty d\eta J_1\left(\eta\right) \ln^m\left(\frac{\eta^2Q^2}{b_0^2p_T^2}\right) = 
	\begin{cases}
        1, \quad \text{if } m = 0 \\
		\ln\frac{Q^2}{p_T^2}, \quad \text{if } m = 1 \\
        \ln^2\frac{Q^2}{p_T^2}, \quad \text{if } m = 2 \\
        \ln^3\frac{Q^2}{p_T^2}-4\zeta(3), \quad \text{if } m = 3 \\     
        \ln^4\frac{Q^2}{p_T^2}-16\zeta(3)\ln\frac{Q^2}{p_T^2}, \quad \text{if } m = 4 \\
        \ln^5\frac{Q^2}{p_T^2}-40\zeta(3)\ln^2\frac{Q^2}{p_T^2}-48\zeta(5), \quad \text{if } m = 5, \\
        \ln^6\frac{Q^2}{p_T^2}-80 \zeta (3) \ln^3\frac{Q^2}{p_T^2}-288 \zeta (5) \ln\frac{Q^2}{p_T^2}+160 \zeta (3)^2, \ \text{if } m = 6, 
	\end{cases}
\end{equation}
where $b_0= 2 e^{-\gamma_E} \simeq 1.123$, and $\gamma_E$ is the Euler constant. Up through $m=2$ is needed for the $\mathcal{O}\left(\alpha_s\right)$ calculations, through $m=4$ is needed for the $\order{2}$ calculations, and all of the above will be needed for the $\order{3}$ calculations.

\section{\texorpdfstring{$\order{3}$}{as3} Asymptotic Expansion coefficients}
\label{app:asym3}

The coefficients for the asymptotic expansion at $\order{3}$ are given as:
\begin{align*}
{}_{3}C^{(i,j)}_{5} & = \frac{1}{4} \left(A^{\text{(1)}}\right)^3 f_i f_j,\\
{}_{3}C^{(i,j)}_{4} & = \left(A^{\text{(1)}}\right)^2\left(\frac{5}{4} B^{\text{(1)}} f_i f_j-\frac{5}{3} \beta_0  f_i f_j+\frac{5}{8}f_i P^{\text{(1)}}\otimes f_j+\frac{5}{8} f_j
   P^{\text{(1)}}\otimes f_i\right),\\
{}_{3}C^{(i,j)}_{3} & = A^{(1)}\left(\left(\left(B^{(1)}\right)^2-\frac{7}{3}\beta_0 B^{(1)}-A^{(2)}+\beta_0^2\right)f_i f_j+2B^{(1)}f_j P^{(1)}\otimes f_i-\frac{7}{3}\beta_0 f_j P^{(1)}\otimes f_i \right. \\
&+\left. \frac{1}{2}P^{(1)}\otimes f_i P^{(1)}\otimes f_j +\frac{1}{2}f_j P^{(1)}\otimes P^{(1)}\otimes f_i \right) \\
&-\left(A^{(1)}\right)^2\left(f_j C^{(1)}\otimes f_i -\frac{1}{2}f_j P^{(1)}\otimes f_i \log\left(\frac{Q^2}{\mu_F^2}\right)-\beta_0 f_i f_j \log\left(\frac{Q^2}{\mu_R^2}\right)\right) + i \leftrightarrow j, \\
{}_{3}C^{(i,j)}_{2} & =\left(2 \beta_0 A^{(2)} +B^{(1)}\left(\beta_0^2-\frac{3}{2}A^{(2)}\right)+A^{(1)}\left(-\frac{3}{2}B^{(2)}+\beta_1-2\beta_0^2\log\left(\frac{Q^2}{\mu_R^2}\right)\right.\right.\\
&\left.\left.+3\beta_0 B^{(1)}\log\left(\frac{Q^2}{\mu_R^2}\right)\right)\right. \\
&\left.-5\left(A^{(1)}\right)^3\zeta_3+\frac{1}{2}\left(B^{(1)}\right)^3-\frac{3}{2}\beta_0\left(B^{(1)}\right)^2\right)f_i f_j-\frac{3}{2} A^{(2)} f_j P^{(1)} \otimes f_i \\
&+ A^{(1)}\left(5\beta_0 f_j C^{(1)}\otimes f_i+B^{(1)}\left(-3 f_j C^{(1)}\otimes f_i -\frac{3}{2}f_j P^{(1)}\otimes f_i \log\left(\frac{Q^2}{\mu_F^2}\right)\right)\right. \\
&\left.-\frac{3}{2}C^{(1)}\otimes f_i P^{(1)}\otimes f_j -\frac{3}{2}f_j C^{(1)}\otimes P^{(1)}\otimes f_i + \beta_0 f_j P^{(1)}\otimes f_i \log\left(\frac{Q^2}{\mu_F^2}\right) \right. \\
&\left.-\frac{3}{4}P^{(1)}\otimes f_iP^{(1)}\otimes f_j \log\left(\frac{Q^2}{\mu_F^2}\right) -\frac{3}{4}P^{(1)}\otimes P^{(1)}\otimes f_i f_j \log\left(\frac{Q^2}{\mu_F^2}\right)\right.\\
&\left.+3\beta_0f_j P^{(1)}\otimes f_i\log\left(\frac{Q^2}{\mu_R^2}\right)-\frac{3}{4}f_j P^{(2)}\otimes f_i \right) \\
&+\beta_0^2 f_j P^{(1)}\otimes f_i-\frac{3}{4}\beta_0P^{(1)}\otimes f_i P^{(1)}\otimes f_j-\frac{3}{4}\beta_0f_jP^{(1)}\otimes P^{(1)}\otimes f_i \\
&+\frac{3}{2}\left(B^{(1)}\right)^2f_j P^{(1)}\otimes f_i \\
&+ B^{(1)}\left(-3\beta_0 f_j P^{(1)}\otimes f_i +\frac{3}{4}P^{(1)}\otimes f_i P^{(1)}\otimes f_j +\frac{3}{4}f_j P^{(1)}\otimes P^{(1)}\otimes f_i\right) \\
& + \frac{3}{8}P^{(1)}\otimes P^{(1)} \otimes f_i P^{(1)}\otimes f_j + \frac{1}{8}f_j P^{(1)}\otimes P^{(1)}\otimes P^{(1)} \otimes f_i \\
&+ i \leftrightarrow j, \\
\end{align*}
\begin{align*}
{}_{3}C^{(i,j)}_{1} & = \left(-2\beta_0 A^{(2)} \log\left(\frac{Q^2}{\mu_R^2}\right) +A^{(3)}-2B^{(1)}B^{(2)}+2B^{(2)}\beta_0+2\left(B^{(1)}\right)^2\beta_0\log\left(\frac{Q^2}{\mu_R^2}\right) \right. \\
&\left.+ B^{(1)}\left(\beta_1-2\beta_0^2\log\left(\frac{Q^2}{\mu_R^2}\right)\right)+A^{(1)}\left(\beta_0^2\log\left(\frac{Q^2}{\mu_R^2}\right)-\beta_1\log\left(\frac{Q^2}{\mu_R^2}\right)\right)\right. \\
&\left.+\left(A^{(1)}\right)^2\left(\frac{40}{3}\beta_0-10B^{(1)}\right)\zeta_3\right)f_i f_j + A^{(2)}\left(2f_j C^{(1)}\otimes f_i +f_j P^{(1)}\otimes f_i \log\left(\frac{Q^2}{\mu_F^2}\right)\right) \\
&-4\beta_0^2f_jC^{(1)}\otimes f_i-2\beta_0^2f_j P^{(1)}\otimes f_i \log\left(\frac{Q^2}{\mu_R^2}\right)-\frac{1}{2}P^{(2)}\otimes f_i P^{(1)}\otimes f_j \\
&-C^{(1)}\otimes P^{(1)}\otimes f_i P^{(1)}\otimes f_j-\frac{1}{2}C^{(1)}\otimes f_i P^{(1)}\otimes P^{(1)}\otimes f_j \\ 
&-\frac{3}{4}P^{(1)}\otimes P^{(1)}\otimes f_i P^{(1)}\otimes f_j \log\left(\frac{Q^2}{\mu_F^2}\right)-2B^{(2)}f_j P^{(1)}\otimes f_i - \frac{1}{4} f_j P^{(1)}\otimes P^{(2)}\otimes f_i \\
&-\frac{1}{4} f_j P^{(2)}\otimes P^{(1)}\otimes f_i-\frac{1}{2}f_j C^{(1)}\otimes P^{(1)}\otimes P^{(1)} \otimes f_i\\
&-\frac{1}{4}f_j P^{(1)}\otimes P^{(1)}\otimes P^{(1)} \otimes f_i \log\left(\frac{Q^2}{\mu_F^2}\right) \\
&+ \left(B^{(1)}\right)^2\left(-2f_j C^{(1)}\otimes f_i-f_j P^{(1)}\otimes f_i \log\left(\frac{Q^2}{\mu_F^2}\right)\right) \\
& +3\beta_0 C^{(1)}\otimes f_i P^{(1)}\otimes f_j +\frac{1}{2}\beta_0 P^{(1)}\otimes f_i P^{(1)}\otimes f_j \left(\log\left(\frac{Q^2}{\mu_F^2}\right)+2\log\left(\frac{Q^2}{\mu_R^2}\right)\right) \\
&+\beta_0 f_j P^{(2)}\otimes f_i + 3f_j C^{(1)}\otimes P^{(1)} \otimes f_i + \frac{1}{2}\beta_0 f_j P^{(1)}\otimes P^{(1)}\otimes f_i \log\left(\frac{Q^2}{\mu_F^2}\right) \\
&+ \beta_0 f_j P^{(1)}\otimes P^{(1)}\otimes f_i \log\left(\frac{Q^2}{\mu_R^2}\right) \\
&+B^{(1)}\left(-2C^{(1)}\otimes f_i P^{(1)}\otimes f_j - P^{(1)}\otimes f_i P^{(1)}\otimes f_j \log\left(\frac{Q^2}{\mu_F^2}\right) \right.\\
&\left.+ \beta_0 f_j P^{(1)}\otimes f_i \log\left(\frac{Q^2}{\mu_F^2}\right) \right. \\
&\left. +4\beta_0 f_j P^{(1)}\otimes f_i \log\left(\frac{Q^2}{\mu_R^2}\right)-f_j P^{(2)}\otimes f_i-2f_j C^{(1)}\otimes P^{(1)}\otimes f_i\right.\\
&\left.- f_j P^{(1)}\otimes P^{(1)}\otimes f_i \log\left(\frac{Q^2}{\mu_F^2}\right) \right. \\
&\left.+6\beta_0 f_j C^{(1)}\otimes f_i +\beta_0 f_j P^{(1)}\otimes f_i \left(\log\left(\frac{Q^2}{\mu_F^2}\right)+4\log\left(\frac{Q^2}{\mu_R^2}\right)\right) \right) \\
&+ A^{(1)} \left(\frac{1}{4} P^{(1)}\otimes f_i P^{(1)}\otimes f_j \log^2\left(\frac{Q^2}{\mu_F^2}\right)+\frac{1}{4}  f_j P^{(1)}\otimes  P^{(1)}\otimes f_i \log^2\left(\frac{Q^2}{\mu_F^2}\right) \right. \\
&\left.+\frac{1}{2}\beta_0 f_j P^{(1)}\otimes f_i \log^2\left(\frac{Q^2}{\mu_F^2}\right) + C^{(1)}\otimes f_i P^{(1)}\otimes f_j \log\left(\frac{Q^2}{\mu_F^2}\right)\right.\\
&\left.+\frac{1}{2}f_j P^{(2)}\otimes f_i \log\left(\frac{Q^2}{\mu_F^2}\right) \right. \\
&\left.+f_j C^{(1)}\otimes P^{(1)} \otimes f_i\log\left(\frac{Q^2}{\mu_F^2}\right) -2\beta_0 f_j P^{(1)}\otimes f_i \log\left(\frac{Q^2}{\mu_F^2}\right)\log\left(\frac{Q^2}{\mu_R^2}\right) \right. \\
&\left. + C^{(1)}\otimes f_i C^{(1)}\otimes f_j + 2 f_j C^{(2)}\otimes f_i - 4 \beta_0 f_j C^{(1)}\otimes f_i \log\left(\frac{Q^2}{\mu_R^2}\right)\right) \\
& + \beta_1 f_j P^{(1)}\otimes f_i - 10\left(A^{(1)}\right)^2 \zeta_3 f_j P^{(1)}\otimes f_i \\
&+ i \leftrightarrow j\,,
 \end{align*}
 \begin{align*}
{}_{3}C^{(i,j)}_{0} & = 2 C^{(1)}\otimes f_j f_i B^{(2)}+P^{(1)}\otimes f_j \log \left(\frac{Q^2}{\mu_F^2}\right) f_i B^{(2)}+2 C^{(1)}\otimes f_i f_j B^{(2)}\\
&+P^{(1)}\otimes f_i \log
   \left(\frac{Q^2}{\mu_F^2}\right) f_j B^{(2)}+\frac{3}{8} P^{(1)}\otimes f_j P^{(1)}\otimes P^{(1)}\otimes f_i \log ^2\left(\frac{Q^2}{\mu_F^2}\right)\\ &+\frac{3}{8}
   P^{(1)}\otimes f_i P^{(1)}\otimes P^{(1)}\otimes f_j \log ^2\left(\frac{Q^2}{\mu_F^2}\right)+P^{(1)}\otimes f_j \log ^2\left(\frac{Q^2}{\mu_R^2}\right) f_i \beta _0^2 \\
 &+4C^{(1)}\otimes f_j \log \left(\frac{Q^2}{\mu_R^2}\right) f_i \beta _0^2+P^{(1)}\otimes f_i \log ^2\left(\frac{Q^2}{\mu_R^2}\right) f_j \beta _0^2\\
 &+4 C^{(1)}\otimes f_i \log
   \left(\frac{Q^2}{\mu_R^2}\right) f_j \beta _0^2+C^{(2)}\otimes f_j P^{(1)}\otimes f_i+C^{(2)}\otimes f_i P^{(1)}\otimes f_j\\
   &+\frac{1}{2} C^{(1)}\otimes f_j
   P^{(2)}\otimes f_i+\frac{1}{2} C^{(1)}\otimes f_i P^{(2)}\otimes f_j\\
   &+C^{(1)}\otimes f_j C^{(1)}\otimes P^{(1)}\otimes f_i+C^{(1)}\otimes f_i C^{(1)}\otimes P^{(1)}\otimes f_j\\
   &+\frac{1}{2}
   P^{(1)}\otimes f_j P^{(2)}\otimes f_i \log \left(\frac{Q^2}{\mu_F^2}\right)+\frac{1}{2} P^{(1)}\otimes f_i P^{(2)}\otimes f_j \log
   \left(\frac{Q^2}{\mu_F^2}\right)\\
   &+P^{(1)}\otimes f_j C^{(1)}\otimes P^{(1)}\otimes f_i \log \left(\frac{Q^2}{\mu_F^2}\right)+P^{(1)}\otimes f_i C^{(1)}\otimes P^{(1)}\otimes f_j
   \log \left(\frac{Q^2}{\mu_F^2}\right)\\
   &+\frac{1}{2} C^{(1)}\otimes f_j P^{(1)}\otimes P^{(1)}\otimes f_i \log \left(\frac{Q^2}{\mu_F^2}\right)+\frac{1}{2} C^{(1)}\otimes f_i
   P^{(1)}\otimes P^{(1)}\otimes f_j \log \left(\frac{Q^2}{\mu_F^2}\right)\\
   &+\frac{1}{8} P^{(1)}\otimes P^{(1)}\otimes P^{(1)}\otimes f_j \log ^2\left(\frac{Q^2}{\mu_F^2}\right) f_i-2
   \beta_1 C^{(1)}\otimes f_j f_i\\
   &+\frac{1}{4} P^{(3)}\otimes f_j f_i+\frac{1}{2} C^{(1)}\otimes P^{(2)}\otimes f_j f_i+C^{(2)}\otimes P^{(1)}\otimes f_j f_i\\
   &+\frac{1}{4}
   P^{(1)}\otimes P^{(2)}\otimes f_j \log \left(\frac{Q^2}{\mu_F^2}\right) f_i+\frac{1}{4} P^{(2)}\otimes P^{(1)}\otimes f_j \log \left(\frac{Q^2}{\mu_F^2}\right) f_i\\
   &+\frac{1}{2}
   C^{(1)}\otimes P^{(1)}\otimes P^{(1)}\otimes f_j \log \left(\frac{Q^2}{\mu_F^2}\right) f_i-\beta_1 P^{(1)}\otimes f_j \log \left(\frac{Q^2}{\mu_R^2}\right)
   f_i\\
   &+\frac{1}{8} P^{(1)}\otimes P^{(1)}\otimes P^{(1)}\otimes f_i \log ^2\left(\frac{Q^2}{\mu_F^2}\right) f_j-2 \beta_1 C^{(1)}\otimes f_i f_j\\
   &+\frac{1}{4} P^{(3)}\otimes f_i
   f_j+\frac{1}{2} C^{(1)}\otimes P^{(2)}\otimes f_i f_j\\
   &+C^{(2)}\otimes P^{(1)}\otimes f_i f_j+\frac{1}{4} P^{(1)}\otimes P^{(2)}\otimes f_i \log \left(\frac{Q^2}{\mu_F^2}\right)
   f_j+\frac{1}{4} P^{(2)}\otimes P^{(1)}\otimes f_i \log \left(\frac{Q^2}{\mu_F^2}\right) f_j\\
   &+\frac{1}{2} C^{(1)}\otimes P^{(1)}\otimes P^{(1)}\otimes f_i \log
   \left(\frac{Q^2}{\mu_F^2}\right) f_j-\beta_1 P^{(1)}\otimes f_i \log \left(\frac{Q^2}{\mu_R^2}\right) f_j\\
   &+\frac{1}{2} P^{(1)}\otimes f_i P^{(1)}\otimes f_j \log
   ^2\left(\frac{Q^2}{\mu_F^2}\right) \beta_0-4 C^{(1)}\otimes f_i C^{(1)}\otimes f_j \beta_0\\
   &-C^{(1)}\otimes f_j P^{(1)}\otimes f_i \log \left(\frac{Q^2}{\mu_F^2}\right) \beta_0-C^{\text{(1)}}\otimes f_i P^{\text{(1)}}\otimes f_j \log \left(\frac{Q^2}{\mu_F^2}\right) \beta_0\\
   &-2 C^{\text{(1)}}\otimes f_j P^{\text{(1)}}\otimes f_i \log \left(\frac{Q^2}{\mu_R^2}\right) \beta_0-2
   C^{\text{(1)}}\otimes f_i P^{\text{(1)}}\otimes f_j \log \left(\frac{Q^2}{\mu_R^2}\right) \beta_0 \\
   &-2 P^{\text{(1)}}\otimes f_i P^{\text{(1)}}\otimes f_j \log \left(\frac{Q^2}{\mu_F^2}\right) \log
   \left(\frac{Q^2}{\mu_R^2}\right) \beta_0\\
   &+\frac{1}{4} P^{\text{(1)}}\otimes P^{\text{(1)}}\otimes f_j \log ^2\left(\frac{Q^2}{\mu_F^2}\right) f_i \beta_0-4 C^{\text{(2)}}\otimes f_j f_i \beta_0-C^{\text{(1)}}\otimes P^{\text{(1)}}\otimes f_j \log \left(\frac{Q^2}{\mu_F^2}\right) f_i \beta_0\\
   &-P^{\text{(2)}}\otimes f_j \log \left(\frac{Q^2}{\mu_R^2}\right) f_i \beta_0-2
   C^{\text{(1)}}\otimes P^{\text{(1)}}\otimes f_j \log \left(\frac{Q^2}{\mu_R^2}\right) f_i \beta_0
   \end{align*}
   \begin{align*}
&-P^{\text{(1)}}\otimes P^{\text{(1)}}\otimes f_j \log \left(\frac{Q^2}{\mu_F^2}\right) \log
   \left(\frac{Q^2}{\mu_R^2}\right) f_i \beta_0+\frac{1}{4} P^{\text{(1)}}\otimes P^{\text{(1)}}\otimes f_i \log ^2\left(\frac{Q^2}{\mu_F^2}\right) f_j \beta_0\\
   &-4 C^{\text{(2)}}\otimes f_i f_j \beta_0-C^{\text{(1)}}\otimes P^{\text{(1)}}\otimes f_i \log \left(\frac{Q^2}{\mu_F^2}\right) f_j \beta_0-P^{\text{(2)}}\otimes f_i \log \left(\frac{Q^2}{\mu_R^2}\right) f_j \beta_0\\
   &-2
   C^{\text{(1)}}\otimes P^{\text{(1)}}\otimes f_i \log \left(\frac{Q^2}{\mu_R^2}\right) f_j \beta_0-P^{\text{(1)}}\otimes P^{\text{(1)}}\otimes f_i \log \left(\frac{Q^2}{\mu_F^2}\right) \log
   \left(\frac{Q^2}{\mu_R^2}\right) f_j \beta_0\\
   &+\left(B^{(1)}\right) \left(\frac{1}{2} P^{\text{(1)}}\otimes f_i P^{\text{(1)}}\otimes f_j \log ^2\left(\frac{Q^2}{\mu_F^2}\right)+\frac{1}{4}
   P^{\text{(1)}}\otimes P^{\text{(1)}}\otimes f_j f_i \log ^2\left(\frac{Q^2}{\mu_F^2}\right) \right.\\
   &\left. +\frac{1}{4} P^{\text{(1)}}\otimes P^{\text{(1)}}\otimes f_i f_j \log
   ^2\left(\frac{Q^2}{\mu_F^2}\right)+\frac{1}{2} P^{\text{(1)}}\otimes f_j f_i \beta_0 \log ^2\left(\frac{Q^2}{\mu_F^2}\right)\right.\\
   &\left.+\frac{1}{2} P^{\text{(1)}}\otimes f_i f_j \beta_0 \log
   ^2\left(\frac{Q^2}{\mu_F^2}\right)+\frac{1}{2} P^{\text{(2)}}\otimes f_j f_i \log \left(\frac{Q^2}{\mu_F^2}\right)\right. \\
   &\left.+C^{\text{(1)}}\otimes f_j P^{\text{(1)}}\otimes f_i \log \left(\frac{Q^2}{\mu_F^2}\right)+C^{\text{(1)}}\otimes f_i P^{\text{(1)}}\otimes f_j \log
   \left(\frac{Q^2}{\mu_F^2}\right)\right.\\
   &\left.+C^{\text{(1)}}\otimes P^{\text{(1)}}\otimes f_j f_i \log
   \left(\frac{Q^2}{\mu_F^2}\right)+C^{\text{(1)}}\otimes P^{\text{(1)}}\otimes f_i f_j \log
   \left(\frac{Q^2}{\mu_F^2}\right)\right.\\
   &\left.+\frac{1}{2} P^{\text{(2)}}\otimes f_i f_j \log \left(\frac{Q^2}{\mu_F^2}\right)+2 C^{\text{(1)}}\otimes f_i C^{\text{(1)}}\otimes f_j\right.\\
   &\left.-2 P^{\text{(1)}}\otimes f_j \log \left(\frac{Q^2}{\mu_R^2}\right) f_i \beta_0 \log \left(\frac{Q^2}{\mu_F^2}\right)-2 P^{\text{(1)}}\otimes f_i \log
   \left(\frac{Q^2}{\mu_R^2}\right) f_j \beta_0 \log \left(\frac{Q^2}{\mu_F^2}\right)\right.\\
   &\left.+2 C^{\text{(2)}}\otimes f_j f_i+2 C^{\text{(2)}}\otimes f_i f_j-4
   C^{\text{(1)}}\otimes f_j \log \left(\frac{Q^2}{\mu_R^2}\right) f_i \beta_0-4 C^{\text{(1)}}\otimes f_i \log \left(\frac{Q^2}{\mu_R^2}\right) f_j \beta_0\right)\\
   &+\left(A^{(1)}\right)^2
   \left(4 C^{\text{(1)}}\otimes f_j \zeta (3) f_i+2 P^{\text{(1)}}\otimes f_j \log \left(\frac{Q^2}{\mu_F^2}\right) \zeta (3) f_i+4 C^{\text{(1)}}\otimes f_i f_j \zeta (3)\right. \\
   &\left.+2 P^{\text{(1)}}\otimes f_i \log
   \left(\frac{Q^2}{\mu_F^2}\right) f_j \zeta (3)\right) \\
   &+\left(A^{(1)}\right) \left(-4 P^{\text{(1)}}\otimes f_i \zeta (3) P^{\text{(1)}}\otimes f_j+\frac{28}{3} f_i \beta _0 \zeta (3)
   P^{\text{(1)}}\otimes f_j\right.\\
   &\left.+\left(B^{(1)}\right) \left(-8 P^{\text{(1)}}\otimes f_j \zeta (3) f_i-8 P^{\text{(1)}}\otimes f_i f_j \zeta (3)\right)-2 P^{\text{(1)}}\otimes P^{\text{(1)}}\otimes f_j f_i \zeta (3)\right.\\
   &\left.-2
   P^{\text{(1)}}\otimes P^{\text{(1)}}\otimes f_i f_j \zeta (3)+\frac{28}{3} P^{\text{(1)}}\otimes f_i f_j \beta_0 \zeta (3)\right)\\
   &+f_i f_j \left(-4 \log \left(\frac{Q^2}{\mu_R^2}\right) \beta_0
   B^{\text{(2)}}+2 B^{\text{(3)}}\right. \\
   &\left.+\left(B^{(1)}\right) \left(2 \log ^2\left(\frac{Q^2}{\mu_R^2}\right) \beta_0^2-2 \beta_1 \log
   \left(\frac{Q^2}{\mu_R^2}\right)\right)+\left(A^{(1)}\right) \left(-8 \zeta (3) \left(B^{(1)}\right)^2\right.\right.\\
   &\left.\left.+\frac{56}{3} \beta _0 \zeta (3) \left(B^{(1)}\right)-8 \beta _0^2 \zeta
   (3)\right)-12 \left(A^{(1)}\right)^3 \zeta (5)+8 A^{\text{(1)}+\text{(2)}} \zeta (3)\right.\\
   &\left.-8 \left(A^{(1)}\right)^2 \log \left(\frac{Q^2}{\mu_R^2}\right) \beta_0 \zeta (3)\right), \\
\end{align*}
\section{Fits to the SIYY form}
\label{app:SIYY}

 \begin{table}[h]
 	\centering
 	\begin{tabular}{|c|c|}
 		\hline
 		Parameter  & SIYY fit values \\
 		\hline
 		\hline
 		$g_1$  & 0.0 $\pm$ 0.004 \\
 		$g_2$  & 0.773 $\pm$ 0.042 \\
 		$g_3$  & 0.122 $\pm$ 0.006 \\
 		\hline
 	\end{tabular}
 	\caption{The values of the fitted parameters $(g_1, g_2, g_3)$ in the SIYY form, when fitting to the complete set of experimental data listed in Table~\ref{tab:nonpert_datasets}. 
 	}
 	\label{tab:SIYY_BestFit}
 \end{table}
 
Here, we summarize the fitted values of the non-perturbative parameters when using the SIYY form. As shown in Ref.~\cite{Sun:2014dqm}, the non-perturbative Sudakov factor was introduced as 
\begin{equation}
    {S_{SIYY}}= g_1 b^2 + g_2 \ln(b/b_*)\ln(Q/Q_0) + g_3 b^2 \left((x_0/x_1)^\lambda+(x_0/x_2)^\lambda\right),
\label{eq:SIYY}
\end{equation}
where $Q_0^2 = 2.4$~GeV$^2$, $x_0=0.01$, $\lambda=0.2$, $x_1=\frac{Q}{\sqrt{s}} e^y$ and $x_2=\frac{Q}{\sqrt{s}} e^{-y}$. Furthermore, the $b_{max}$ is set to 1.5~GeV$^{-1}$, same as the value used in the original SIYY fit~\cite{Sun:2014dqm}. 

Table~\ref{tab:SIYY_BestFit} lists the 
values of the fitted parameters of the SIYY form, when fitting to the complete set of experimental data listed in Table~\ref{tab:nonpert_datasets}. We note that a non-zero value of the $g_3$ parameter in the SIYY form implies that the fitted data set favors a rapidity-dependent non-perturbative function. 
The quality of the fits for each experiment included in the SIYY fit is summarized in Table~\ref{tab:chi2}.
Furthermore, the correlation matrix of the fitted parameters $(g_1, g_2, g_3)$ is found to be  
\begin{equation}
    C = 
    \begin{pmatrix}
    1 & 0.198 & -0.748 \\
    0.198 & 1 & -0.708  \\
    -0.748 & -0.708 & 1 \\
    \end{pmatrix}\,.
\end{equation}

\bibliographystyle{apsrev4-1}
\bibliography{bibliography}

\end{document}